\documentclass[showpacs,aps,prd,reprint,superscriptaddress,nofootinbib,longbibliography]{revtex4-2}
\usepackage[colorlinks=true, pdfstartview=FitV,
bookmarks=true, bookmarksnumbered=true, breaklinks]{hyperref}
\usepackage[dvipdfmx]{graphicx}
\usepackage{amsmath,amssymb,bm,color,longtable,mathrsfs,slashed,comment,tikz}
\usepackage{soul}

\definecolor{darkviolet}{rgb}{0.58, 0.0, 0.83}
\definecolor{electricultramarine}{rgb}{0.25, 0.0, 1.0}
\definecolor{brightpink}{rgb}{1.0, 0.0, 0.5}
\hypersetup{linkcolor=brightpink, citecolor=darkviolet, urlcolor=electricultramarine}

\definecolor{lime}{HTML}{A6CE39}
\DeclareRobustCommand{\orcidicon}{
	\hspace{-3mm}
	\begin{tikzpicture}
	\draw[lime, fill=lime] (0,0) 
	circle [radius=0.16] 
	node[white] {{\fontfamily{qag}\selectfont \tiny ID}};
	\draw[white, fill=white] (-0.0625,0.095) 
	circle [radius=0.007];
	\end{tikzpicture}
	\hspace{-3mm}
}

\foreach \x in {A, ..., Z}{\expandafter\xdef\csname orcid\x\endcsname{\noexpand\href{https://orcid.org/\csname orcidauthor\x\endcsname}
			{\noexpand\orcidicon}}
}

\begin{document}

\title{Casimir effect in magnetic dual chiral density waves}

\author{Daisuke Fujii\orcidA{}}
\email[]{daisuke@rcnp.osaka-u.ac.jp}
\affiliation{Advanced Science Research Center, Japan Atomic Energy Agency (JAEA), Tokai, 319-1195, Japan}
\affiliation{Research Center for Nuclear Physics, Osaka University, Ibaraki 567-0048, Japan}

\author{Katsumasa~Nakayama\orcidB{}}
\email[]{katsumasa.nakayama@riken.jp}
\affiliation{RIKEN Center for Computational Science, Kobe, 650-0047, Japan}

\author{Kei~Suzuki\orcidC{}}
\email[]{k.suzuki.2010@th.phys.titech.ac.jp}
\affiliation{Advanced Science Research Center, Japan Atomic Energy Agency (JAEA), Tokai, 319-1195, Japan}

\begin{abstract}
We theoretically investigate the Casimir effect originating from Dirac fields in finite-density matter under a magnetic field.
In particular, we focus on quark fields in the magnetic dual chiral density wave phase as a possible inhomogeneous ground state of interacting Dirac-fermion systems.
In this system, the distance dependence of Casimir energy shows a complex oscillatory behavior by the interplay between the chemical potential, magnetic field, and inhomogeneous ground state.
By decomposing the total Casimir energy into contributions of each Landau level, we elucidate what types of Casimir effects are realized from each Landau level: The lowest or some types of higher Landau levels lead to different behaviors of Casimir energies.
Furthermore, we point out characteristic behaviors due to level splitting between different fermion flavors, i.e., up and down quarks.
These findings provide new insights into Dirac-fermion (or quark) matter with a finite thickness.
\end{abstract}

\maketitle

%\tableofcontents
\section{Introduction}

The Casimir effect proposed by Casimir~\cite{Casimir:1948dh} is of great importance for understanding small-volume physics in quantum field theory.
Casimir predicted that a reduction in the zero-point energy of the photon field by two parallel conducting plates in a vacuum would induce an attractive force, known as the Casimir energy or Casimir force.
This theoretical prediction was experimentally confirmed several decades later~\cite{Lamoreaux:1996wh,Bressi:2002fr} (for reviews, see Refs.~\cite{Plunien:1986ca,Mostepanenko:1988bs,Bordag:2001qi,Milton:2001yy,Klimchitskaya:2009cw,Woods:2015pla,Gong:2020ttb,Lu:2021jvu}).

While the original Casimir effect means an attractive force from the photon field in a vacuum, other types of Casimir (or Casimir-like) effects have also been explored.
For example, one can consider counterparts induced by fermion fields~\cite{Johnson:1975zp,Mamaev:1980jn} such as quarks or in systems filled by a medium~\cite{Dzyaloshinskii:1961sfr}.
Such an unusual setup sometimes leads to anomalous phenomena, such as sign-flipping and oscillating behaviors of Casimir energy as a function of distance.
In particular, in fermionic systems,  (i) external magnetic fields\footnote{
Since the photon is not directly coupled to an external magnetic field, the magnetic response of the photonic Casimir effect in quantum electrodynamics (QED) vacuum is described as a higher-order correction due to electron-positron loops~\cite{Robaschik:1986vj}.
On the other hand, the Casimir effect originated from Dirac fields~\cite{Cougo-Pinto:1998jwo,Cougo-Pinto:2001kks,Elizalde:2002kb,Ostrowski:2005rm,Miltao:2008zza,Sitenko:2014kza,Sitenko:2015eza,Sitenko:2015wzd,Nakayama:2022fvh,Rohim:2023tmy,Erdas:2023wzy,Flachi:2024ztd} (as well as charged scalar fields~\cite{Cougo-Pinto:1998jun,Cougo-Pinto:1998mdw,Cougo-Pinto:1998fpo,Cougo-Pinto:1998zge,Elizalde:2002kb,Ostrowski:2005rm,Erdas:2013jga,Erdas:2013dha,Sitenko:2014wwa,Sitenko:2015eza,Erdas:2015yac,Erdas:2020ilo,Haridev:2021jwi, Erdas:2021xvv,Erdas:2024gkq}) is directly affected by magnetic fields.
Therefore, quark fields which we consider in this work can be a realistic testing ground for the Casimir effect coupled to magnetic fields.} and (ii) chemical potentials can be experimentally tunable, so that these parameters should be useful for the controllability and versatility of the fermionic Casimir effect.

\begin{figure}[b!]
    \centering
    \begin{minipage}[t]{1.0\columnwidth}
    \includegraphics[clip,width=1.0\columnwidth]{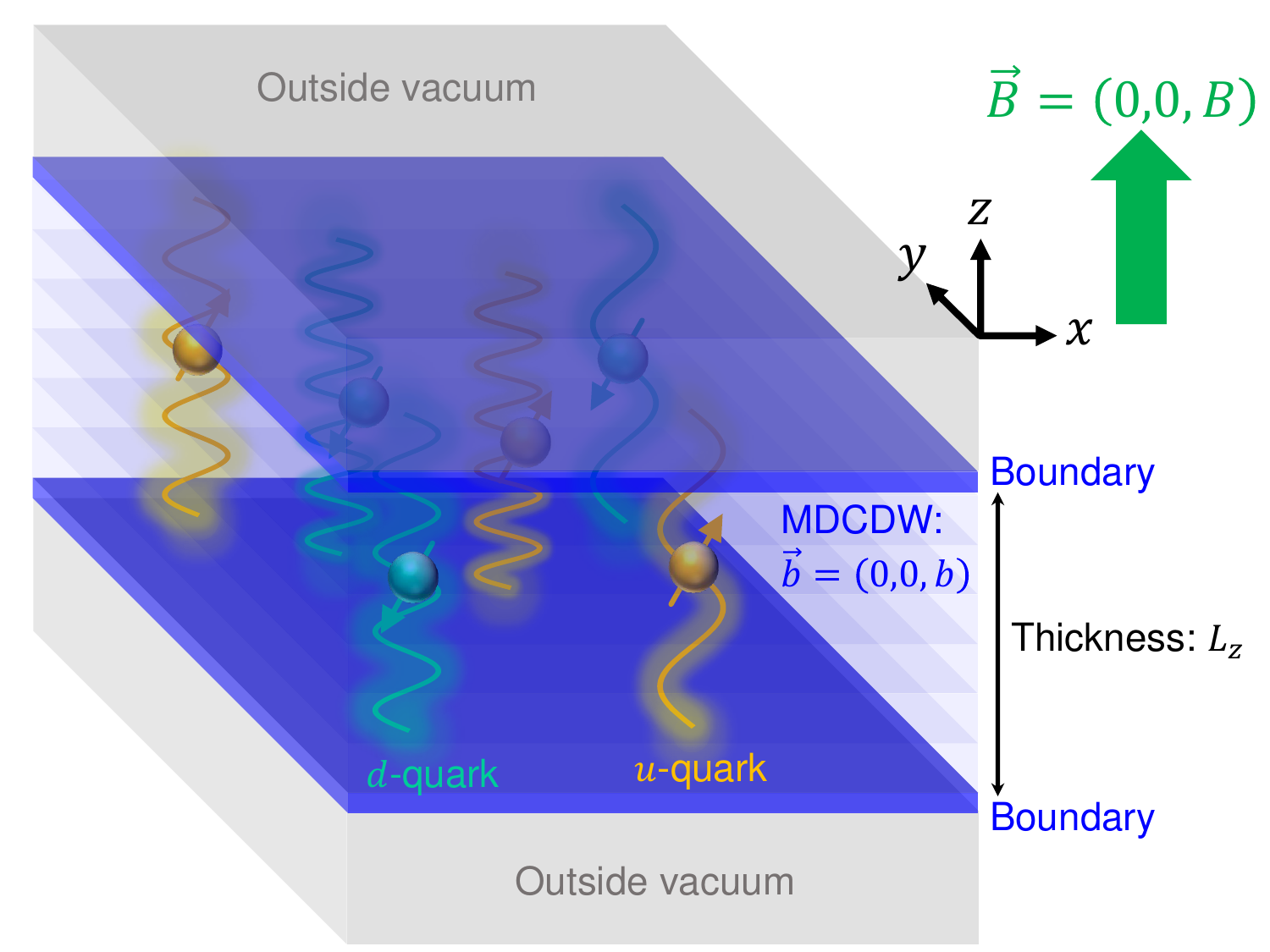}
    \end{minipage}
\caption{
Schematic picture of the Casimir effect in the two-flavor MDCDW phase, where the MDCDW phase is sandwiched by two boundary conditions at $z=0$ and $z=L_z$.
The magnetic field $\vec{B}$ and the wave number  $\vec{b}$ of density waves are parallel to the $z$ direction.
}
\label{fig:schematic}
\end{figure}

In this paper, we investigate the Casimir effect for Dirac fields under both (i) and (ii) for the first time, where we particularly focus on thin quark matter under a magnetic field (see Fig.~\ref{fig:schematic}).
Our previous work~\cite{Fujii:2024fzy} found that an {\it oscillating Casimir effect} is induced not only by the quark Fermi sea, but also by the dynamics of quark fields modified in the {\it dual chiral density wave} (DCDW) phase~\cite{Dautry:1979bk,Tatsumi:2004dx,Nakano:2004cd} which is a possible ground state of quantum chromodynamics (QCD) at finite quark chemical potential.
However, the DCDW phase (more generally, an inhomogeneous phase) can exist only in a narrow density region.
Therefore, in this study, we consider a more feasible situation, that is, the DCDW phase under a magnetic field.
When a magnetic field is imposed on the DCDW phase, the property of the DCDW is modified, which may be called the {\it magnetic dual chiral density wave} (MDCDW) \cite{Frolov:2010wn} (see Refs.~\cite{Tatsumi:2014wka,Nishiyama:2015fba,Carignano:2015kda,Ferrer:2015iop,Ferrer:2016toh,Abuki:2018iqp,Ferrer:2019zfp,Feng:2020vxn,Ferrer:2020ulz,Ferrer:2021vuy,Anzuini:2021gwi,Gyory:2022hnv,Ferrer:2022afu,Ghalati:2023npr,Ferrer:2023xvl,Ferrer:2024xwu} for related studies and Ref.~\cite{Ferrer:2021mpq} for a review).
A magnetic field makes the DCDW phase more robust, and, in particular, at finite temperature, it removes the Landau-Peierls instability~\cite{Ferrer:2019zfp}.
Such robustness might support the formation of DCDW in fireballs produced by heavy-ion collision experiments and in the interior of neutron stars. 

This paper is organized as follows.
In Sec.~\ref{Sec:model}, we review the MDCDW phase described by an effective model of interacting Dirac fermions and explain how to calculate Casimir energy in this phase.
In Sec.~~\ref{Sec:result}, we show our results.
Section~\ref{Sec:summary} is devoted to the conclusion.

\section{Model construction} \label{Sec:model}
\subsection{The model}
To investigate quark matter in a magnetic field, we use the Nambu--Jona-Lasinio (NJL) model~\cite{Nambu:1961tp,Nambu:1961fr}, which is an effective model of QCD (see Refs.~\cite{Vogl:1991qt,Klevansky:1992qe,Hatsuda:1994pi,Buballa:2003qv} for reviews).
The Lagrangian density of the NJL model in a magnetic field is written as
\begin{align}
    \mathcal{L}_{\rm NJL}=\bar\psi(i\slashed{D}+\mu\gamma_0)\psi+G[(\bar\psi\psi)^2+(\bar\psi i\gamma_5\vec{\tau}\psi)^2], \label{eq:NJL}
\end{align}
where the quark field $\psi$ has two-flavor ($N_f=2$) and three-color ($N_c=3$) components and $\mu>0$ is the chemical potential of quarks.
$G$ is the coupling constant for the four-point interactions, $\vec{\tau}$ is the Pauli matrix in the isospin (up or down quark) space, and $\gamma^\mu$ and $\gamma_5 \equiv i\gamma^0\gamma^1\gamma^2\gamma ^3$ are the gamma matrix in the $3+1$-dimensional spacetime.
The covariant derivative is defined as $\slashed{D} \equiv \gamma^\mu(\partial_\mu+iQA_\mu)$ using the gauge field $A_\mu$ and the electric-charge matrix $Q \equiv {\rm diag}(q_u,q_d) = {\rm diag}(\frac{2}{3}e,-\frac{1}{3}e)$ with the elementary charge $e>0$.
Note that, in a vanishing electromagnetic field, the model~(\ref{eq:NJL}) satisfies the $SU(2)_L\times SU(2)_R$ chiral symmetry.
When an external magnetic field is switched on, this symmetry is broken to $U(1)_L\times U(1)_R$.

As the mean-field ansatz for the DCDW phase, we adopt 
\begin{align}
\begin{aligned}
    &\langle \bar{\psi}\psi \rangle=\Delta\cos(\vec{q}\cdot\vec{r}),
    &&\langle \bar{\psi}i\gamma_5 \tau_3 \psi\rangle=\Delta\sin(\vec{q}\cdot\vec{r}), \\
    &\langle \bar{\psi}i\gamma_5 \tau_1 \psi\rangle=0,
    &&\langle \bar{\psi}i\gamma_5 \tau_2 \psi\rangle=0,
\end{aligned} \label{eq:MFansatz}
\end{align}
where $\Delta$, $\vec{q}=(0,0,q)$, and $\vec{r}=(x,y,z)$ are the amplitude of DCDW, the wave number of DCDW propagating in the $z$ direction, and the position vector, respectively.

Using the mean-field ansatz~(\ref{eq:MFansatz}), we obtain the mean-field Lagrangian
\begin{align}
    \mathcal{L}_{\rm MF}=\bar\psi[i \slashed{D}+\mu\gamma_0-M(\cos {qz}+i\gamma_5\tau_3\sin{qz})]\psi-\frac{M^2}{4G},
\end{align}
where $M=-2G\Delta$.

By performing a local chiral transformation
\begin{align}
    \psi\rightarrow e^{i\gamma_5\tau_3qz/2}\psi , \ \ \ \bar\psi\rightarrow \bar\psi e^{i\gamma_5\tau_3qz/2}
\end{align}
(called the Weinberg transformation), we can eliminate the position dependence in the Lagrangian. The Lagrangian is then rewritten as 
\begin{align}
    \mathcal{L}_{\rm MF}=\bar\psi(i\slashed{D}+\mu\gamma_0-M+\gamma_5\tau_3\gamma^\mu q_\mu/2)\psi-\frac{M^2}{4G}. \label{eq:L_MF}
\end{align}
Assuming also that the background electromagnetic field is $A_\mu=(0,0,Bx,0)$ using the Landau gauge, we obtain a magnetic field $\vec{B}$ parallel to $\vec{q}$ (see Fig.~\ref{fig:schematic}).

From the Lagrangian (\ref{eq:L_MF}), by diagonalizing the inverse of the quark propagator in momentum space, we obtain the following energy eigenvalues~\cite{Frolov:2010wn}:
\begin{align}
    &\omega_{l=0}=E_{l=0}+b-\mu,
    \ \ \ \tilde{\omega}_{l=0}=-E_{l=0}+b-\mu, \label{EV_LLLs}\\
    &\omega_{\zeta,l}=E_{\zeta,l}-\mu, \ \ \ \tilde{\omega}_{\zeta,l}=-E_{\zeta,l}-\mu, \label{EV_HLLs}\\
    &E_{l=0} \equiv \sqrt{M^2+k_z^2}, \notag \\
    &E_{\zeta,l} \equiv \sqrt{(\zeta\sqrt{M^2+k^2_z}+b)^2+2|q_fB|l} \ \ \ (l=1,2,\dots), \notag
\end{align}
where we redefined $b_\mu \equiv q_\mu/2 =  (0,0,0,q/2) = (0,0,0,b)$.
$\omega$ and $\tilde{\omega}$ denote the positive and negative energy solutions (at $\mu=0$), respectively.
$\zeta=\pm$ is a spin polarization index, and $q_f =q_u$ or $q_d$.
The presence of a magnetic field splits the eigenmodes of quarks into an infinite number of Landau levels (LLs) labeled by $l$.
The two modes with $l=0$, Eq.~\eqref{EV_LLLs}, are called lowest Landau levels (LLLs), and an infinite set of four modes with $l\geq1$, Eq.~\eqref{EV_HLLs}, is called higher Landau levels (HLLs).
The LLLs have no spin index because only one spin component is chosen.\footnote{In the LLLs, Eq.~(\ref{EV_LLLs}), the asymmetry between $\omega_{l=0}$ and $\tilde{\omega}_{l=0}$ due to $b\neq 0$ is called the {\it spectral asymmetry}~\cite{Frolov:2010wn,Tatsumi:2014wka}.}
On the other hand, HLLs have both modes with $\zeta=\pm$.

\subsection{Casimir energy}

From the partition function $Z$ of the mean-field Lagrangian (\ref{eq:L_MF}), the grand potential $\Omega \equiv -\frac{T}{V}\ln Z$ per unit volume $V=L_xL_yL_z$ at temperature $T=1/\beta$ is written as~\cite{Frolov:2010wn}
\begin{align}
    &\Omega=\Omega_{\rm LLL}+\Omega_{\rm HLL}+\frac{M^2}{4G}, \label{eq:omega} \\
    &\Omega_{\rm LLL}=-N_c \sum_{q_f}\frac{|q_f B|}{2\pi}\int\frac{dk_z}{2\pi} \Big[\frac{1}{2}|\omega_{l=0}|+\frac{1}{2}|\tilde{\omega}_{l=0}| \notag \\
    &\hspace{13mm}+\frac{1}{\beta} \ln\left\{\left(1+e^{-\beta|\omega_{l=0}|}\right)\left(1+e^{-\beta|\tilde{\omega}_{l=0}|}\right)\right\}\Big], \notag \\
    &\Omega_{\rm HLL}=-N_c \sum_{q_f,\zeta}\frac{|q_f B|}{2\pi}\int\frac{dk_z}{2\pi}  \sum_{l=1}^\infty \Big[\frac{1}{2}|\omega_{\zeta,l}|+\frac{1}{2}|\tilde{\omega}_{\zeta,l}| \notag \\
    &\hspace{13mm}+\frac{1}{\beta} \ln\left\{\left(1+e^{-\beta|\omega_{\zeta,l}|}\right)\left(1+e^{-\beta|\tilde{\omega}_{\zeta,l}|}\right)\right\}\Big], \notag
\end{align}
where $|q_fB|/2\pi$ is called the Landau degeneracy factor which is the remnant of the ($k_x$, $k_y$) integrals at $B=0$.
In the zero-temperature limit, we obtain~\cite{Frolov:2010wn}
\begin{align}
    &\Omega(T\rightarrow0)=\Omega_{\rm LLL}^{T\rightarrow 0}+\Omega_{\rm HLL}^{T\rightarrow 0}+\frac{M^2}{4G} \equiv \frac{E_0^{\rm int}}{L_z}, \label{E0int} \\
    &\Omega_{\rm LLL}^{T\rightarrow 0}=-N_c\sum_{q_f}\frac{|q_f B|}{2\pi}\int\frac{dk_z}{2\pi}\left(\frac{1}{2}|\omega_{l=0}|+\frac{1}{2}|\tilde{\omega}_{l=0}|\right), \notag \\
    &\Omega_{\rm HLL}^{T\rightarrow 0}=-N_c\sum_{q_f,\zeta}\frac{|q_f B|}{2\pi}\int\frac{dk_z}{2\pi}\sum_{l=1}^\infty\left(\frac{1}{2}|\omega_{\zeta,l}|+\frac{1}{2}|\tilde{\omega}_{\zeta,l}|\right), \notag 
\end{align}
where we defined a new notation $E_0^\mathrm{int}$ as the zero-point energy per unit area $L_xL_y$ (not per unit volume).

The Casimir energy appears as a finite volume effect of the zero-point energy.
In this work, we impose the periodic boundary conditions (PBCs) at $z=0$ and $z=L_z$ (see Fig.~\ref{fig:schematic}).
Then, the momentum in the $z$ direction is discretized as $k_z\rightarrow 2n\pi/L_z,\ (n=0,\pm1,\dots ,\pm \infty)$, the momentum integral of Eq.~\eqref{E0int} is replaced by the corresponding momentum sum.
Thus, the zero-point energy at finite $L_z$ is written as 
\begin{align}
    &E_0^{\rm sum}\equiv E_0^{\rm int}(k_z\rightarrow \frac{2n\pi}{L_z})=E_{0,{\rm LLL}}^{\rm sum}+E_{0,{\rm HLL}}^{\rm sum}+\frac{M^2}{4G}L_z, \label{E0sum} \\
    &E_{0,{\rm LLL}}^{\rm sum}=-N_c\sum_{q_f}\frac{|q_f B|}{2\pi}\sum_{n=-\infty}^\infty\Big(\frac{1}{2}|\omega_{l=0,n}|+\frac{1}{2}|\tilde{\omega}_{l=0,n}|\Big), \notag \\
    &E_{0,{\rm HLL}}^{\rm sum}=-N_c\sum_{q_f,\zeta}\frac{|q_f B|}{2\pi}\sum_{l=1}^\infty\sum_{n=-\infty}^\infty\Big(\frac{1}{2}|\omega_{\zeta,l,n}|+\frac{1}{2}|\tilde{\omega}_{\zeta,l,n}|\Big). \notag
\end{align}

In this work, unlike the conventional mean-field approach in the NJL model, we do not minimize the thermodynamic potential (or solve the gap equation) at finite $L_z$.\footnote{In the NJL model at finite $(B,L_z)$~\cite{Ferrer:1999gs,Abreu:2019czp,Abreu:2020uxc,Abreu:2021btt,Abreu:2022cgm,Correa:2023ebh}, it is known that when $L_z$ is sufficiently short, the PBC tends to enhance the chiral symmetry breaking (i.e., the order parameter $M$), whereas the antiperiodic boundary condition (APBC) suppresses it.
Therefore, we naively expect the enhancement of the DCDW phase by the PBC and its suppression by the APBC.
At least, at a sufficiently long $L_z$ for both the boundaries, the DCDW phase should survive, and a magnetic field also enhances the DCDW phase.
Furthermore, on the phase diagram at finite $(\mu,L_z)$~\cite{Ebert:2010eq}, phase boundaries between different phases (e.g., the $M \neq0 $ and $M =0 $ phases) oscillate.
Therefore, also at finite $(\mu,B,L_z)$, we can expect oscillations of phase boundaries between the DCDW and other phases.}
Instead, we fix the values of the order parameters and then investigate what types of behaviors of the Casimir energy can be realized under a parameter set.
Note that the Casimir energy appears also from the $L_z$ dependence of the term with $M^2/4G$, but it will be neglected in our definition of Casimir energy.
This is because this contribution is just the free energy shift by the $L_z$ dependence of the order parameter $M$ and is not directly regarded as the fermionic Casimir effect.

The infinite sum (\ref{E0sum}) contains an ultraviolet divergence, but it becomes finite by using a regularization scheme.
Here we apply the Lifshitz formula~\cite{Lifshitz:1956zz}, which was first proposed for the conventional photonic Casimir effect and is well known nowadays.
Our previous works established analogous formulas for the DCDW-type dispersion relation ($M\neq0,b\neq0$)~\cite{Fujii:2024fzy} and at finite chemical potential ($\mu \neq 0$)~\cite{Fujii:2024ixq}.
Using this formula, the Casimir energy for fermion fields in the MDCDW phase (i.e., $M\neq0,b\neq0,\mu \neq 0, B\neq 0$) is written as\footnote{If $b=\mu=0$, $M \neq0$, and $B\neq0$, Eq.~(\ref{1dimLifshitz}) is equivalent to the known formula for the massive Dirac field in a magnetic field~\cite{Cougo-Pinto:1998jwo} obtained from the proper-time regularization, except for the color and flavor factors.}
\begin{align}
    E_\mathrm{Cas} =&-2N_c\int_{-\infty}^\infty \frac{d\xi}{2\pi} \sum_{q_f}\frac{|q_f B|}{2\pi} \ln \left[1-e^{-L_z \tilde{k}^{[l=0]}_z} \right] \notag \\ 
    &-2N_c\int_{-\infty}^\infty \frac{d\xi}{2\pi} \sum_{q_f,\zeta}\sum_{l=1}^\infty\frac{|q_f B|}{2\pi} \ln \left[1-e^{-L_z \tilde{k}^{[l,\zeta]}_z} \right],\label{1dimLifshitz} \\ 
    \tilde{k}^{[l=0]}_z =& \sqrt{M^2-\left(b-\mu-i\xi\right)^2}, \notag \\
    \tilde{k}^{[l,\zeta]}_z =& \sqrt{M^2-\Big(b+\zeta\sqrt{(i\xi+\mu)^2-2|q_fB|l}\Big)^2}, \notag 
\end{align}
where the overall factor of 2 means the factor from the PBCs.
The integration variable $\xi$ is the imaginary part of the imaginary energy $i\xi$.
The first and second terms come from the contributions of the LLLs and the HLLs, respectively. 
This formula is regarded as the infinite sum of a (quasi-)one-dimensional analog of the Lifshitz formula because of the absence of the transverse-momentum integral.

Also, taking the $B\to0$ limit of Eq.~(\ref{1dimLifshitz}) leads to the Casimir energy in the usual DCDW phase (i.e., $M\neq0,b\neq0,\mu \neq 0$),\footnote{Note that Eq.~(\ref{3dimLifshitz}) was not given in our previous paper~\cite{Fujii:2024fzy}.
By substituting $\mu=0$ into Eq.~(\ref{3dimLifshitz}), we obtain Eq.~(11) in Ref.~\cite{Fujii:2024fzy}.} 
\begin{align}
    &E_\mathrm{Cas}(B\rightarrow0) = \notag \\ 
    &-2N_fN_c \int_{-\infty}^\infty \frac{d\xi}{2\pi} \sum_{\zeta=\pm}\int \frac{dk_xdk_y}{(2\pi)^2} \ln \left[1-e^{-L_z \tilde{k}_z^{[\zeta]}} \right], \label{3dimLifshitz}  \\ 
    &\tilde{k}_z^{[\zeta]} = \sqrt{M^2-\Big(b+\zeta\sqrt{(i\xi+\mu)^2-k_x^2-k_y^2}\Big)^2}. \notag 
\end{align}

Here, we comment on the relationship between Eqs.~(\ref{E0sum}) and (\ref{1dimLifshitz}).
Although Eq.~(\ref{E0sum}) contains both the vacuum and finite-density parts, they can be rewritten separately~\cite{Frolov:2010wn}: e.g., $E_{0,\mathrm{LLL}}^\mathrm{sum}=E^\mathrm{vac}_\mathrm{LLL}+E^\mathrm{den}_\mathrm{LLL}$.
From the vacuum parts $E^\mathrm{vac}_\mathrm{LLL/HLL}$, we can derive the corresponding Lifshitz formula, i.e., Eq.~(\ref{1dimLifshitz}) at $\mu=0$, which is the Casimir energy in vacuum (i.e., inside the Dirac sea).
The finite-density part $ E^\mathrm{den}_\mathrm{LLL/HLL}$ (i.e.,  the contribution below the Fermi level and above the Dirac sea) leads to the Casimir(-like) energy from the finite-density effect (in detail, see Refs.~\cite{Fujii:2024fzy,Fujii:2024ixq}).
In addition, $E^\mathrm{den}_\mathrm{LLL/HLL}$ can be written as  $E_{0,\mathrm{LLL/HLL}}^\mathrm{sum} - E^\mathrm{vac}_\mathrm{LLL/HLL}$, which is the subtraction between the two divergent quantities. 
After using a regularization trick, for the LLLs, one can find not only a regularized density part, but also an anomalous term proportional to $\mu b$~\cite{Frolov:2010wn}, which is significant for the stability of the DCDW phase.
In Appendix~\ref{App:anom}, we discuss the behavior of this term in finite $L_z$.
This contribution is roughly proportional to $L_z$, and its Casimir-energy-like quantity is roughly constant, so that it might be called the finite-volume effect of a thermodynamic quantity rather than a Casimir effect. 
In this sense, Eq.~(\ref{1dimLifshitz}) includes the regularized $E_{0,\mathrm{LLL}}^\mathrm{sum} - E_\mathrm{LLL}^\mathrm{vac}$ (as well as the regularized $E_\mathrm{LLL}^\mathrm{vac}$) but does not include the anomalous $\mu b$ term.

As another approach to calculate the Casimir energy, we use the lattice regularisation scheme~\cite{Actor:1999nb,Pawellek:2013sda,Ishikawa:2020ezm,Ishikawa:2020icy,Nakayama:2022ild,Nakata:2022pen,Mandlecha:2022cll,Nakayama:2022fvh,Swingle:2022vie,Nakata:2023keh,Flores:2023whr,Nakayama:2023zvm,Beenakker:2024yhq,Fujii:2024fzy,Fujii:2024ixq}.
By replacing the continuous momentum $k_z$ in the zero-point energy~\eqref{E0int} per unit area as $k_z\rightarrow(2-2\cos ak_z)/a^2$, where $a$ is the lattice spacing in the $z$ direction, the Casimir energy on the lattice is defined as 
\begin{align}
    E^{\rm Lat}_{\rm Cas}=E^{\rm sum}_0-E^{\rm int}_0, \label{eq:lat}
\end{align}
where the sum of $n$ is restricted within the first Brillouin zone (BZ): $n=0,1,\dots ,N_z-1$ with the number of lattice cells $N_z \equiv L_z/a$.
By taking the continuum limit $a\to 0$ (i.e., by using a sufficiently small $a$), we can get the correct Casimir energy.

Finally, we define a Casimir coefficient:
\begin{align}
    C^{[d]}_{\rm Cas}=L_z^dE_{\rm Cas}/\Lambda^{3-d}.
\end{align}
Since now $E_{\rm Cas}$ is defined as a quantity with the mass dimension $3$,
$L_z^dE_{\rm Cas}$ is dimensionless only when $d=3$. 
Therefore, we define a dimensionless quantity divided by $\Lambda^{3-d}$, where $\Lambda$ is a parameter with mass dimension one. 
On the other hand, in a nonzero magnetic field, since each LL is regarded as a particle moving in the quasi-one-dimensional space, the corresponding Casimir energy may scale as $E_{\rm Cas} \sim 1/L_z$.
Therefore, in the following, $C^{[1]}_{\rm Cas}$ will be used to characterize the Casimir energies decomposed in each LL, while $C^{[3]}_{\rm Cas}$ will be used to characterize the total Casimir energy summed over all the LLs.

\subsection{Classification of dispersion relations}

In the presence of a magnetic field, the quark energy levels split into the infinite number of Landau levels.
If the homogeneous chiral condensate phase is realized, all the HLLs behave as quasi-one-dimensional massive (or gapped) dispersion relations.
In the case of the MDCDW phase, the low-energy dispersion relations of HLLs are distorted and different from the usual massive one, which is a remnant of dispersion relations with the two Weyl points characterizing the DCDW phase in a zero magnetic field.
In a weak magnetic field, lower-$l$ (i.e., occupied) modes in the HLLs produce Fermi points (FPs).
As the magnetic field increases, the spacing between $\omega_{\pm,l}$ and $\tilde{\omega}_{\pm,l}$ with the same $l$ becomes large.
$\omega_{\pm,l}$ near the Fermi level begin to exceed the Fermi level sequentially, and eventually at a strong magnetic field, all $\omega_{\pm,l}$ exceed the Fermi level, which means that there are no FPs.

\begin{figure}
\includegraphics[scale=0.17]{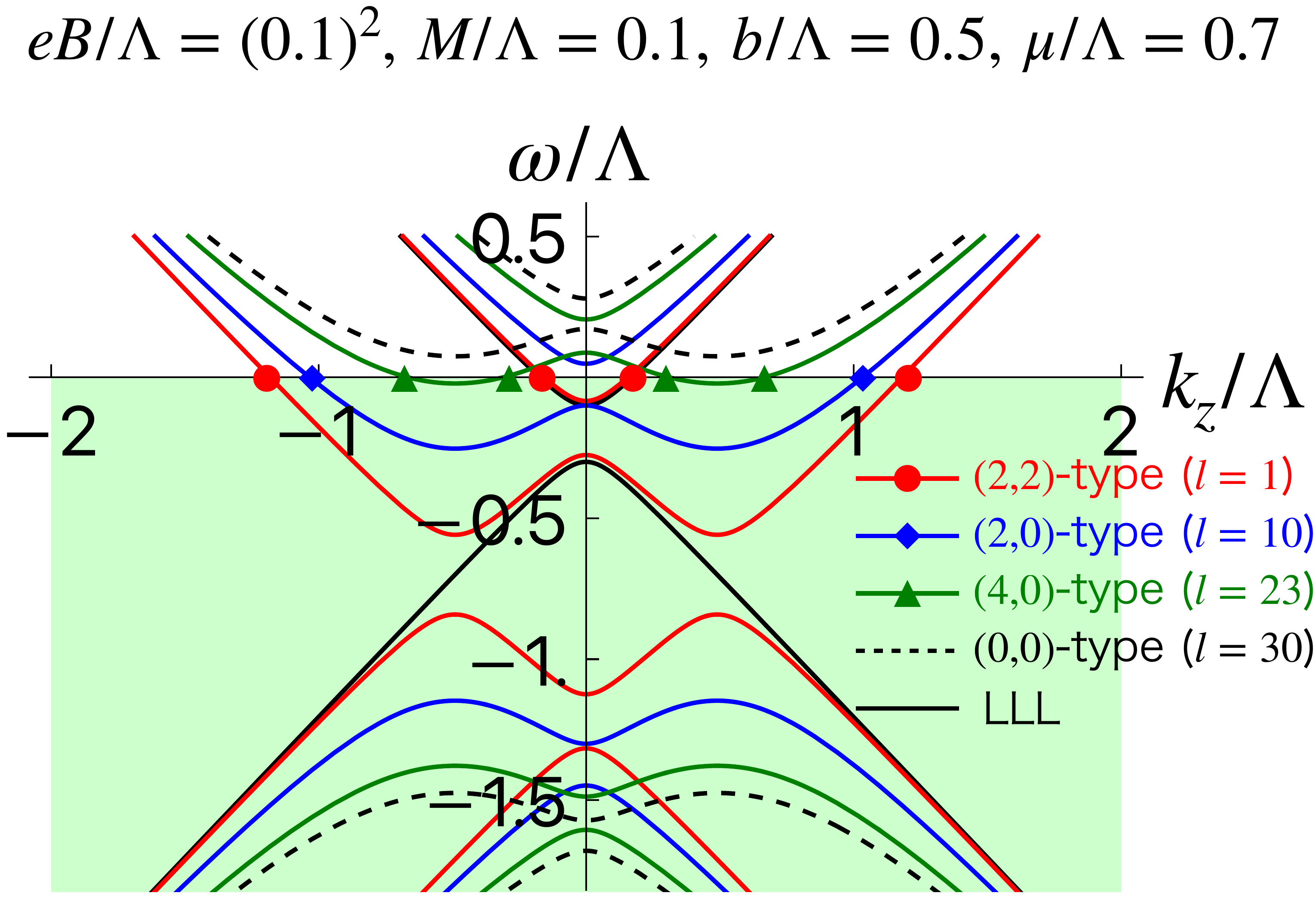}
\caption{\label{disp_type} Typical examples of dispersion relations for the LLLs (solid black line) and HLLs (red, blue, green, or dashed black line) of fermion fields in the MDCDW phase.
Filled symbols stand for crossing points with the Fermi level, namely the FPs.}
\end{figure}

In Fig.~\ref{disp_type}, we show the dispersion relations~\eqref{EV_LLLs} and \eqref{EV_HLLs} at $M/\Lambda=0.1$, $ b/\Lambda=0.5$, and $ \mu/\Lambda=0.7$ under a magnetic field $eB/\Lambda^2=(0.1)^2$.
Here, for simplicity, we consider the one-flavor case with an electric charge $q_f=\pm e$, which can be regarded as a (quasi)electron or positron system.
The choice of these parameters realizes the DCDW phase in a magnetic field, i.e., the MDCDW phase.
The two solid black lines are the upper and lower modes of the LLLs.
The other colored lines and the black dashed lines are the HLLs, which split the eigenmodes of the HLLs for the two spin degrees of freedom.
The circles, diamonds, and triangles represent FPs.
In the case of the black dashed line, there is no FP because the gap between the positive and negative energy modes spreads as $l$ increases.

\begin{table*}[t!]
\centering
\caption{Classification of dispersion relations possible in the MDCDW phase at $\mu>0$: two types of LLLs and four types of HLLs, where we also list the modes with FPs, the number of induced oscillations, the classification of Casimir effect, and the oscillation period.
$\omega_{l=0}$ and $\omega_{\pm,l}$ are defined as Eqs.~(\ref{EV_LLLs}) and (\ref{EV_HLLs}).
The form of $L_z^\mathrm{osc}$ is given in Eqs.~(\ref{period_LLL}) and (\ref{period1}).
}
\begin{tabular}{llcccl}
\hline\hline
Dispersion type & Condition & With FPs & No. of osc.& Casimir effect  &$L_z^\mathrm{osc}$ for PBC\\
\hline
LLL (``metal") & $M^2<(\mu-b)^2$ &$\omega_{l=0}$ & 1 & Singly oscillating & $\frac{2\pi}{|k^{\rm FP}_{z}|}$ \\
LLL (``insulator") & $(\mu-b)^2 < M^2$ &No & 0 & Nonoscillating & No \\
\hline
$(2,2)$ or ``metals" &
$2|q_f B|l<\mu^2-(M+b)^2$
&$\omega_{-,l},\omega_{+,l}$ &  2 & Dually oscillating &  $\frac{2\pi}{|k^{\rm FP}_{z,\pm}|}$ for $\omega_{\mp,l}$ \\
$(2,0)$ or ``metal-insulator" & $\mu^2-(M+b)^2< 2|q_fB|l< \mu^2-(M-b)^2$& $\omega_{-,l}$  & 1 &Singly oscillating & $\frac{2\pi}{|k^{\rm FP}_{z,+}|}$ \\
$(4,0)$ or ``island"& $\mu^2-(M-b)^2<2|q_f B|l<\mu^2$& $\omega_{-,l}$  & 2 & Dually oscillating & $\frac{2\pi}{|k^{\rm FP}_{z,+}|},\frac{2\pi}{|k^{\rm FP}_{z,-}|}$\\
$(0,0)$ or ``insulators"&$\mu^2< 2|q_f B|l$&No & 0 &Sign flipping & No \\
\hline\hline
\end{tabular}
  \label{Tab:type}
\end{table*}

\begin{table*}[t!]
\centering
\caption{Classification of dispersion relations of the LLLs and HLLs in the {\it homogeneous-chiral-condensate} phase (or massive-Dirac-fermion vacuum or matter) in a magnetic field and at $\mu>0$.
The notations are the same as those in Table~\ref{Tab:type}.
}
\begin{tabular}{llcccl}
\hline\hline
Dispersion type & Condition & With FPs & No. of osc.& Casimir effect  &$L_z^\mathrm{osc}$ for PBC\\
\hline
LLL (metal) & $M^2<\mu^2$ &$\omega_{l=0}$ & 1 & Singly oscillating & $\frac{2\pi}{|k^{\rm FP}_{z}|}$ at $b=0$\\
LLL (insulator) & $\mu^2 < M^2$ &No & 0 & Nonoscillating & No \\
\hline
HLL (metal) &
$M^2+2|q_f B|l<\mu^2$
&$\omega_{\pm,l}$ (degenerate) &  1 & singly oscillating & $\frac{2\pi}{|k^{\rm FP}_{z,\pm}|}$ at $b=0$\\
HLL (insulator) &$\mu^2< M^2+2|q_f B|l$&No & 0 &non-oscillating & No \\
\hline\hline
\end{tabular}
  \label{Tab:type_homo}
\end{table*}

We have shown in the previous work~\cite{Fujii:2024fzy} that FPs induce an oscillation in the Casimir energy.
Under magnetic fields, the presence of FPs also results in the oscillating Casimir energy in the same way. 
In the case of current parameters, the LLLs have FPs and induce an oscillation of Casimir energy.
Its oscillation period is determined by the momenta of FPs:
\begin{align}
    L^{\rm osc}_z=\frac{2\pi}{|k_z^{\rm FP}|} \ \ \  \Big(k_z^{\rm FP}=\pm\sqrt{(\mu- b)^2 - M^2}\Big) \label{period_LLL}
\end{align}
when the PBC is applied.\footnote{For the MIT bag boundary condition leading to $k_z\to (n+1/2)\pi/L_z$ ($n=0,1,2,\dots$), the oscillation period becomes the half of that with the PBC.}
Thus, the necessary condition for this oscillation is the presence of FPs, i.e., $(\mu- b)^2>M^2$. 

For the HLLs in the MDCDW phase, we can classify possible dispersion relations into four types.
Their typical forms are shown in Fig.~\ref{disp_type}.

\begin{figure*}[t!]
\includegraphics[scale=0.16]{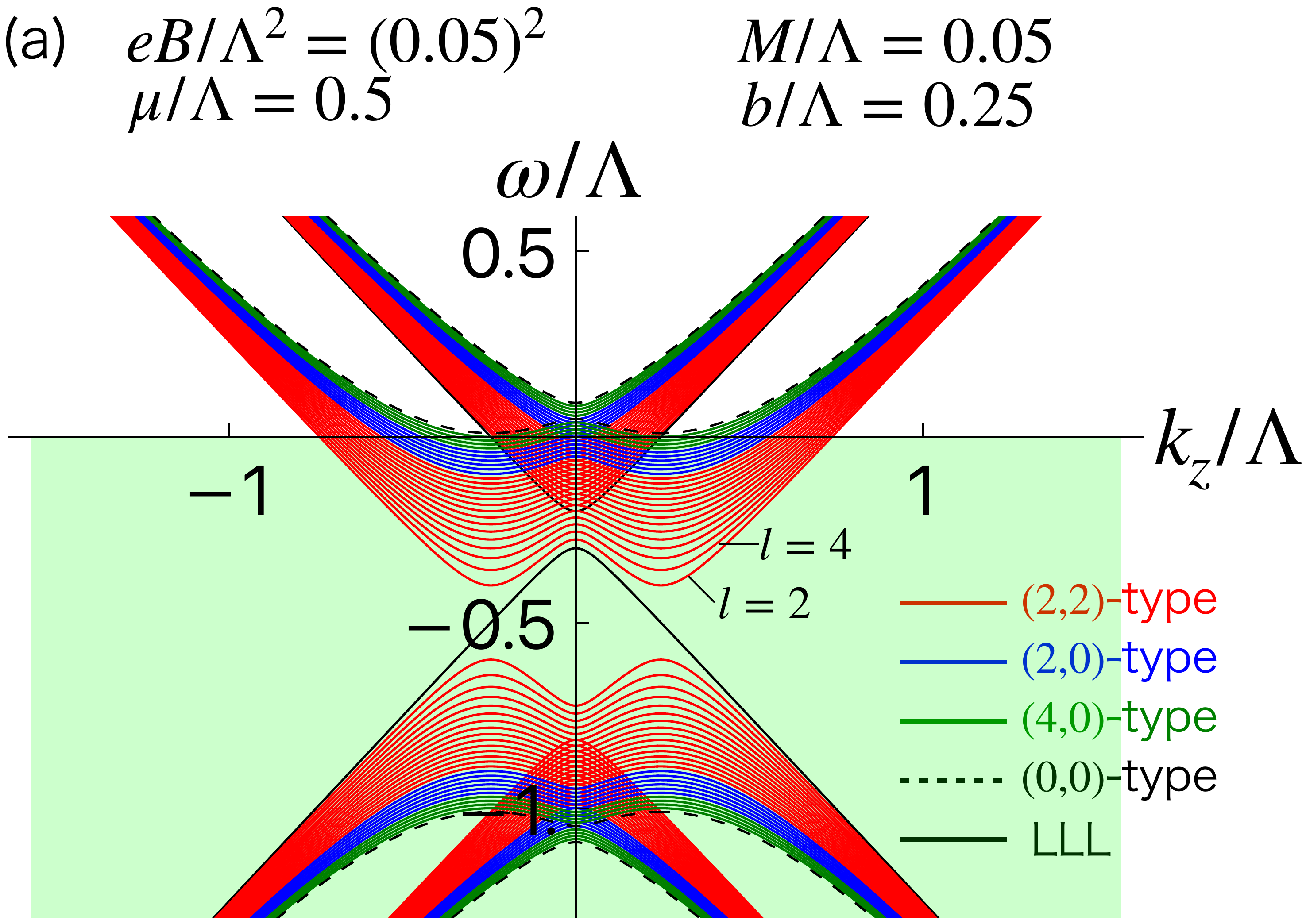}
\includegraphics[scale=0.15]{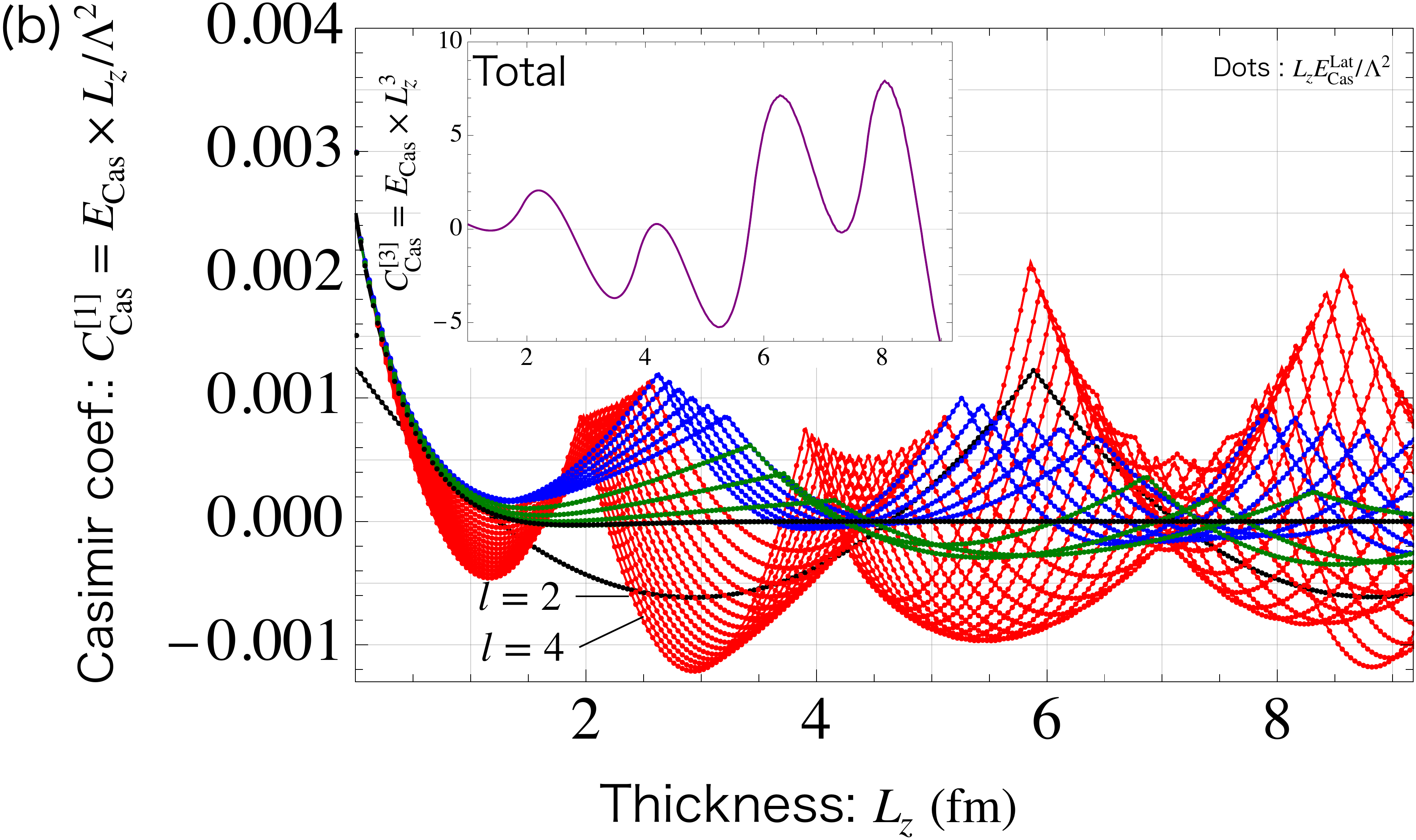}
\caption{\label{CCas_MDCDW_one} (a) Dispersion relations of fermion fields in the {\it one-flavor} MDCDW phase.
(b) Thickness dependence of Casimir coefficients $C^{[1]}_{\rm Cas}$ for each LL.
Inset: the total Casimir coefficients $C^{[3]}_{\rm Cas}$.}
\end{figure*}

\begin{enumerate}
    \item[(a)] $(2,2)$ {\it type}. This type is realized when the condition $2|q_f B|l<(\mu^2-(M+b)^2)$ is satisfied: The magnetic field and the label of LLs are small enough.
    In this type, the upper energy mode $\omega_+$ makes two FPs, and the lower energy mode $\omega_-$ also makes two FPs.
    We call this type of dispersion relation the $(2,2)$ type, since both the modes make two FPs.
    The momentum of FPs is given as
    \begin{align}
        |k_{z,\pm}^{\rm FP}|=&\sqrt{(\sqrt{\mu^2-2|q_fB|l}\pm b)^2-M^2}. \label{period1}
    \end{align}
    We note that $|k_{z,+}^{\rm FP}|>|k_{z,-}^{\rm FP}|$, where $|k_{z,+}^{\rm FP}|$ and $|k_{z,-}^{\rm FP}|$ correspond to $\omega_-$ and $\omega_+$, respectively.
    The resulting Casimir energy is a superposition of two oscillations with different periods ($L_z^{\rm osc}=2\pi/|k^{\rm FP}_{z,+}|$ and $L_z^{\rm osc}=2\pi/|k^{\rm FP}_{z,-}|$). 
    \item[(b)] $(2,0)$ {\it type}. This type is realized when the condition $(\mu^2-(M+b)^2)< 2|q_fB|l< (\mu^2-(M-b)^2)$ is satisfied, where only the lower mode $\omega_-$ makes two FPs.
    We call this type of dispersion relation the $(2,0)$ type, since only $\omega_-$ make two FPs.
The resulting Casimir energy show an oscillation with a period ($L_z^{\rm osc}=2\pi/|k^{\rm FP}_{z,+}|$) due to the FPs $|k_{z,+}^{\rm FP}|$
    made by $\omega_-$.
    \item[(c)] $(4,0)$ {\it or island type}.\footnote{When $M \geq b$, this type is forbidden by the property of quartic functions.
We will see this situation later. \label{no(4,0)type}} This type is realized when the condition $(\mu^2-(M-b)^2)<2|q_f B|l<\mu^2$ is satisfied, where the lower mode $\omega_-$ has four FPs.
    We call this type of dispersion relation the $(4,0)$-type or island type\footnote{Island means a bump structure located on the Fermi sea.} since only the $\omega_-$ mode has the four FPs.
    The resulting Casimir energy is a superposition of oscillations of two different periods due to the FPs $|k_{z,\pm}^{\rm FP}|$ with $|k_{z,+}^{\rm FP}|>|k_{z,-}^{\rm FP}|$ 
    made by $\omega_-$.
    \item[(d)] $(0,0)$ {\it type}. This type is realized when the condition $2|q_f B|l> \mu^2$ is satisfied: The magnetic field and/or the label of LLs is large enough.
    In this type, the lower mode $\omega_-$ is located above the Fermi level, so that there is no FP.
    Therefore, we call this type of dispersion relation the $(0,0)$ type.
    The resulting Casimir energy does not oscillate.
\end{enumerate}

Finally, in Table~\ref{Tab:type}, we summarize the dispersion relations of LLLs and HLLs in the MDCDW phase and the properties of corresponding Casimir energy.
As a comparison, in Table~\ref{Tab:type_homo}, we summarize the case of the homogeneous-chiral-condensate phase (which is also regarded as just the massive-Dirac-fermion vacuum or matter) under a magnetic field.\footnote{In the case of the homogeneous-chiral-condensate phase, the metal-type dispersion relations induce the Casimir energy originating from both the Dirac and Fermi seas, while the insulator-type dispersion relations lead to the Casimir energy from only the Dirac sea.
For the Casimir effect from the metal type, an analysis separating the contributions from the Dirac and Fermi seas was done in Refs.~\cite{Fujii:2024fzy,Fujii:2024ixq} (e.g., see Fig.~8 in Ref.~\cite{Fujii:2024fzy}).
Also in the DCDW phase, such an analysis is possible~\cite{Fujii:2024fzy}.}
Thus, the only difference in the LLLs is the factor $b$, while the Casimir effects from the HLLs are drastically modified by the MDCDW.

\section{Results} \label{Sec:result}

In this section, we discuss the Casimir energy of the MDCDW phase.
In our previous work~\cite{Fujii:2024fzy}, we discussed the Casimir effect in the DCDW phase (without a magnetic field) and proposed that oscillating Casimir energy is produced in this phase. The origin of this effect is that the dispersion relations of quarks have Fermi points.
The presence of Fermi points means that the absolute value of a dispersion relation, i.e., $|\omega|$ or $|\tilde{\omega}|$, has a nondifferentiable point in momentum space.

In this section, we adopt $\Lambda=860$ MeV and $\hbar c \sim 197.327$ ${\rm MeV\cdot fm}$, where $\hbar$ is the reduced Planck constant and $c$ the speed of light.

\subsection{Casimir energy in MDCDW for one-flavor case} \label{sec_one}

In this subsection, we consider the one-flavor case as a demonstration, where we use Eqs.~\eqref{EV_LLLs} and \eqref{EV_HLLs} with $q_f=\pm e$ as the energy eigenvalues. Based on the understanding of the dispersion relation described here and the corresponding Casimir energy, we will discuss the more realistic two-flavor case in the next subsections.

In Fig.~\ref{CCas_MDCDW_one}(a), we show the dispersion relations in a DCDW phase characterized by
$(M/\Lambda, b/\Lambda, \mu/\Lambda)=(0.05,0.25,0.5)$ at $eB/\Lambda^2=(0.05)^2$.
Figure~\ref{CCas_MDCDW_one}(b) shows the corresponding Casimir coefficients $C^{[1]}_{\rm Cas}$, where we separately plot the contributions from each Landau level.\footnote{In this plot, the contribution of the LLLs means the sum of the Casimir energies from the two modes, $\omega_{l=0}$ and $\tilde{\omega}_{l=0}$.
The contribution from the HLLs with $l\neq0$ means the sum of the Casimir energies from the four modes, $\omega_{+,l}$, $\omega_{-,l}$, $\tilde{\omega}_{+,l}$, and $\tilde{\omega}_{-,l}$.}
The inset of Fig.~\ref{CCas_MDCDW_one}(b) shows the total Casimir coefficients $C^{[3]}_{\rm Cas}$ obtained by summing sufficiently many LLs.
The solid lines and the dots are the results from the Lifshitz formula~\eqref{1dimLifshitz} and the lattice regularization~\eqref{eq:lat}, respectively.

For the LLLs, the dispersion relations and the Casimir coefficient are shown as the solid black lines in Figs.~\ref{CCas_MDCDW_one}(a) and \ref{CCas_MDCDW_one}(b), respectively. 
We find that the $C_{\rm Cas}^{[1]}$ shown in Fig.~\ref{CCas_MDCDW_one}(b) oscillate with respect to $L_z$.
This is due to the presence of FPs created by $\omega_{l=0}$, as explained in the previous section.
Its oscillation period is determined by the position of the FPs, and from Eq.~\eqref{period_LLL}, we can estimate $L_z^{\rm osc}=5.89 \ {\rm fm}$.

For the HLLs, the behaviors of the Casimir coefficients are classified as the following four patterns. 
\begin{enumerate}
    \item $l=1$--$31$ [$(2,2)$ {\it type}].
    In this type, the Casimir coefficient behaves as the superposition of oscillations with two different periods.
    In Fig.~\ref{CCas_MDCDW_one} (a), we show the quark dispersion relations of $l=2,4,\ldots,30$ as the solid red lines. 
    We can see that each mode of $\omega_+$ or $\omega_-$ has two FPs.
    The periods of oscillations read from Fig.~\ref{CCas_MDCDW_one} (b) coincide with Eq.~\eqref{period1} using FPs positions in Fig.~\ref{CCas_MDCDW_one} (a). 
    \item $l=32$--$42$ [$(2,0)$ {\it type}].
    In this case, the Casimir coefficient oscillates with one period.
    We show the quark dispersion relations of $l=32,34,\ldots,42$
    %every four from $l=32$ to $l=40$ 
    as the blue solid lines in Fig.~\ref{CCas_MDCDW_one} (a).
    \item $l=43$--$50$ [$(4,0)$ {\it or island type}].
    In this case, the Casimir coefficient is a superposition of oscillations with two different periods.
    We show the quark dispersion relations of $l= 44,46,48,50$ as the green solid lines in Fig.~\ref{CCas_MDCDW_one} (a).
    We can see that $\omega_-$ has four FPs.
    \item $l\geq51$ [$(0,0)$ {\it type}].
    In this case, the Casimir coefficient does not oscillate and damps (see Sec.~\ref{subsec:strong} for its sign-flipping behavior).
    This is because every $\omega_{\pm}$ is located above the Fermi level.
\end{enumerate}

\begin{figure*}[tbh!]
\includegraphics[scale=0.16]{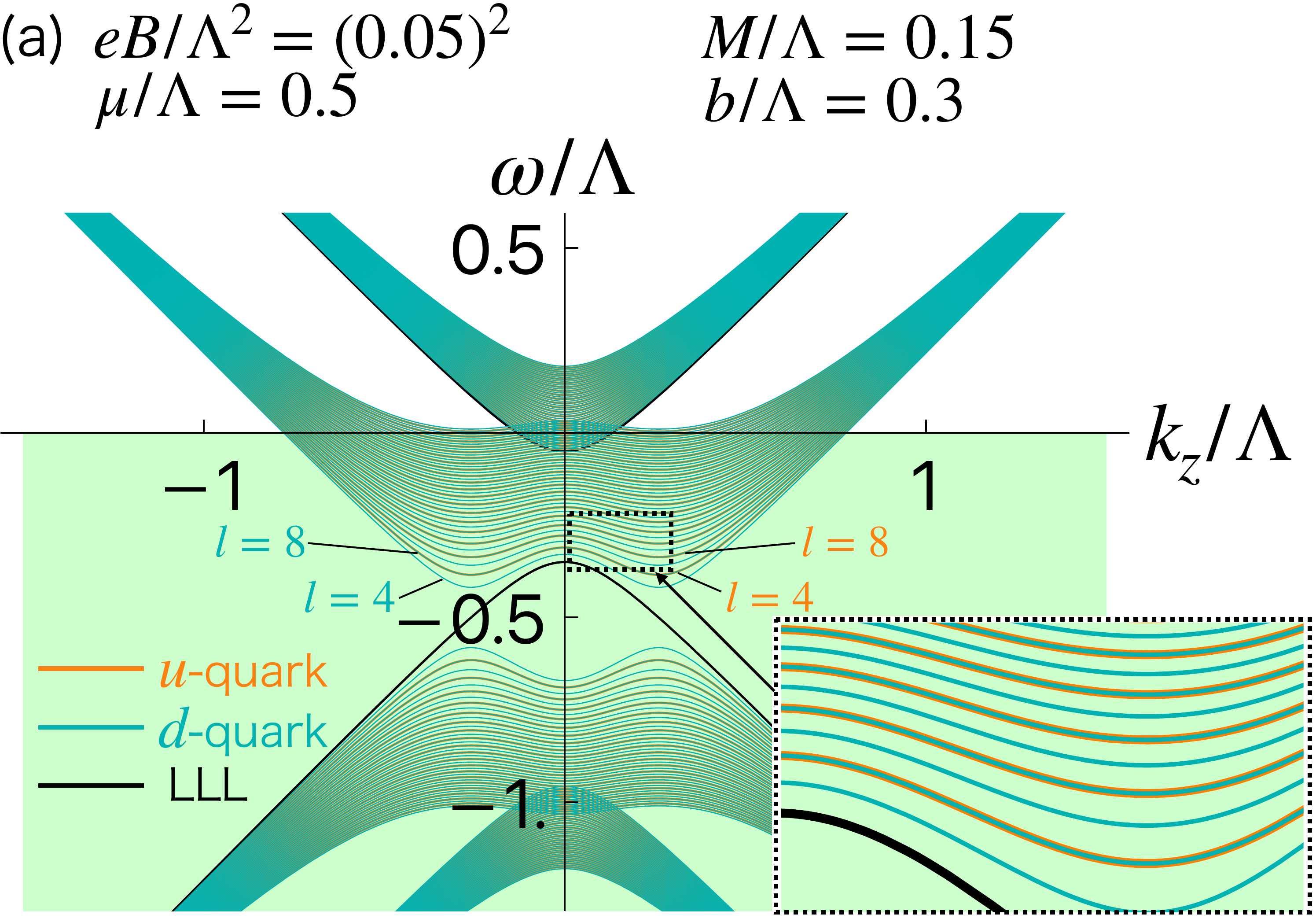}
\includegraphics[scale=0.15]{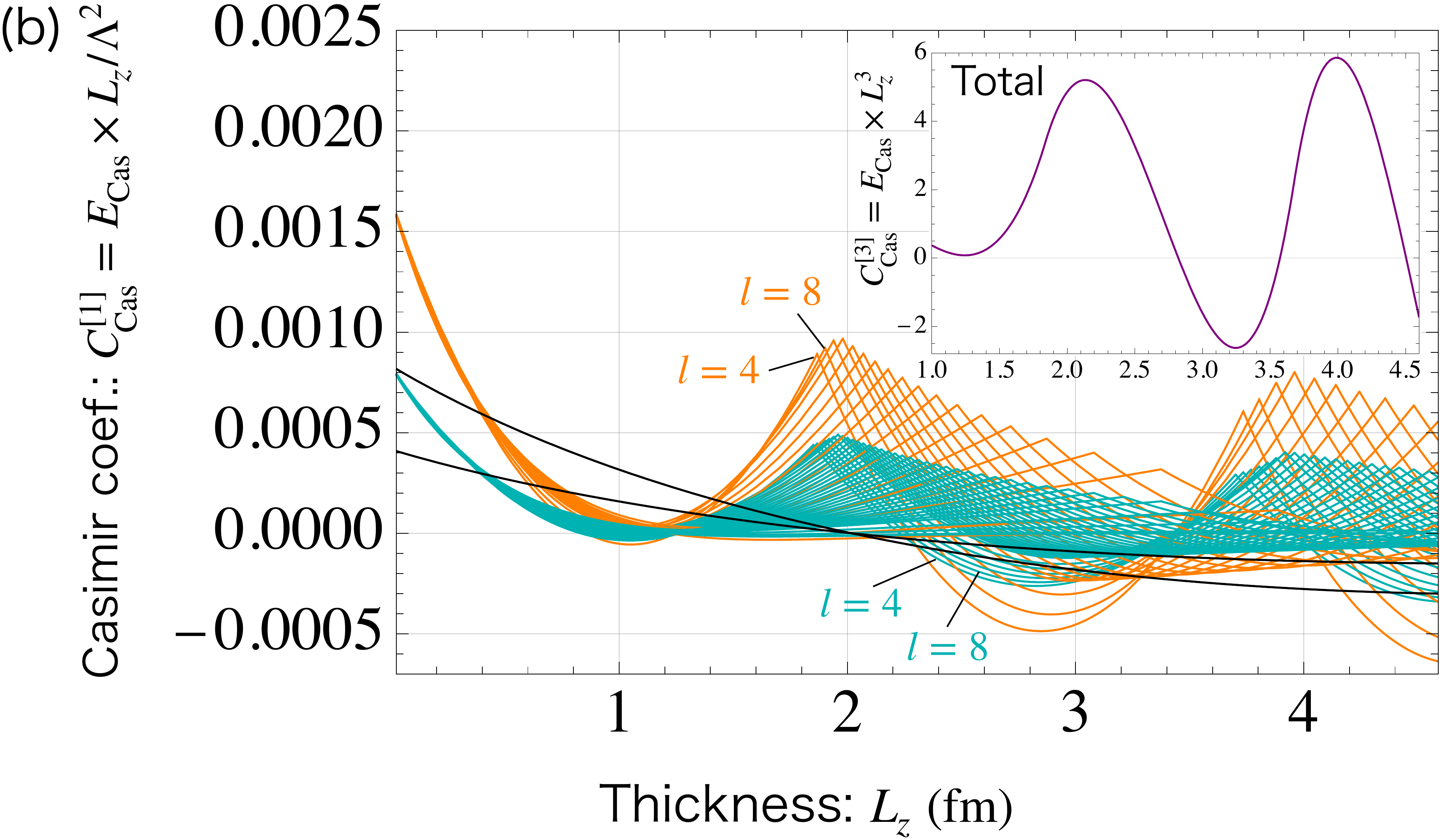}
\caption{\label{CCas_MDCDW_weak} (a) Dispersion relations of fermion fields in the two-flavor MDCDW phase under a {\it weak} magnetic field $eB/\Lambda^2=(0.05)^2$.
(b) Thickness dependence of Casimir coefficients $C^{[1]}_{\rm Cas}$ for each LL.
Inset: the total Casimir coefficient $C^{[3]}_{\rm Cas}$.}
\end{figure*}

Furthermore, we find that at $L_z\sim0$ the amplitude of the Casimir energy from the LLLs is half of that from HLLs.
This is because the eigenmodes of the LLLs pick up one spin component whereas the HLLs have two spin components.
Thus, in this strength of magnetic field, the contribution from the HLLs is more dominant than that from the LLLs.
Also, the result for each LL approaches to a constant
\begin{align}
&C_\mathrm{Cas,LLL}^{[1]} (L_z\to0)=N_c \times \frac{|q_fB|}{2\pi \Lambda^2} \times \frac{\pi}{3}, \\
&C_\mathrm{Cas,HLL}^{[1]} (L_z\to0)= N_c \times \frac{|q_fB|}{2\pi \Lambda^2} \times \frac{2\pi}{3},
\end{align}
where $|q_fB|/2\pi$ is the Landau degeneracy factor [appearing as the coefficient in the thermodynamic potential~(\ref{eq:omega}) or the Lifshitz formula~(\ref{1dimLifshitz})] and $\pi/3$ or $2\pi/3$ is the factor known in the Casimir energy from the massless Dirac field (without or with the spin degrees of freedom $2$) in the $1+1$-dimensional spacetime with the PBC in one direction.
In the current parameters, we get $C_\mathrm{Cas,LLL}^{[1]}\to 0.00125$ and $C_\mathrm{Cas,HLL}^{[1]} \to 0.0025$.

The total Casimir coefficient $C_{\rm Cas}^{[3]}$, which is calculated by summing a sufficient number of LLs ($l_\mathrm{max}=1000$), is shown in the inset of Fig.~\ref{CCas_MDCDW_one} (b).
In a sufficiently weak magnetic field, the total Casimir energy roughly agrees with that obtained from the three-dimensional Lifshitz formula~\eqref{3dimLifshitz}.
On the other hand, when $L_z$ is small, the two Casimir energies do not coincide.
This is because the HLLs become non-negligible at small $L_z$, meaning that the number of LLs summed up is insufficient. 
In Appendix \ref{App:LL}, we examine the $l_\mathrm{max}$ dependence in the short-$L_z$ region.

Finally, we comment on the energy scale of our Casimir energy.
As a reference, in the massless-quark vacuum (i.e., at $M=b=\mu=B=0$),
we know $C_\mathrm{Cas}^{[3]}=N_fN_c\times2\pi^2/45 \sim 2.63$ at any $L_z$.
Then, at $L_z=1$ fm, $E_\mathrm{Cas}=C_\mathrm{Cas}^{[3]}\times \hbar c/L_z^3 \sim 519$ $\mathrm{MeV}/\mathrm{fm}^2 \sim 8.32 \times 10^4$ $\mathrm{N}/\mathrm{fm}$, and at $L_z=10$ fm, $E_\mathrm{Cas}\sim 0.519 \,\mathrm{MeV}/\mathrm{fm}^2$.
The total Casimir energy in Fig.~\ref{CCas_MDCDW_one} (b) is about $C_\mathrm{Cas}^{[3]} \sim 5$ (even in the longer $L_z$), which is comparable with that in the massless-quark vacuum.
The emergence of the energy scale in the longer $L_z$ is caused by the combination of the quantum effect and the finite-density effect (for the zero-density case, see Sec.~\ref{subsec:strong}).
In addition, it may be instructive to discuss the energy scale of the Casimir energy {\it from each LL} in Fig.~\ref{CCas_MDCDW_one} (b).
Then, we can estimate $C_\mathrm{Cas}^{[1]} \sim 0.001$.
This value is transformed to $E_\mathrm{Cas}\sim  0.001\times \Lambda^2/L_z \hbar c =3.75$ MeV/fm$^2$ at $L_z=1$ fm and $E_\mathrm{Cas}\sim 0.375$ MeV/fm$^2$ at $L_z=10$ fm.
Therefore, $E_\mathrm{Cas}$ of each LL at $L_z \sim 10$ fm is comparable to that in the massless-quark vacuum at $L_z \sim 10$ fm.

\subsection{Casimir energy in MDCDW (weak $eB$)} \label{sec:weak}

\begin{figure*}[tbh!]
\includegraphics[scale=0.16]{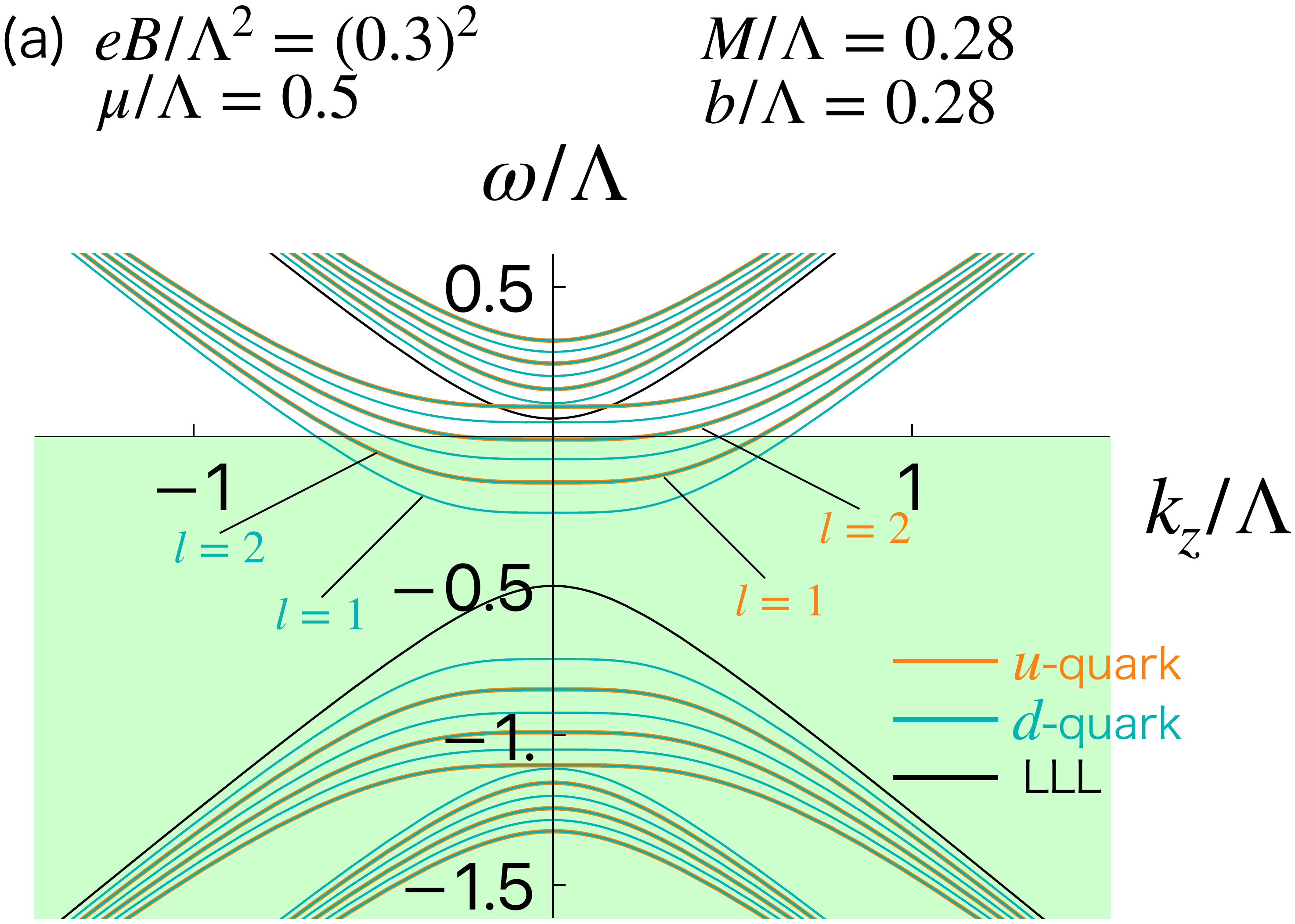}
\includegraphics[scale=0.15]{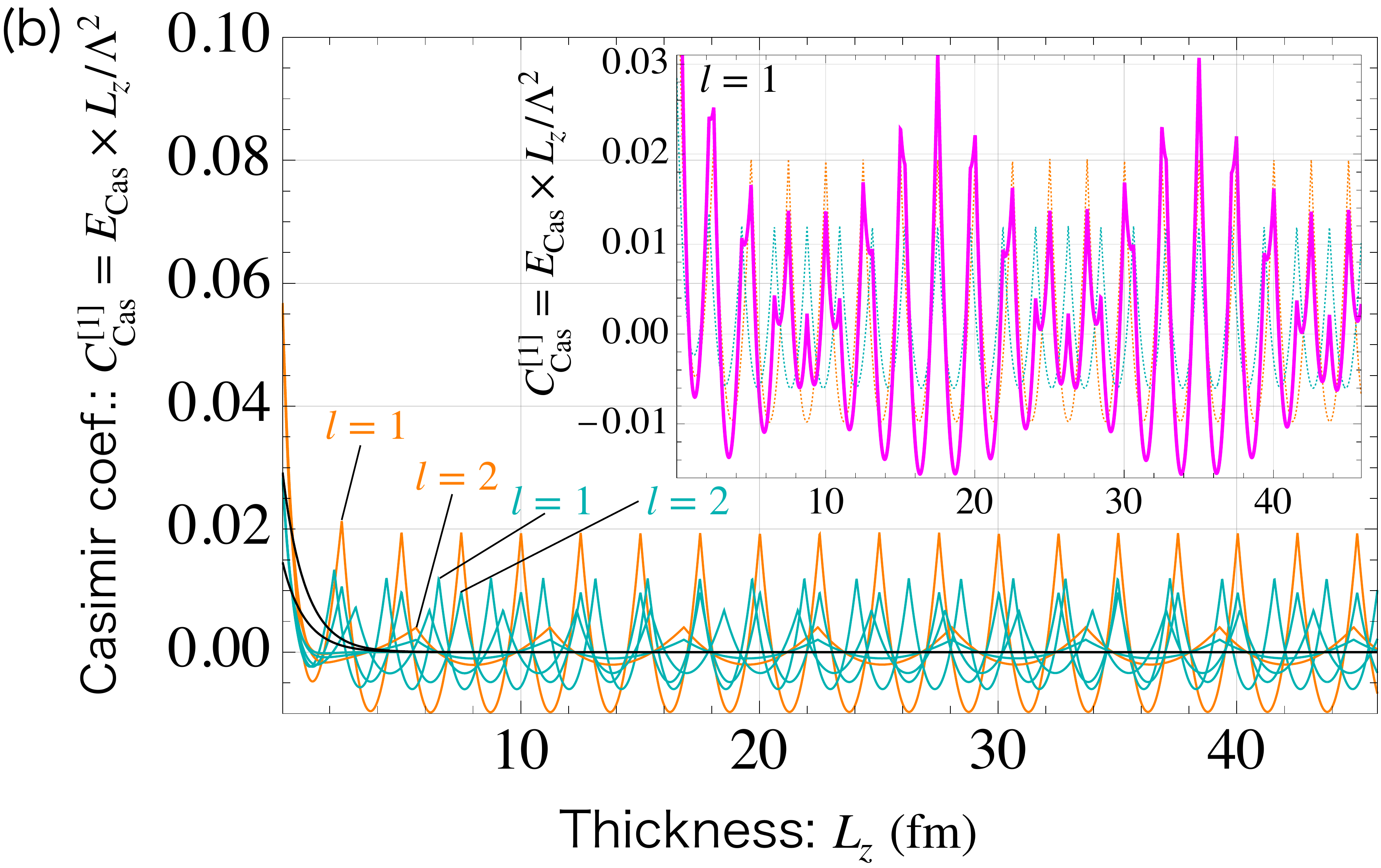}
\caption{\label{CCas_MDCDW_beat} (a) Dispersion relations of fermion fields in the two-flavor MDCDW phase under an {\it intermediate} magnetic field $eB/\Lambda^2=(0.3)^2$.
(b) Thickness dependence of Casimir coefficients $C^{[1]}_{\rm Cas}$ from each LL.
 Inset: the sum of Casimir coefficients $C^{[1]}_{\rm Cas}$ from $u$ and $d$ quarks at $l=1$.}
\end{figure*}

In the following, we consider the two-flavor case, which is more realistic as quark matter containing $u$ and $d$ quarks. 
According to Ref.~\cite{Frolov:2010wn}, the DCDW phase (characterized by $M\neq0$ and $b\neq0$) is more stable than the homogeneous chiral-condensate phase ($M\neq0$ and $b=0$) if both the chemical potential $\mu$ and the magnetic field $B$ are nonzero. 
Therefore, in this work, we consider only the MDCDW phase at $\mu \neq 0$ and $B \neq 0$ and do not consider the homogeneous chiral-condensate phase at $\mu \neq 0$ and $B \neq 0$.

First, we discuss the DCDW phase under a weak magnetic field.
As a parameter set characterizing the MDCDW phase, we adopt $(M/\Lambda, b/\Lambda, \mu/\Lambda)=(0.15,0.3,0.5)$ and $eB/\Lambda^2=(0.05)^2$ which is determined by solving a gap equation in Ref.~\cite{Frolov:2010wn}.\footnote{See Fig.~2 (c) in Ref~\cite{Frolov:2010wn}, where the coupling constant of the NJL model is supercritical: At $B=0$ and a smaller $\mu$, the chiral symmetry is broken ($M\neq0$).
In the nonzero-$B$ and intermediate-$\mu$ region, the MDCDW phase is realized.}
In this case, the dispersion relations for the LLLs and HLLs ($l=4,8,\ldots$) are shown in Fig.~\ref{CCas_MDCDW_weak} (a).
From this figure, in the case of the $u$ quark (plotted as the orange lines), the LLs of $l=1$--$47$ belong to the $(2,2)$ type, $l=48$--$62$ to the $(2,0)$ type, $l=63$--$75$ to the $(4,0)$ or island type, and $l>75$ to the $(0,0)$ type. 
From Eq.~\eqref{EV_HLLs}, in general, for the two-flavor case, an even number of $l$ in the HLL dispersion relations for $d$ quarks completely coincides with the dispersion relation for $u$ quarks.
In the case of $d$ quarks (the green lines), the LL of $l=1$--$95$ belongs to the $(2,2)$ type, $l=96$--$126$ to the $(2,0)$ type, $l=127$--$150$ to the $(4,0)$ or island type, and $l>150$ to the $(0,0)$ type.
Figure.~\ref{CCas_MDCDW_weak}(b) shows the $C^{[1]}_{\rm Cas}$ for each of LLs and the total $C^{[3]}_{\rm Cas}$ summing the sufficient number of LLs.
Here, each line of $C^{[1]}_{\rm Cas}$ corresponds to the dispersion relation plotted in Fig.~\ref{CCas_MDCDW_weak} (a).

In the case of the LLLs, the dispersion relations of the $u$ and $d$ quarks are degenerate.
The period of each $C^{[1]}_{\rm Cas}$ is estimated to be $L_z^{\rm osc}=10.90 \ {\rm fm}$ from Eq.~\eqref{period_LLL}.
Thus, the oscillation period is independent of the flavor, but due to the difference between the electric charges in the Landau degeneracy factor $|q_fB|/2\pi$, the amplitude of Casimir energy is different: The Casimir energy created by the $u$ quark is exactly twice as large as that by the $d$ quark.

In the case of the HLLs, the Casimir energy created by each of the $u,d$ quarks behaves as explained in Sec.~\ref{sec_one}.
Due to the difference in the electric charge, the position of the FPs, $|k^{\rm FP}_z|$, for the $u$ quarks labeled by $l$ is always smaller than that for the $d$ quarks labeled by the same $l$.
Consequently, the oscillation period becomes longer.
Also, similar to the discussion for the LLLs, due to the difference in the Landau degeneracy factor, the Casimir energy produced by the $u$ quark is roughly twice as large as that produced by the $d$ quark.

\begin{figure*}[tbh!]
\includegraphics[scale=0.16]{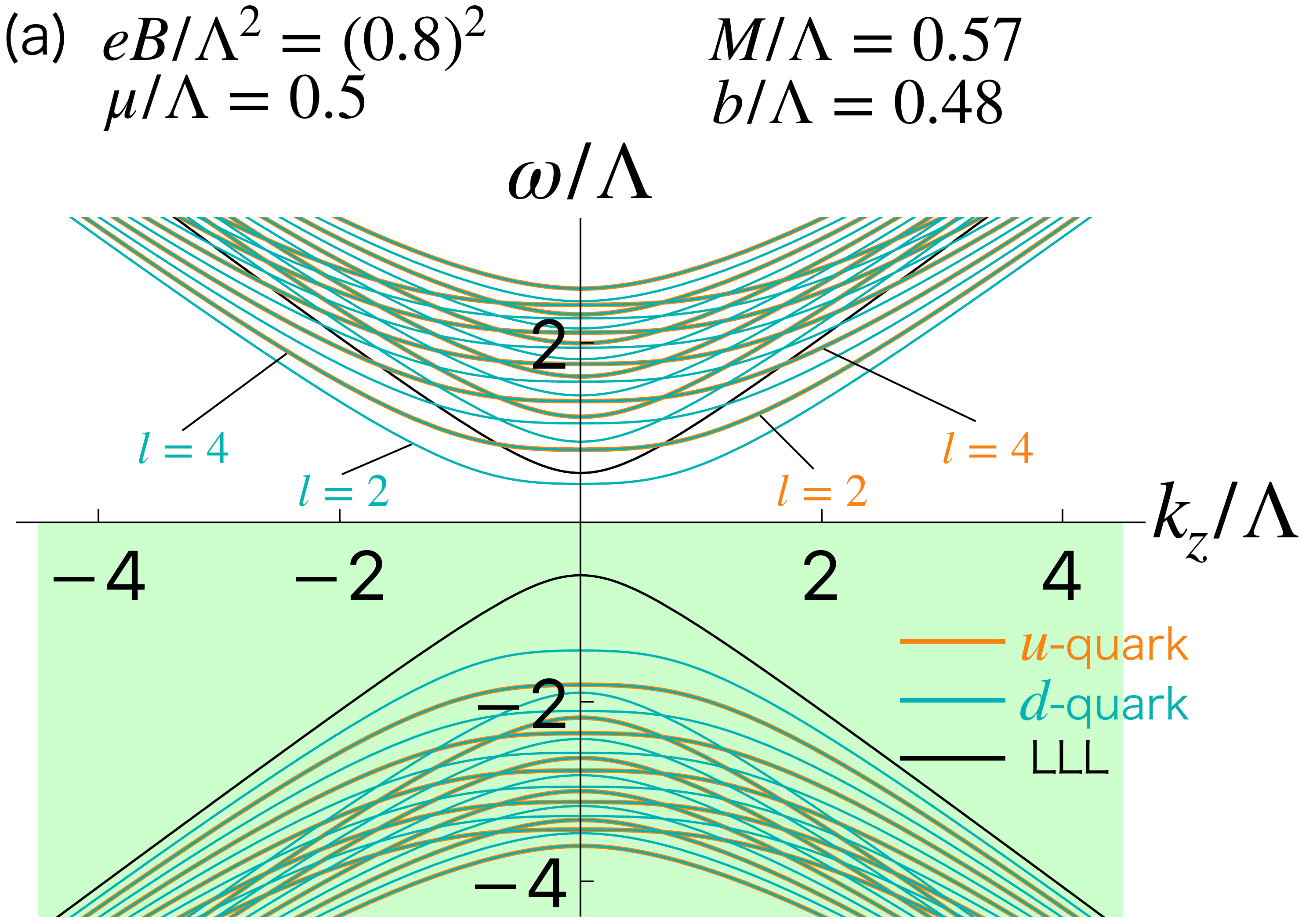}
\includegraphics[scale=0.15]{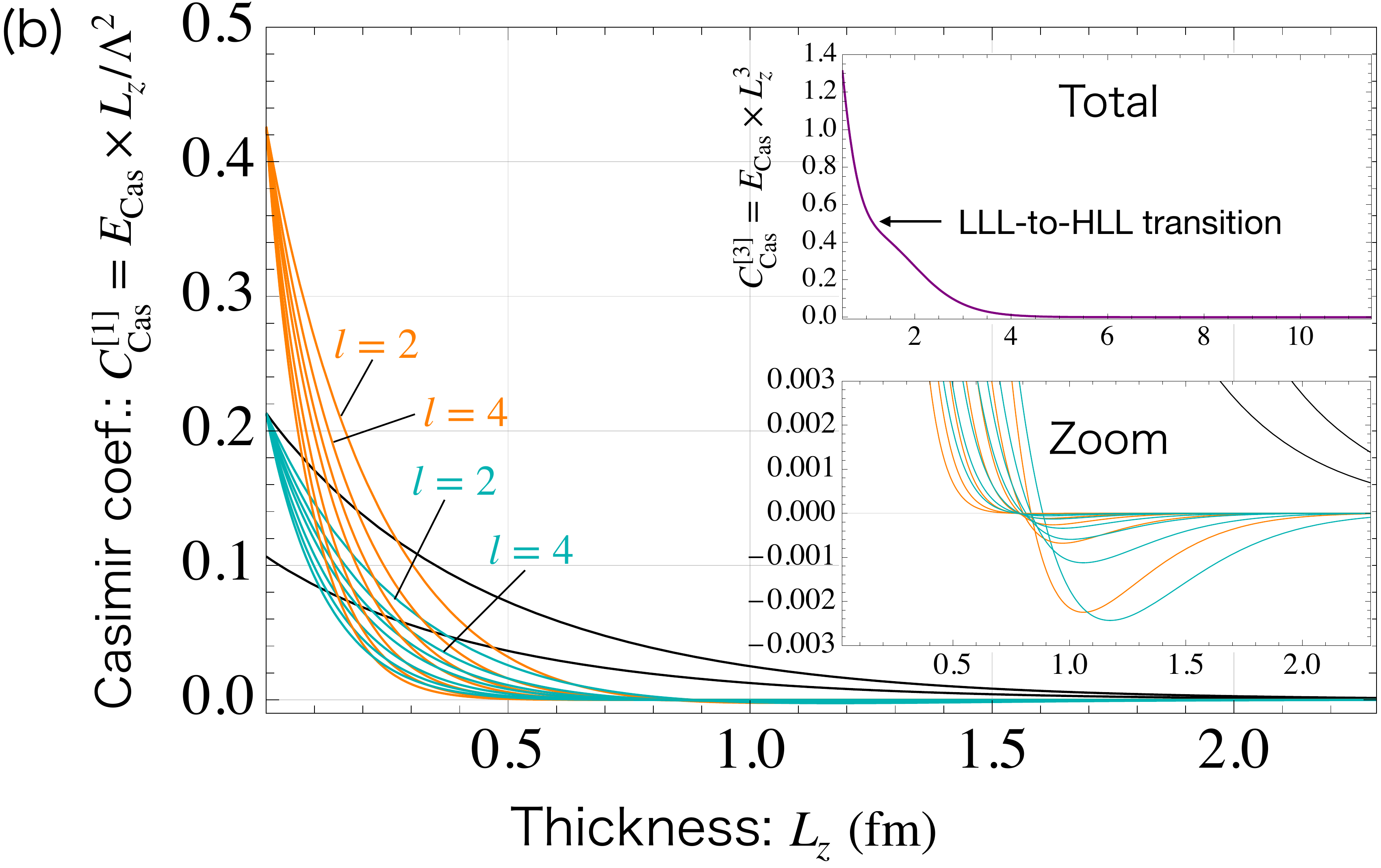}
\caption{(a) Dispersion relations of fermion fields in the two-flavor MDCDW phase under a {\it strong} magnetic field $eB/\Lambda^2=(0.8)^2$.
(b) Thickness dependence of Casimir coefficients $C^{[1]}_{\rm Cas}$ for each LL.
Upper inset: the total Casimir coefficients $C^{[3]}_{\rm Cas}$.
Lower inset: a close-up view near the $C^{[1]}_{\rm Cas}=0$.
}
\label{CCas_MDCDW_strong}
\end{figure*}

\subsection{Casimir energy in MDCDW (intermediate $eB$)}

As an intermediate magnetic field, we adopt $(M/\Lambda, b/\Lambda,\mu/\Lambda)=(0.28,0.28,0.5)$ and $eB/\Lambda^2=(0.3)^2$ as a solution obtained in Ref.~\cite{Frolov:2010wn}.
As the magnetic field increases, the upper dispersion relations $\omega_{\pm,l}$ of HLLs with higher $l$ are above the Fermi level (the lower dispersion relations $\tilde{\omega}_{\pm,l}$ are still below the Fermi level).
Since $\omega_{+,l}$ does not have FPs, there is no $(2,2)$-type dispersion relation in the current parameter set.

In this case, the dispersion relations for the LLLs and the HLLs ($l=1,2,\ldots$) are shown in Fig.~\ref{CCas_MDCDW_beat} (a).
From this figure, in the case of $u$ quark, the LLs of $l=1,2$ belong to the $(2,0)$ type, no LL to the $(4,0)$ or island type, and $l>2$ to the $(0,0)$ type.
In the case of $d$ quark, the LL of $l=1$--$4$ belongs to the $(2,0)$ type, the no LL to the $(4,0)$ or island type and the LL of $l>4$ to the $(0,0)$ type. 
As mentioned in footnote \ref{no(4,0)type}, when $M \geq b$, the $(4,0)$ or island type is forbidden by the property of quartic functions. 
The Casimir coefficient $C^{[1]}_{\rm Cas}$ from each of LLs at $l=1,2$ is shown in Fig.~\ref{CCas_MDCDW_beat} (b).

For the current parameters, the dispersion relations of the LLLs do not intersect with the Fermi level, and hence no oscillation of Casimir energy arises, which is the same as the Casimir energy for a one-dimensional massive Dirac field (with no spin degrees of freedom $2$).\footnote{Its Casimir coefficient is given as the well-known formula
\begin{align}
C_\mathrm{Cas,LLL}^{[1]} = N_c \times \frac{|q_fB|}{2\pi \Lambda^2} \times \frac{2ML_z}{\pi} \sum_{m=1}^\infty \frac{K_1(mML_z)}{m},
\end{align}
where $K_1$ is the modified Bessel function.}
This can be understood from the fact that the eigenvalue~\eqref{EV_LLLs} is formally the same as that of the ordinary one-dimensional massive Dirac field shifted by $b-\mu$.

Here, we focus on only $l=1$ of the HLLs.
In the inset of Fig.~\ref{CCas_MDCDW_beat} (b), we show the Casimir energy from each of the $u,d$ quarks at $l=1$ and the sum of them (the magenta line).
For the current parameters, the oscillation periods of Casimir energies produced by the $u,d$ quarks are very close to each other.
Then, the superposition of the two different oscillations leads to a longer periodicity of the Casimir energy, i.e., a beating behavior with a period $L_z^\mathrm{beat}$ estimated by $1/L_z^\mathrm{beat} = 1/|L_z^{\mathrm{osc},u}|-1/|L_z^{\mathrm{osc},d}|$.
This is a new type of {\it beating Casimir effect} in the sense that it originates from the flavor-dependent splitting by a magnetic field.\footnote{
As other examples of the beating Casimir effect, it is induced also by spin-split Dirac points~\cite{Nakayama:2022fvh}, multiple exceptional points~\cite{Nakata:2023keh}, or multiple chemical potentials~\cite{Fujii:2024ixq}.
}
Note that this beating behavior is realized when we focus on a LL index (now, $l=1$), but in general, the oscillatory behavior of the total Casimir energy should be more complex due to contamination of oscillations from other LLs.
In the current parameters, since the LLL and HLLs of $l>3$ for the $u$ quark and $l>5$ for the $d$ quark do not induce any oscillations, the total oscillation consists of the superposition of oscillatory behaviors of lower six levels.

\subsection{Casimir energy in MDCDW (strong $eB$)} \label{subsec:strong}

Finally, we discuss the behavior in the strong-magnetic-field region.
In this subsection, we adopt $(M/\Lambda,b/\Lambda,\mu/\Lambda)  =(0.57, 0.48, 0.5)$ and $eB/\Lambda^2=(0.8)^2$ as a solution obtained in Ref.~\cite{Frolov:2010wn}. 
In this case, the dispersion relations for the LLLs and the HLLs with $l=2,4,\ldots$ are shown in Fig.~\ref{CCas_MDCDW_strong} (a). 
For both the $u,d$ quarks, the dispersion relations for all HLLs belong to the $(0,0)$ type, which means that all the HLLs do not cross the Fermi level.
Figure~\ref{CCas_MDCDW_strong} (b) shows $C^{[1]}_{\rm Cas}$ from each of the LLs.

In this parameter set, the dispersion relations of all the LLs do not cross the Fermi level, and, hence, do not induce any oscillating behavior of Casimir energy.
In particular, the functional form of Casimir energies for the LLLs is the same as that from ordinary one-dimensional massive Dirac fields.

We find that, in the short-$L_z$ region, the contribution of HLLs is not negligible.
On the other hand, in the long-$L_z$ region, the Casimir energy for the HLLs rapidly decreases, so that the total Casimir energy is dominated by the LLLs. 
Also, we find that the signs of Casimir energy for HLLs change near $L_z \sim 0.8$ fm and then damps, as shown in the lower inset of Fig.~\ref{CCas_MDCDW_strong} (b).
Such a {\it sign-flipping Casimir effect} is distinct from the case of the LLLs, which originates from the functional form of the eigenvalues~(\ref{EV_HLLs}) of HLLs.\footnote{A similar system is the axion electrodynamics where the eigenvalues of modified photon fields in the $3+1$-dimensional spacetime are given as $\omega_{\zeta=\pm} = \sqrt{(\zeta\sqrt{\tilde{b}^2+k_z^2}+\tilde{b})^2+k_x^2+k_y^2}$ with a parameter $\tilde{b}$.
The sign-flipping Casimir effect originating from such photon fields was studied in Refs.~\cite{Fukushima:2019sjn,Brevik:2021ivj,Canfora:2022xcx,Oosthuyse:2023mbs,Favitta:2023hlx,Ema:2023kvw,Nakayama:2023zvm}.
These dispersion relations correspond to $b=M \to \tilde{b}$ and $2|q_fB|l \to k_x^2+k_y^2$ in our eigenvalues~(\ref{EV_HLLs}) at $\mu=0$.
Therefore, our finding is a new type of sign-flipping Casimir effect, which is induced by HLLs in the MDCDW phase.
}

In addition, we can find a transition of the total $C^{[3]}_{\rm Cas}$ at $L_z \sim 1$ fm, where the Casimir energy switches from the region dominated by the LLLs to that dominated by HLLs, which may be called the {\it LLL-to-HLL transition}.
This transition was first discovered by the formula for the massive Dirac field under a magnetic field in the early study~\cite{Elizalde:2002kb}.
Since the LLL-to-HLL transition is a property for the normal massive Dirac field, there is no direct relationship to the MDCDW.

\section{Summary and Outlook} \label{Sec:summary}

In this paper, we have discussed the Casimir energy produced by the quark field in the DCDW phase of quark matter under a magnetic field.
We extended the Lifshitz formula to Dirac-fermion fields in the MDCDW phase: Our main formula is Eq.~(\ref{1dimLifshitz}).
Using this formula, we have calculated the Casimir energy in some parameter regions realizing the MDCDW phase.

In particular, we have examined the LL decomposition of the Casimir energy in a magnetic field.
This analysis clarifies how each LL (among the LLLs and some types of HLLs) contributes to the total Casimir energy.
The HLLs are classified by the presence or absence of FPs, as in Table~\ref{Tab:type}.
Our findings may be summarized as, in the one-flavor case,
\begin{enumerate}
\item[(i)] LLL (weak $eB$): singly oscillating Casimir effect;
\item[(ii)] LLL (strong $eB$): nonoscillating Casimir effect;
\item[(iii)] $(2,2)$ type: dually oscillating Casimir effect;
\item[(iv)] $(2,0)$ type: singly oscillating Casimir effect;
\item[(v)] $(4,0)$ or island type: dually oscillating Casimir effect; and
\item[(vi)] $(0,0)$ type: sign-flipping Casimir effect.
\end{enumerate}
Here, the ``dual" effect in the $(2,2)$ type is caused by the spin splitting, while that in the $(4,0)$ or island type is attributed to the low-energy distortion of one dispersion relation.
The situation of the two-flavor case is more complex, where the magnetic field splits the dispersion relations of $u,d$ quarks with different charges.
As a result, we have observed the beating Casimir energy produced by $u,d$ quarks.

Finally, we summarize our possible outlooks.
\begin{enumerate}
\item[(1)] {\it Lattice simulations}. To examine the Casimir effect of interacting fermion systems, one can utilize numerical simulations of a lattice field theory (e.g., see Refs.~\cite{Chernodub:2018pmt,Chernodub:2018aix,Kitazawa:2019otp,Chernodub:2023dok} for the Casimir effect from Yang-Mills fields).
In particular, the existence (or absence) of the MDCDW phase can be numerically tested by lattice NJL or QCD simulations.
Although Monte Carlo simulations are usually difficult at finite chemical potential due to the sign problem, the chemical potential required for the MDCDW phase is smaller than that of the DCDW phase in a zero magnetic field.
Therefore, the lattice simulations of the Casimir effect in the MDCDW phase would be easier.
\item[(2)] {\it Real kink crystals}.
As another possible ground state within the NJL model, one can consider a solitonic modulation, the so-called real kink crystal (RKC) phase.
The early references~\cite{Nishiyama:2015fba,Abuki:2018iqp,Anzuini:2021gwi} predict that sufficiently strong magnetic fields tend to favor the MDCDW over the RKC (see Ref.~\cite{Cao:2016fby} for analysis under only the RKC).
The stability of the MDCDW phase is attributed to the spectral asymmetry of the LLLs in the MDCDW phase.
However, in smaller magnetic fields (not dominated by only the LLLs), the RKC phase (or a hybridized phase~\cite{Nishiyama:2015fba}) may survive, and the Casimir effect in such a phase might be interesting.
\item[(3)] {\it Chiral soliton lattice}.
Apart from the NJL model, the chiral perturbation theory, which is an effective field theory of low-energy QCD based on mesonic degrees of freedom, predicts a ground state of magnetized finite-density QCD: the chiral soliton lattice (CSL) in QCD~\cite{Son:2007ny,Brauner:2016pko}, which is a stack of parallel $\pi^0$ domain walls.
The MDCDW phase may be consistent with the CSL in the sense that it is induced by the Wess-Zumino-Witten-type anomaly~\cite{Tatsumi:2014wka,Nishiyama:2015fba}, while in the viewpoint of the Casimir effect, comparing the descriptions based on mesonic and quark degrees of freedom would be important.
\item[(4)] {\it Dirac or Weyl semimetals}.
In the typical dispersion relations of relativistic fermions in the three-dimensional Dirac or Weyl semimetals, Dirac or Weyl points are located at finite momenta (see Refs.~\cite{Armitage:2017cjs,Lv:2021oam} for reviews).
When a magnetic field is switched on, the low-energy spectrum of LLs can be distorted like those in Fig.~\ref{disp_type}, which is a remnant of Dirac or Weyl points. 
Since the typical behavior of the Casimir effect is characterized by the form of the dispersion relations, the classifications we suggested in this paper will be also useful for understanding the fermionic Casimir effect inside thin films of Dirac or Weyl semimetals under a magnetic field (for analysis with only the LLL, see Ref.~\cite{Nakayama:2022fvh}).
\end{enumerate}

\acknowledgments
This work was supported by the Japan Society for the Promotion of Science (JSPS) KAKENHI (Grants No. JP20K14476, No. JP24K07034, No. JP24K17054, and No. JP24K17059).

\appendix

\section{Anomalous term in finite volume} \label{App:anom}
The thermodynamic potential in the MDCDW phase contains an anomalous term from the spectral asymmetry of LLLs.
This term stabilizes the MDCDW phase~\cite{Frolov:2010wn} and leads to the anomalous quark number density~\cite{Tatsumi:2014wka} and the chiral-anomaly-induced Hall current~\cite{Ferrer:2015iop}.
In this appendix, we derive a finite-volume effect for the anomalous contribution of LLLs in the zero-point energy~(\ref{E0sum}).
In the main text, we do not include this contribution in the definition of the Casimir energy.
However, this is relevant for solving the gap equation in finite volume.

\subsection{Derivation}
The derivation for the MDCDW phase in infinite volume was provided in Ref.~\cite{Frolov:2010wn}, and we extend its regularization trick to a finite-volume situation.
From Eq.~(\ref{E0sum}), the zero-point energy for the LLL under the PBC is
\begin{align}
&E_{0,{\rm LLL}}^{\rm sum}=-N_c\sum_{q_f}\frac{|q_f B|}{2\pi}\sum_{n=-\infty}^\infty\Big(\frac{1}{2}|\omega_{l=0,n}|+\frac{1}{2}|\tilde{\omega}_{l=0,n}|\Big), \notag\\
&\omega_{l=0,n}=E_{l=0,n}+b-\mu,
    \ \ \ \tilde{\omega}_{l=0,n}=-E_{l=0,n}+b-\mu, \notag\\
&E_{l=0,n} \equiv \sqrt{M^2+\left(\frac{2n\pi}{L_z}\right)^2}. \notag
\end{align}
Its medium part is defined by subtracting the vacuum part (at $\mu=0$) from the total one (at $\mu \neq 0$):
\begin{align}
E_0^\mu &\equiv E_{0,\mathrm{LLL}}^\mathrm{sum}(\mu)-E_{0,\mathrm{LLL}}^\mathrm{sum}(\mu=0) \notag\\
&= -N_c \sum_{q_f} \frac{|q_f B|}{2\pi} {\sum_{n=0}^\infty}^\prime \left( |\omega_{l=0,n}| + |\tilde{\omega}_{l=0,n}| -(\mu\to0) \right), \label{eq:omega_med}
\end{align}
where $(\mu \to0)$ means $|\omega_{l=0,n}|+|\tilde{\omega}_{l=0,n}|$ at $\mu=0$.
The factor $1/2$ of the zero-point energy is canceled by folding the summation range, $\sum_{n=-\infty}^\infty \to 2 {\sum_{n=0}^{\infty}}^\prime$, where the prime means that the factor $1/2$ is multiplied only in the term at $n=0$.
Because all the terms are divergent due to the limit of $n \to \infty$, we need a regularization.

Here, we consider a general form of dispersion relation with a constant shift $a$, where $a=b$ or $b-\mu$.
We prepare the two types of zero-point energies, $E_0^\mathrm{asym} $ with asymmetric cutoffs $N_\mathrm{max}^{\pm}$ and $E_0^\mathrm{sym}$ with a symmetric cutoff $N_\mathrm{max}$:
\begin{align}
E_0^\mathrm{asym} =& -N_c \sum_{q_f} \frac{|q_f B|}{2\pi} \left[\sum_{n=0}^{N_\mathrm{max}^{+} \prime} \left|\sqrt{M^2+ \left(\frac{2n\pi}{L_z} \right)^2}+a \right| \right. \notag \\
&\left. +\sum_{n=0}^{N_\mathrm{max}^{-}  \prime} \left| -\sqrt{M^2+ \left(\frac{2n\pi}{L_z} \right)^2}+a \right|  \right], \\
E_0^\mathrm{sym} =& -N_c \sum_{q_f} \frac{|q_f B|}{2\pi} \left[\sum_{n=0}^{N_\mathrm{max}  \prime} \left|\sqrt{M^2+ \left(\frac{2n\pi}{L_z} \right)^2}+a \right| \right. \notag\\
 & \left. + \sum_{n=0}^{N_\mathrm{max} \prime} \left| -\sqrt{M^2+ \left(\frac{2n\pi}{L_z} \right)^2}+a \right|  \right],
\end{align}
where the asymmetric cutoffs $N_\mathrm{max}^\pm$ are defined as
\begin{align}
N_\mathrm{max}^+ &= \mathrm{floor} \left[  \sqrt{\left(N_\mathrm{max} - \frac{bL_z}{2\pi} \right)^2 - \left(\frac{ML_z}{2\pi}\right)^2} \right], \\
N_\mathrm{max}^- &= \mathrm{ceil} \left[  \sqrt{\left(N_\mathrm{max} + \frac{bL_z}{2\pi} \right)^2 - \left(\frac{ML_z}{2\pi}\right)^2} \right].
\end{align}
The form of these asymmetric cutoffs is analogous to the large-momentum cutoffs introduced in infinite volume~\cite{Frolov:2010wn}, while the floor and ceiling functions are crucial.
These functions are required to adjust the difference in the numbers of discrete $\pm$ modes to be a nonzero even number in the limit $N_\mathrm{max} \to \infty$ for $bL_z>0$ (for $bL_z<0$, the floor and ceiling functions in the cutoffs are replaced with each other). 

By subtracting $E_0^\mathrm{sym}$ from $E_0^\mathrm{asym}$, we can correctly extract the contribution from the anomalous term.
Since the medium part is obtained by subtracting the vacuum part ($a=b$) from the total one ($a= b-\mu$),
\begin{align}
&(E_0^\mathrm{asym} - E_0^\mathrm{sym}) |_{a=b-\mu} -(E_0^\mathrm{asym} - E_0^\mathrm{sym}) |_{a=b} \notag\\
%&= -N_c \sum_{q_f} \frac{|q_f B|}{2\pi} (N_\mathrm{max}^+ - N_\mathrm{max}^-) [(b-\mu) - b] \notag\\
&=-N_c \sum_{q_f} \frac{|q_f B|}{2\pi} (N_\mathrm{max}^+ - N_\mathrm{max}^-) (-\mu). \label{eq:reg1}
\end{align}
This calculation is valid when $N_\mathrm{max}$ is large enough.
Note that only the terms proportional to $N_\mathrm{max}^\pm \times \mu$ survive, and the other terms are exactly canceled.
By taking the limit of $N_\mathrm{max} \to \infty$,
\begin{align}
\lim_{N_\mathrm{max} \to \infty} (N_\mathrm{max}^+ - N_\mathrm{max}^-) = -2 \,\mathrm{ceil} \left[\frac{bL_z}{2\pi} \right]. \label{eq:ceil}
\end{align}

In addition, the terms depending on only $b$ and $\mu$ are obtained as
\begin{align}
&E_0^{\mu,\mathrm{reg,sum}} \equiv E_0^\mathrm{sym} |_{a=b-\mu} - E_0^\mathrm{sym} |_{a=b} \notag\\
&= -N_c \sum_{q_f} \frac{|q_fB|}{2\pi} {\sum_{n=0}^\infty}^\prime \left( |\omega_{l=0,n}| + |\tilde{\omega}_{l=0,n}| -(\mu\to0) \right)_\mathrm{reg}. \label{eq:reg2}
\end{align}

Finally, using Eqs.~(\ref{eq:reg1}) and (\ref{eq:ceil}), the similar equations for $bL_z < 0$, and Eq.~(\ref{eq:reg2}), the medium part obtained by regularizing the divergent zero-point energy~(\ref{eq:omega_med}) is
\begin{align}
&E_0^{\mu,\mathrm{reg}} = E_0^{\mu,\mathrm{reg,sum}} \notag\\
& + \left\{
\begin{aligned}
&-N_c \sum_{q_f} \frac{|q_fB|}{2\pi} 2\mu \,\mathrm{ceil} \left[\frac{bL_z}{2\pi} \right] \ \ (bL_z  \geq 0)\\
&-N_c \sum_{q_f} \frac{|q_fB|}{2\pi} 2\mu \,\mathrm{floor} \left[\frac{bL_z}{2\pi} \right] \ \ (bL_z < 0)
\end{aligned}
\right. .\label{eq:E0mureg}
\end{align}
The first term is the regularized momentum sum, where the $\mu=0$ contribution (i.e., the vacuum part with $b\neq0$) is subtracted from the total one.
The second term is the anomalous contribution in finite $L_z$, which is our new finding.
By taking the limit $L_z \to \infty$, this term is reduced to $\lim_{L_z\to \infty} \frac{2\mu}{L_z} \mathrm{ceil} \left[\frac{bL_z}{2\pi} \right] =\lim_{L_z\to \infty} \frac{2\mu}{L_z} \mathrm{floor} \left[\frac{bL_z}{2\pi} \right]= \frac{2\mu b}{2\pi}$, which coincides with the form derived in infinite volume~\cite{Frolov:2010wn}.

\subsection{Casimir-energy-like quantity}
Equation~(\ref{eq:E0mureg}) is useful for solving the gap equation in the usual framework of the NJL model.
On the other hand, the Casimir-energy-like quantity is defined by subtracting the quantity in infinite $L_z$ from that in finite $L_z$.
From the first term of Eq.~(\ref{eq:E0mureg}),
\begin{align}
&E_0^{\mu,\mathrm{reg,sum}} - E_0^{\mu,\mathrm{reg,int}} =E_\mathrm{Cas}^\mathrm{LLL}(\mu) - E_\mathrm{Cas}^\mathrm{LLL}(\mu=0), \label{eq:ECas_mu}
\end{align}
where we defined
\begin{align}
&E_0^{\mu,\mathrm{reg,int}} \equiv \notag\\
&-N_c \sum_{q_f} \frac{|q_fB|}{2\pi}  L_z \int_0^\infty \frac{dk_z}{2\pi} \left( |\omega_{l=0}| + |\tilde{\omega}_{l=0}| -(\mu\to0) \right)_\mathrm{reg}.
\end{align}
Note that $E_0^{\mu,\mathrm{reg,sum}} - E_0^{\mu,\mathrm{reg,int}}$ is equivalent to the finite-$\mu$ contribution of the Casimir energy (for the LLLs) calculated as the difference between two Lifshitz formulas~(\ref{1dimLifshitz}) (see Appendix \ref{App:finitepart} for another expression).

From the second term of Eq.~(\ref{eq:E0mureg}),
\begin{align}
&E_\mathrm{anom}= \notag\\
&\left\{
\begin{aligned}
&-N_c \sum_{q_f} \frac{|q_fB|}{2\pi} \left(2\mu \, \mathrm{ceil}\left[\frac{bL_z}{2\pi} \right] - \frac{\mu b L_z}{\pi} \right) 
\ \ (bL_z \geq 0),\\
&-N_c \sum_{q_f} \frac{|q_fB|}{2\pi} \left(2\mu \,\mathrm{floor}\left[\frac{bL_z}{2\pi} \right] - \frac{\mu b L_z}{\pi} \right) \ \ (bL_z < 0).
\end{aligned}
\right. \label{eq:E_anom}
\end{align}

\subsection{Typical behavior}
\begin{figure}
\includegraphics[clip,width=1.0\columnwidth]{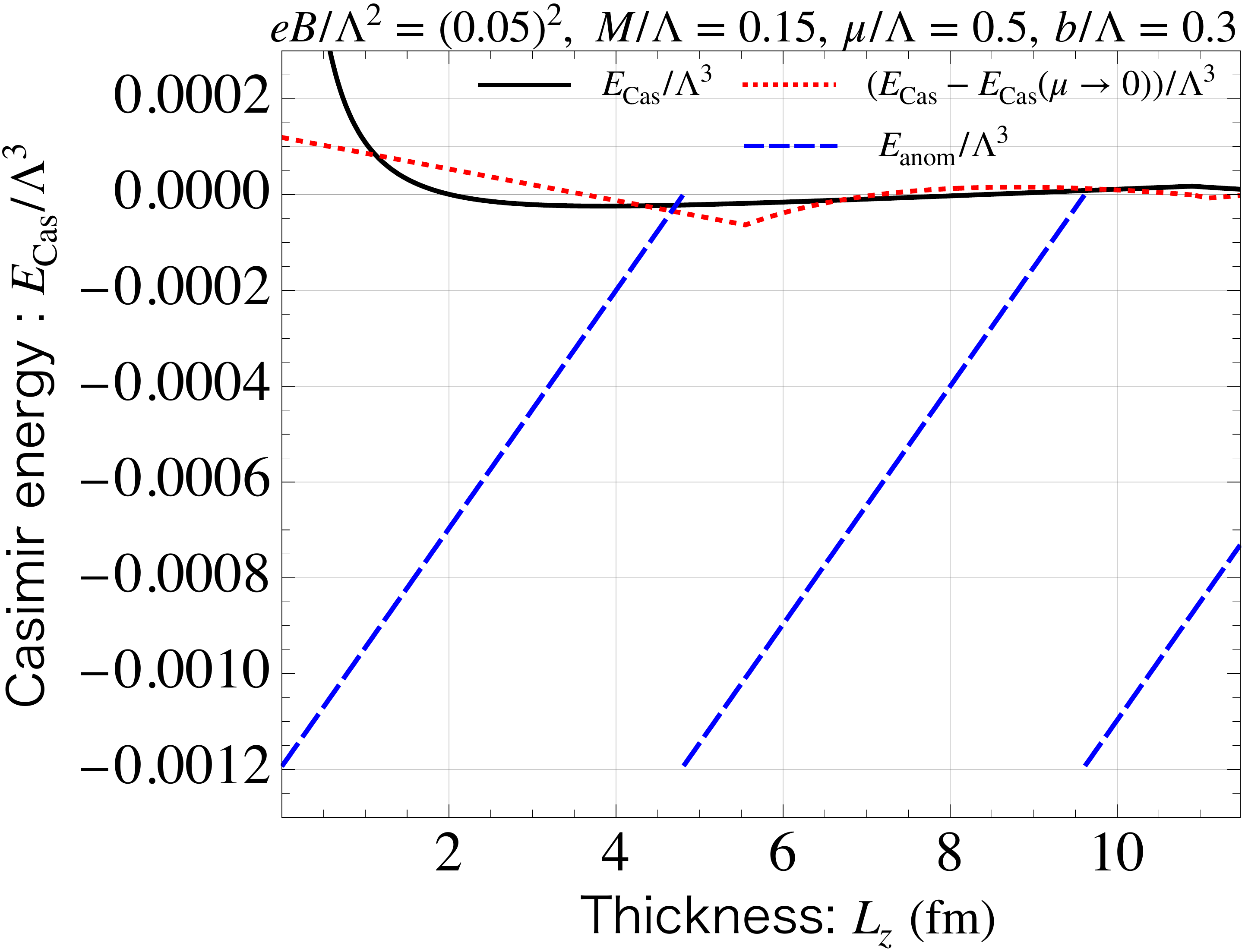}
\caption{Comparison between the contribution (\ref{eq:E_anom}) from the anomalous term and the finite-$\mu$ contribution (\ref{eq:ECas_mu}) of the Casimir energy for the LLLs.}
\label{ceil}
\end{figure}

Finally, we compare the typical behaviors of the anomalous-term contribution and the Casimir energy.
For example, Fig.~\ref{ceil} shows the Casimir energy from the LLLs in the two-flavor MDCDW phase under a weak magnetic field, as discussed in Sec.~\ref{sec:weak}. 
Here, the black solid, red dotted, and blue dashed lines correspond to the total Casimir energy $E_{\rm Cas}$, its finite-$\mu$ contribution (\ref{eq:ECas_mu}), and the contribution~(\ref{eq:E_anom}) from the anomalous term, respectively.
Thus, the anomalous term leads to the $L_z$ dependence of the sawtooth function with a constant amplitude, and its contribution is not damped as a function of $L_z$, which is different from the typical behavior of the Casimir energy.

From Eq.~(\ref{eq:E_anom}), the period from the anomalous term is
\begin{align}
L^{\rm osc, anom}_z = \frac{2\pi}{|b|}.
\end{align}
Using the current parameter set, we can estimate $L^{\rm osc, anom}_z \sim 4.81\ \text{fm}$.
Thus, the oscillation period induced by the anomalous term is determined by only $b$.
This property is different from that in the Casimir energy, where the period (\ref{period_LLL}) is determined by the momentum of FPs.
In this sense, the two oscillations can be distinguished from each other.
Also, the minimum values of the amplitude is calculated as $E_{\rm anom}=-2\mu N_c (|(2/3)eB|+|(-1/3)eB|)/2\pi\sim-0.00119 \times \Lambda^3$ (the maximum value is $E_{\rm anom}=0$).

\section{Relations between finite part and Lifshitz formula} \label{App:finitepart}
Since the finite-$\mu$ contribution (\ref{eq:ECas_mu}) of the Casimir energy is regularized, it can be obtained by directly performing the momentum integration and summation. For the sum term, using the step function,
\begin{align}
&{\sum_{n=0}^\infty}^\prime \left( |\omega_{l=0,n}| + |\tilde{\omega}_{l=0,n}| -(\mu\to0) \right)_\mathrm{reg} \notag\\
=& \sum_{n=-\infty}^\infty \left[-\omega_{l=0,n} \theta(-\omega_{l=0,n}) +\tilde{\omega}_{l=0,n}\theta(\tilde{\omega}_{l=0,n}) -(\mu\to 0)\right],
\end{align}
where the $(\mu\to 0)$ term means $-\omega_{l=0,n} \theta(-\omega_{l=0,n}) +\tilde{\omega}_{l=0,n}\theta(\tilde{\omega}_{l=0,n})$ at $\mu=0$.
For the integral term, we apply the replacement $ \sum_{n=-\infty}^\infty \to L_z \int_{-\infty}^\infty \frac{dk_z}{2\pi}$.

The Casimir-energy-like quantities are equivalent to those obtained by subtracting two different Lifshitz formulas~(\ref{1dimLifshitz}) (i.e., two Casimir energies). 
In the following, we show some relations between the direct calculations and the Lifshitz formulas, where we focus on the LLLs.

As shown in Appendix~\ref{App:anom}, the finite-$\mu$ contribution is obtained by subtracting the $\mu = 0$ part from the total Casimir energy $E_\mathrm{Cas}^\mathrm{LLL}(b,\mu)$:
\begin{align}
&\hspace{-3mm}E_\mathrm{Cas}^\mathrm{LLL}(b,\mu) - E_\mathrm{Cas}^\mathrm{LLL}(\mu= 0) \notag\\
=& -N_c \sum_{q_f} \frac{|q_fB|}{2\pi} \notag\\
&\times \left\{ \sum_{n=-\infty}^\infty \left[-\omega_{l=0,n} \theta(-\omega_{l=0,n}) +\tilde{\omega}_{l=0,n}\theta(\tilde{\omega}_{l=0,n}) \right]  \right. \notag\\
& \left.
- L_z\int \frac{dk_z}{2\pi} \left[-\omega_{l=0} \theta(-\omega_{l=0}) +\tilde{\omega}_{l=0}\theta(\tilde{\omega}_{l=0}) \right] \right\} \notag\\
&
-(\mu\to 0).
\end{align}
Similarly, the finite-$b$ contribution is also obtained by subtracting the $b = 0$ part from the total one.

Furthermore, the contribution from the energy shift of $b-\mu$ is obtained by subtracting the $b-\mu= 0$ part (i.e., the ``pure" Dirac-sea contribution) from the total one:
\begin{align}
&\hspace{-3mm}E_\mathrm{Cas}^\mathrm{LLL}(b,\mu) - E_\mathrm{Cas}^\mathrm{LLL}(b-\mu= 0) \notag\\
=&-N_c \sum_{q_f} \frac{|q_fB|}{2\pi} \notag \\
&\times\left\{\sum_{n=-\infty}^\infty \left[-\omega_{l=0,n} \theta(-\omega_{l=0,n}) +\tilde{\omega}_{l=0,n}\theta(\tilde{\omega}_{l=0,n}) \right]\right. \notag\\
&\left.
-L_z \int \frac{dk_z}{2\pi} \left[-\omega_{l=0} \theta(-\omega_{l=0}) +\tilde{\omega}_{l=0}\theta(\tilde{\omega}_{l=0}) \right]\right\}.
\end{align}
Note that, in the right-hand side, the $(b-\mu \to 0)$ term is not needed because it is zero due to $\omega_{l=0}\geq0$ and $\tilde{\omega}_{l=0}\leq0$.

\begin{figure}[b!]
\includegraphics[clip,width=1.0\columnwidth]{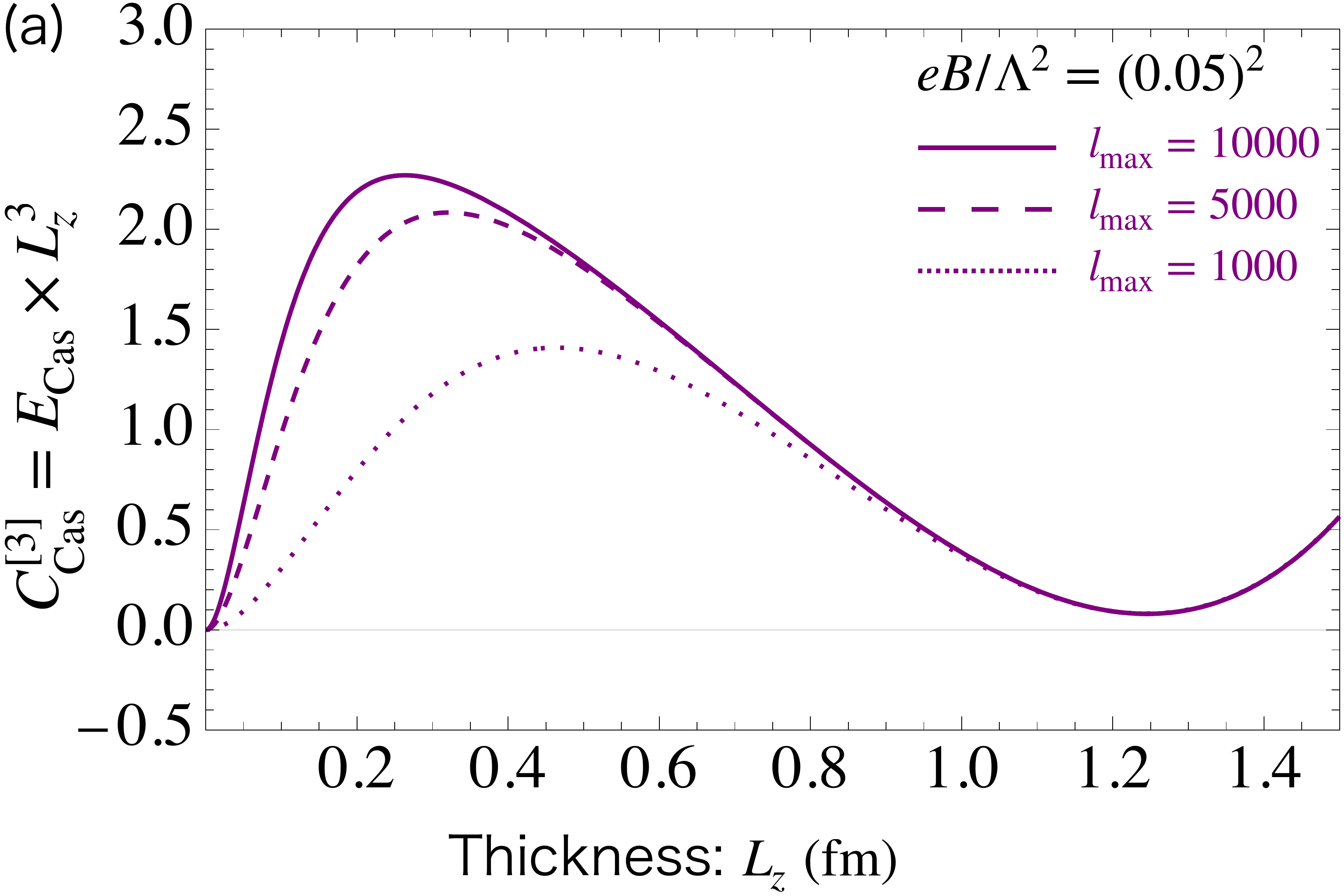}
\includegraphics[clip,width=1.0\columnwidth]{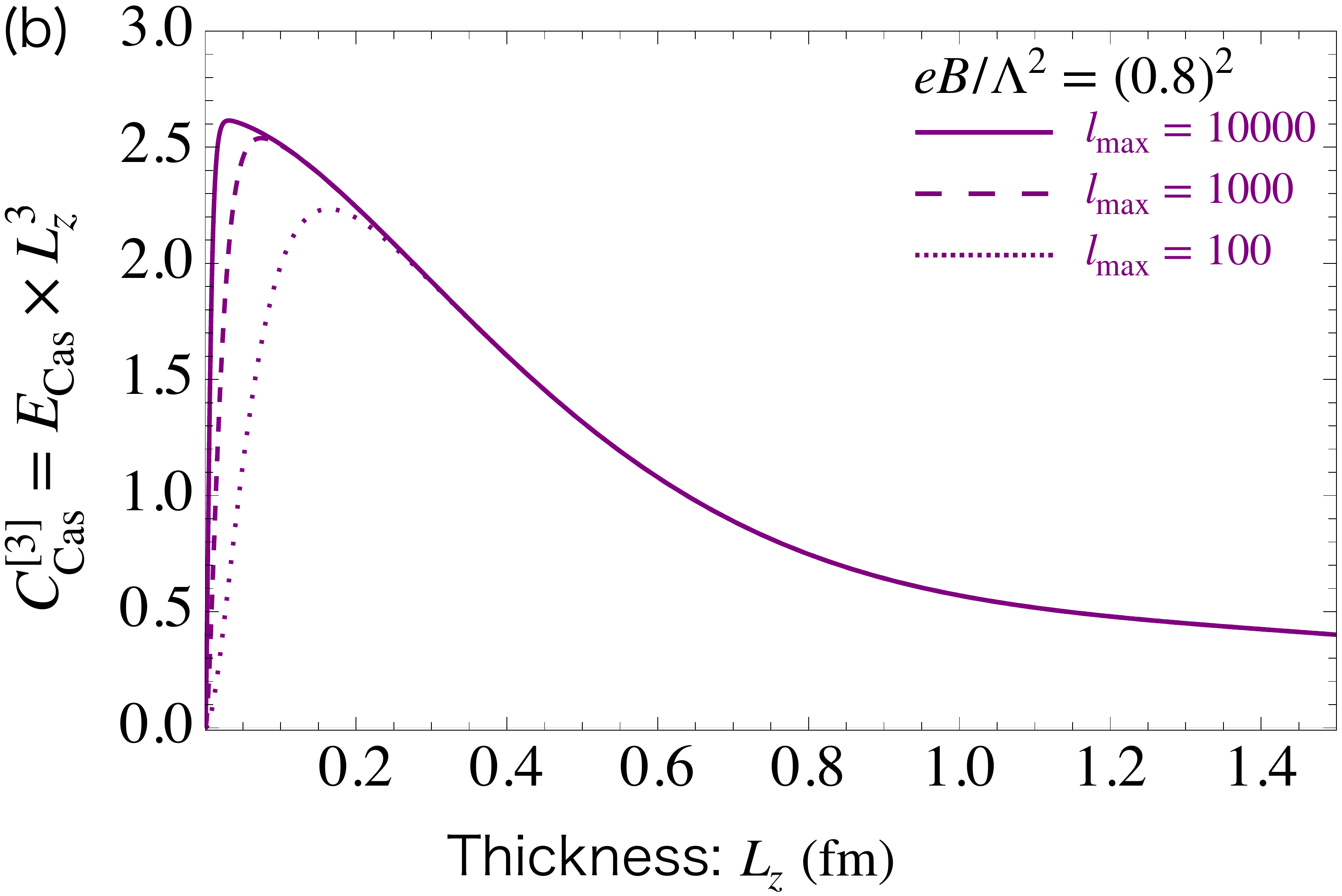}
\caption{
$l_{\rm max}$ dependence of Casimir coefficient, where $l_{\rm max}$ is the maximum number of summed Landau levels.
(a) The case in a weak magnetic field as in Fig.~\ref{CCas_MDCDW_weak}.
(b) The case in a strong magnetic field as in Fig.~\ref{CCas_MDCDW_strong}.
}
\label{ldep}
\end{figure}

\section{Dependence on summation of LLs} \label{App:LL}
In our numerical analysis in Sec.~\ref{Sec:result}, we summed LLs up to a maximum number $l_{\rm max}$, which is large enough, and ignored higher levels.
For the calculation of Casimir energy, this approximation is valid when $L_z$ is long enough but is not sufficient when $L_z$ is very short.  
In this appendix, we examine the dependence of Casimir energy on $l_{\rm max}$.

In Fig.~\ref{ldep}, we compare some results by different $l_{\rm max}$.
Here, we show the total Casimir energy in the two-flavor MDCDW phase under a weak magnetic field [as in Fig.~\ref{CCas_MDCDW_weak} (b)] or under a strong magnetic field [as in Fig.~\ref{CCas_MDCDW_strong} (b)].
As shown in Fig.~\ref{ldep} (a), in a weak magnetic field, we can see that $l_{\rm max}=1000$ is sufficient up to $L_z\sim1.2$ fm while $l_{\rm max}=5000$ is up to $L_z\sim0.7$ fm.
As shown in Fig.~\ref{ldep} (b), in a strong magnetic field, we can see that $l_{\rm max}=5000$ and $10000$ are sufficient up to $L_z\sim 0.1$ fm and near $L_z\sim 0$, respectively.
Thus, if the magnetic field is strong enough, we can obtain almost exact results by summing many Landau levels.

Also, we find that, at $L_z \sim 0$, the results using a sufficient $l_\mathrm{max}$ approach to $C_\mathrm{Cas}^{[3]} = N_fN_c \times 2\pi^2/45$, where $2\pi^2/45$ is the factor known as the Casimir energy from the massless Dirac field in the $3+1$-dimensional spacetime with the PBC in the $z$ direction.

\bibliography{ref}

%apsrev4-2.bst 2019-01-14 (MD) hand-edited version of apsrev4-1.bst
%Control: key (0)
%Control: author (8) initials jnrlst
%Control: editor formatted (1) identically to author
%Control: production of article title (0) allowed
%Control: page (0) single
%Control: year (1) truncated
%Control: production of eprint (0) enabled
\begin{thebibliography}{103}%
\makeatletter
\providecommand \@ifxundefined [1]{%
 \@ifx{#1\undefined}
}%
\providecommand \@ifnum [1]{%
 \ifnum #1\expandafter \@firstoftwo
 \else \expandafter \@secondoftwo
 \fi
}%
\providecommand \@ifx [1]{%
 \ifx #1\expandafter \@firstoftwo
 \else \expandafter \@secondoftwo
 \fi
}%
\providecommand \natexlab [1]{#1}%
\providecommand \enquote  [1]{``#1''}%
\providecommand \bibnamefont  [1]{#1}%
\providecommand \bibfnamefont [1]{#1}%
\providecommand \citenamefont [1]{#1}%
\providecommand \href@noop [0]{\@secondoftwo}%
\providecommand \href [0]{\begingroup \@sanitize@url \@href}%
\providecommand \@href[1]{\@@startlink{#1}\@@href}%
\providecommand \@@href[1]{\endgroup#1\@@endlink}%
\providecommand \@sanitize@url [0]{\catcode `\\12\catcode `\$12\catcode
  `\&12\catcode `\#12\catcode `\^12\catcode `\_12\catcode `\%12\relax}%
\providecommand \@@startlink[1]{}%
\providecommand \@@endlink[0]{}%
\providecommand \url  [0]{\begingroup\@sanitize@url \@url }%
\providecommand \@url [1]{\endgroup\@href {#1}{\urlprefix }}%
\providecommand \urlprefix  [0]{URL }%
\providecommand \Eprint [0]{\href }%
\providecommand \doibase [0]{https://doi.org/}%
\providecommand \selectlanguage [0]{\@gobble}%
\providecommand \bibinfo  [0]{\@secondoftwo}%
\providecommand \bibfield  [0]{\@secondoftwo}%
\providecommand \translation [1]{[#1]}%
\providecommand \BibitemOpen [0]{}%
\providecommand \bibitemStop [0]{}%
\providecommand \bibitemNoStop [0]{.\EOS\space}%
\providecommand \EOS [0]{\spacefactor3000\relax}%
\providecommand \BibitemShut  [1]{\csname bibitem#1\endcsname}%
\let\auto@bib@innerbib\@empty
%</preamble>
\bibitem [{\citenamefont {Casimir}(1948)}]{Casimir:1948dh}%
  \BibitemOpen
  \bibfield  {author} {\bibinfo {author} {\bibfnamefont {H.~B.~G.}\
  \bibnamefont {Casimir}},\ }\bibfield  {title} {\bibinfo {title} {{On the
  Attraction Between Two Perfectly Conducting Plates}},\ }\href@noop {}
  {\bibfield  {journal} {\bibinfo  {journal} {Proc. Kon. Ned. Akad. Wetensch.}\
  }\textbf {\bibinfo {volume} {51}},\ \bibinfo {pages} {793} (\bibinfo {year}
  {1948})}\BibitemShut {NoStop}%
%%CITATION = IMTHB,10,261;%%
\bibitem [{\citenamefont {Lamoreaux}(1997)}]{Lamoreaux:1996wh}%
  \BibitemOpen
  \bibfield  {author} {\bibinfo {author} {\bibfnamefont {S.~K.}\ \bibnamefont
  {Lamoreaux}},\ }\bibfield  {title} {\bibinfo {title} {{Demonstration of the
  Casimir Force in the 0.6 to $6\ensuremath{\mu}m$ Range}},\ }\href
  {https://doi.org/10.1103/PhysRevLett.78.5} {\bibfield  {journal} {\bibinfo
  {journal} {Phys. Rev. Lett.}\ }\textbf {\bibinfo {volume} {78}},\ \bibinfo
  {pages} {5} (\bibinfo {year} {1997})},\ \bibinfo {note} {[Erratum:
  \href{https://doi.org/10.1103/PhysRevLett.81.5475}{Phys. Rev. Lett. {\bf 81},
  5475 (1998)}]}\BibitemShut {NoStop}%
%%CITATION = PRLTA,78,5;%%
\bibitem [{\citenamefont {Bressi}\ \emph {et~al.}(2002)\citenamefont {Bressi},
  \citenamefont {Carugno}, \citenamefont {Onofrio},\ and\ \citenamefont
  {Ruoso}}]{Bressi:2002fr}%
  \BibitemOpen
  \bibfield  {author} {\bibinfo {author} {\bibfnamefont {G.}~\bibnamefont
  {Bressi}}, \bibinfo {author} {\bibfnamefont {G.}~\bibnamefont {Carugno}},
  \bibinfo {author} {\bibfnamefont {R.}~\bibnamefont {Onofrio}},\ and\ \bibinfo
  {author} {\bibfnamefont {G.}~\bibnamefont {Ruoso}},\ }\bibfield  {title}
  {\bibinfo {title} {{Measurement of the Casimir force between parallel
  metallic surfaces}},\ }\href {https://doi.org/10.1103/PhysRevLett.88.041804}
  {\bibfield  {journal} {\bibinfo  {journal} {Phys. Rev. Lett.}\ }\textbf
  {\bibinfo {volume} {88}},\ \bibinfo {pages} {041804} (\bibinfo {year}
  {2002})},\ \Eprint {https://arxiv.org/abs/quant-ph/0203002}
  {arXiv:quant-ph/0203002} \BibitemShut {NoStop}%
\bibitem [{\citenamefont {Plunien}\ \emph {et~al.}(1986)\citenamefont
  {Plunien}, \citenamefont {M\"uller},\ and\ \citenamefont
  {Greiner}}]{Plunien:1986ca}%
  \BibitemOpen
  \bibfield  {author} {\bibinfo {author} {\bibfnamefont {G.}~\bibnamefont
  {Plunien}}, \bibinfo {author} {\bibfnamefont {B.}~\bibnamefont {M\"uller}},\
  and\ \bibinfo {author} {\bibfnamefont {W.}~\bibnamefont {Greiner}},\
  }\bibfield  {title} {\bibinfo {title} {{The Casimir Effect}},\ }\href
  {https://doi.org/10.1016/0370-1573(86)90020-7} {\bibfield  {journal}
  {\bibinfo  {journal} {Phys. Rep.}\ }\textbf {\bibinfo {volume} {134}},\
  \bibinfo {pages} {87} (\bibinfo {year} {1986})}\BibitemShut {NoStop}%
\bibitem [{\citenamefont {Mostepanenko}\ and\ \citenamefont
  {Trunov}(1988)}]{Mostepanenko:1988bs}%
  \BibitemOpen
  \bibfield  {author} {\bibinfo {author} {\bibfnamefont {V.~M.}\ \bibnamefont
  {Mostepanenko}}\ and\ \bibinfo {author} {\bibfnamefont {N.~N.}\ \bibnamefont
  {Trunov}},\ }\bibfield  {title} {\bibinfo {title} {{The Casimir Effect and
  Its Applications}},\ }\href {https://doi.org/10.1070/PU1988v031n11ABEH005641}
  {\bibfield  {journal} {\bibinfo  {journal} {Sov. Phys. Usp.}\ }\textbf
  {\bibinfo {volume} {31}},\ \bibinfo {pages} {965} (\bibinfo {year}
  {1988})}\BibitemShut {NoStop}%
\bibitem [{\citenamefont {Bordag}\ \emph {et~al.}(2001)\citenamefont {Bordag},
  \citenamefont {Mohideen},\ and\ \citenamefont
  {Mostepanenko}}]{Bordag:2001qi}%
  \BibitemOpen
  \bibfield  {author} {\bibinfo {author} {\bibfnamefont {M.}~\bibnamefont
  {Bordag}}, \bibinfo {author} {\bibfnamefont {U.}~\bibnamefont {Mohideen}},\
  and\ \bibinfo {author} {\bibfnamefont {V.~M.}\ \bibnamefont {Mostepanenko}},\
  }\bibfield  {title} {\bibinfo {title} {{New developments in the Casimir
  effect}},\ }\href {https://doi.org/10.1016/S0370-1573(01)00015-1} {\bibfield
  {journal} {\bibinfo  {journal} {Phys. Rep.}\ }\textbf {\bibinfo {volume}
  {353}},\ \bibinfo {pages} {1} (\bibinfo {year} {2001})},\ \Eprint
  {https://arxiv.org/abs/quant-ph/0106045} {arXiv:quant-ph/0106045 [quant-ph]}
  \BibitemShut {NoStop}%
%%CITATION = QUANT-PH/0106045;%%
\bibitem [{\citenamefont {Milton}(2001)}]{Milton:2001yy}%
  \BibitemOpen
  \bibfield  {author} {\bibinfo {author} {\bibfnamefont {K.~A.}\ \bibnamefont
  {Milton}},\ }\href {https://doi.org/10.1142/4505} {\emph {\bibinfo {title}
  {{The Casimir Effect: Physical Manifestations of Zero-Point Energy}}}}\
  (\bibinfo  {publisher} {World Scientific, Singapore},\ \bibinfo {year}
  {2001})\BibitemShut {NoStop}%
\bibitem [{\citenamefont {Klimchitskaya}\ \emph {et~al.}(2009)\citenamefont
  {Klimchitskaya}, \citenamefont {Mohideen},\ and\ \citenamefont
  {Mostepanenko}}]{Klimchitskaya:2009cw}%
  \BibitemOpen
  \bibfield  {author} {\bibinfo {author} {\bibfnamefont {G.~L.}\ \bibnamefont
  {Klimchitskaya}}, \bibinfo {author} {\bibfnamefont {U.}~\bibnamefont
  {Mohideen}},\ and\ \bibinfo {author} {\bibfnamefont {V.~M.}\ \bibnamefont
  {Mostepanenko}},\ }\bibfield  {title} {\bibinfo {title} {{The Casimir force
  between real materials: Experiment and theory}},\ }\href
  {https://doi.org/10.1103/RevModPhys.81.1827} {\bibfield  {journal} {\bibinfo
  {journal} {Rev. Mod. Phys.}\ }\textbf {\bibinfo {volume} {81}},\ \bibinfo
  {pages} {1827} (\bibinfo {year} {2009})},\ \Eprint
  {https://arxiv.org/abs/0902.4022} {arXiv:0902.4022 [cond-mat.other]}
  \BibitemShut {NoStop}%
\bibitem [{\citenamefont {Woods}\ \emph {et~al.}(2016)\citenamefont {Woods},
  \citenamefont {Dalvit}, \citenamefont {Tkatchenko}, \citenamefont
  {Rodriguez-Lopez}, \citenamefont {Rodriguez},\ and\ \citenamefont
  {Podgornik}}]{Woods:2015pla}%
  \BibitemOpen
  \bibfield  {author} {\bibinfo {author} {\bibfnamefont {L.~M.}\ \bibnamefont
  {Woods}}, \bibinfo {author} {\bibfnamefont {D.~A.~R.}\ \bibnamefont
  {Dalvit}}, \bibinfo {author} {\bibfnamefont {A.}~\bibnamefont {Tkatchenko}},
  \bibinfo {author} {\bibfnamefont {P.}~\bibnamefont {Rodriguez-Lopez}},
  \bibinfo {author} {\bibfnamefont {A.~W.}\ \bibnamefont {Rodriguez}},\ and\
  \bibinfo {author} {\bibfnamefont {R.}~\bibnamefont {Podgornik}},\ }\bibfield
  {title} {\bibinfo {title} {{Materials perspective on Casimir and van der
  Waals interactions}},\ }\href {https://doi.org/10.1103/RevModPhys.88.45003}
  {\bibfield  {journal} {\bibinfo  {journal} {Rev. Mod. Phys.}\ }\textbf
  {\bibinfo {volume} {88}},\ \bibinfo {pages} {045003} (\bibinfo {year}
  {2016})},\ \Eprint {https://arxiv.org/abs/1509.03338} {arXiv:1509.03338
  [cond-mat.mtrl-sci]} \BibitemShut {NoStop}%
\bibitem [{\citenamefont {Gong}\ \emph {et~al.}(2021)\citenamefont {Gong},
  \citenamefont {Corrado}, \citenamefont {Mahbub}, \citenamefont {Shelden},\
  and\ \citenamefont {Munday}}]{Gong:2020ttb}%
  \BibitemOpen
  \bibfield  {author} {\bibinfo {author} {\bibfnamefont {T.}~\bibnamefont
  {Gong}}, \bibinfo {author} {\bibfnamefont {M.~R.}\ \bibnamefont {Corrado}},
  \bibinfo {author} {\bibfnamefont {A.~R.}\ \bibnamefont {Mahbub}}, \bibinfo
  {author} {\bibfnamefont {C.}~\bibnamefont {Shelden}},\ and\ \bibinfo {author}
  {\bibfnamefont {J.~N.}\ \bibnamefont {Munday}},\ }\bibfield  {title}
  {\bibinfo {title} {{Recent progress in engineering the Casimir effect
  \textendash{} applications to nanophotonics, nanomechanics, and chemistry}},\
  }\href {https://doi.org/10.1515/nanoph-2020-0425} {\bibfield  {journal}
  {\bibinfo  {journal} {Nanophotonics}\ }\textbf {\bibinfo {volume} {10}},\
  \bibinfo {pages} {523} (\bibinfo {year} {2021})}\BibitemShut {NoStop}%
\bibitem [{\citenamefont {Lu}(2021)}]{Lu:2021jvu}%
  \BibitemOpen
  \bibfield  {author} {\bibinfo {author} {\bibfnamefont {B.-S.}\ \bibnamefont
  {Lu}},\ }\bibfield  {title} {\bibinfo {title} {{The Casimir Effect in
  Topological Matter}},\ }\href {https://doi.org/10.3390/universe7070237}
  {\bibfield  {journal} {\bibinfo  {journal} {Universe}\ }\textbf {\bibinfo
  {volume} {7}},\ \bibinfo {pages} {237} (\bibinfo {year} {2021})},\ \Eprint
  {https://arxiv.org/abs/2105.11059} {arXiv:2105.11059 [cond-mat.mes-hall]}
  \BibitemShut {NoStop}%
\bibitem [{\citenamefont {Johnson}(1975)}]{Johnson:1975zp}%
  \BibitemOpen
  \bibfield  {author} {\bibinfo {author} {\bibfnamefont {K.}~\bibnamefont
  {Johnson}},\ }\bibfield  {title} {\bibinfo {title} {{The M.I.T. Bag Model}},\
  }\href@noop {} {\bibfield  {journal} {\bibinfo  {journal} {Acta Phys. Pol.
  B}\ }\textbf {\bibinfo {volume} {6}},\ \bibinfo {pages} {865} (\bibinfo
  {year} {1975})}\BibitemShut {NoStop}%
\bibitem [{\citenamefont {Mamaev}\ and\ \citenamefont
  {Trunov}(1980)}]{Mamaev:1980jn}%
  \BibitemOpen
  \bibfield  {author} {\bibinfo {author} {\bibfnamefont {S.~G.}\ \bibnamefont
  {Mamaev}}\ and\ \bibinfo {author} {\bibfnamefont {N.~N.}\ \bibnamefont
  {Trunov}},\ }\bibfield  {title} {\bibinfo {title} {{Vacuum expectation values
  of the energy-momentum tensor of quantized fields on manifolds with different
  topologies and geometries. III}},\ }\href
  {https://doi.org/10.1007/BF00891938} {\bibfield  {journal} {\bibinfo
  {journal} {Sov. Phys. J.}\ }\textbf {\bibinfo {volume} {23}},\ \bibinfo
  {pages} {551} (\bibinfo {year} {1980})}\BibitemShut {NoStop}%
%%CITATION = SOPJA,23,551;%%
\bibitem [{\citenamefont {Dzyaloshinskii}\ \emph {et~al.}(1961)\citenamefont
  {Dzyaloshinskii}, \citenamefont {Lifshitz},\ and\ \citenamefont
  {Pitaevskii}}]{Dzyaloshinskii:1961sfr}%
  \BibitemOpen
  \bibfield  {author} {\bibinfo {author} {\bibfnamefont {I.~E.}\ \bibnamefont
  {Dzyaloshinskii}}, \bibinfo {author} {\bibfnamefont {E.~M.}\ \bibnamefont
  {Lifshitz}},\ and\ \bibinfo {author} {\bibfnamefont {L.~P.}\ \bibnamefont
  {Pitaevskii}},\ }\bibfield  {title} {\bibinfo {title} {{The general theory of
  van der Waals forces}},\ }\href {https://doi.org/10.1080/00018736100101281}
  {\bibfield  {journal} {\bibinfo  {journal} {Adv. Phys.}\ }\textbf {\bibinfo
  {volume} {10}},\ \bibinfo {pages} {165} (\bibinfo {year} {1961})}\BibitemShut
  {NoStop}%
\bibitem [{\citenamefont {Robaschik}\ \emph {et~al.}(1987)\citenamefont
  {Robaschik}, \citenamefont {Scharnhorst},\ and\ \citenamefont
  {Wieczorek}}]{Robaschik:1986vj}%
  \BibitemOpen
  \bibfield  {author} {\bibinfo {author} {\bibfnamefont {D.}~\bibnamefont
  {Robaschik}}, \bibinfo {author} {\bibfnamefont {K.}~\bibnamefont
  {Scharnhorst}},\ and\ \bibinfo {author} {\bibfnamefont {E.}~\bibnamefont
  {Wieczorek}},\ }\bibfield  {title} {\bibinfo {title} {{Radiative corrections
  to the Casimir pressure under the influence of temperature and external
  fields}},\ }\href {https://doi.org/10.1016/0003-4916(87)90034-0} {\bibfield
  {journal} {\bibinfo  {journal} {Annals Phys.}\ }\textbf {\bibinfo {volume}
  {174}},\ \bibinfo {pages} {401} (\bibinfo {year} {1987})}\BibitemShut
  {NoStop}%
\bibitem [{\citenamefont {Cougo-Pinto}\ \emph
  {et~al.}(1999{\natexlab{a}})\citenamefont {Cougo-Pinto}, \citenamefont
  {Farina},\ and\ \citenamefont {Tort}}]{Cougo-Pinto:1998jwo}%
  \BibitemOpen
  \bibfield  {author} {\bibinfo {author} {\bibfnamefont {M.~V.}\ \bibnamefont
  {Cougo-Pinto}}, \bibinfo {author} {\bibfnamefont {C.}~\bibnamefont
  {Farina}},\ and\ \bibinfo {author} {\bibfnamefont {A.~C.}\ \bibnamefont
  {Tort}},\ }\bibfield  {title} {\bibinfo {title} {{Fermionic Casimir effect in
  an external magnetic field}},\ }\href@noop {} {\bibfield  {journal} {\bibinfo
   {journal} {Conf. Proc. C}\ }\textbf {\bibinfo {volume} {9809142}},\ \bibinfo
  {pages} {235} (\bibinfo {year} {1999}{\natexlab{a}})},\ \Eprint
  {https://arxiv.org/abs/hep-th/9809215} {arXiv:hep-th/9809215} \BibitemShut
  {NoStop}%
\bibitem [{\citenamefont {Cougo-Pinto}\ \emph {et~al.}(2001)\citenamefont
  {Cougo-Pinto}, \citenamefont {Farina},\ and\ \citenamefont
  {Tort}}]{Cougo-Pinto:2001kks}%
  \BibitemOpen
  \bibfield  {author} {\bibinfo {author} {\bibfnamefont {M.~V.}\ \bibnamefont
  {Cougo-Pinto}}, \bibinfo {author} {\bibfnamefont {C.}~\bibnamefont
  {Farina}},\ and\ \bibinfo {author} {\bibfnamefont {A.~C.}\ \bibnamefont
  {Tort}},\ }\bibfield  {title} {\bibinfo {title} {{The influence of an
  external magnetic field on the fermionic Casimir effect}},\ }\href
  {https://doi.org/10.1590/S0103-97332001000100016} {\bibfield  {journal}
  {\bibinfo  {journal} {Braz. J. Phys.}\ }\textbf {\bibinfo {volume} {31}},\
  \bibinfo {pages} {84} (\bibinfo {year} {2001})}\BibitemShut {NoStop}%
\bibitem [{\citenamefont {Elizalde}\ \emph {et~al.}(2002)\citenamefont
  {Elizalde}, \citenamefont {Santos},\ and\ \citenamefont
  {Tort}}]{Elizalde:2002kb}%
  \BibitemOpen
  \bibfield  {author} {\bibinfo {author} {\bibfnamefont {E.}~\bibnamefont
  {Elizalde}}, \bibinfo {author} {\bibfnamefont {F.~C.}\ \bibnamefont
  {Santos}},\ and\ \bibinfo {author} {\bibfnamefont {A.~C.}\ \bibnamefont
  {Tort}},\ }\bibfield  {title} {\bibinfo {title} {{Confined quantum fields
  under the influence of a uniform magnetic field}},\ }\href
  {https://doi.org/10.1088/0305-4470/35/34/311} {\bibfield  {journal} {\bibinfo
   {journal} {J. Phys. A}\ }\textbf {\bibinfo {volume} {35}},\ \bibinfo {pages}
  {7403} (\bibinfo {year} {2002})},\ \Eprint
  {https://arxiv.org/abs/hep-th/0206143} {arXiv:hep-th/0206143} \BibitemShut
  {NoStop}%
\bibitem [{\citenamefont {Ostrowski}(2006)}]{Ostrowski:2005rm}%
  \BibitemOpen
  \bibfield  {author} {\bibinfo {author} {\bibfnamefont {M.}~\bibnamefont
  {Ostrowski}},\ }\bibfield  {title} {\bibinfo {title} {{Casimir effect in
  external magnetic field}},\ }\href
  {https://www.actaphys.uj.edu.pl/R/37/6/1753} {\bibfield  {journal} {\bibinfo
  {journal} {Acta Phys. Polon. B}\ }\textbf {\bibinfo {volume} {37}},\ \bibinfo
  {pages} {1753} (\bibinfo {year} {2006})},\ \Eprint
  {https://arxiv.org/abs/hep-th/0504112} {arXiv:hep-th/0504112} \BibitemShut
  {NoStop}%
\bibitem [{\citenamefont {Miltao}\ and\ \citenamefont
  {Farias}(2008)}]{Miltao:2008zza}%
  \BibitemOpen
  \bibfield  {author} {\bibinfo {author} {\bibfnamefont {M.~S.~R.}\
  \bibnamefont {Miltao}}\ and\ \bibinfo {author} {\bibfnamefont {F.~A.}\
  \bibnamefont {Farias}},\ }\bibfield  {title} {\bibinfo {title} {{The Casimir
  energy of Dirac field under a general boundary condition using the zeta
  function method}},\ }\href {https://www.actaphys.uj.edu.pl/R/39/8/1931}
  {\bibfield  {journal} {\bibinfo  {journal} {Acta Phys. Polon. B}\ }\textbf
  {\bibinfo {volume} {39}},\ \bibinfo {pages} {1931} (\bibinfo {year}
  {2008})}\BibitemShut {NoStop}%
\bibitem [{\citenamefont {Sitenko}(2015{\natexlab{a}})}]{Sitenko:2014kza}%
  \BibitemOpen
  \bibfield  {author} {\bibinfo {author} {\bibfnamefont {Y.~A.}\ \bibnamefont
  {Sitenko}},\ }\bibfield  {title} {\bibinfo {title} {{Casimir effect with
  quantized charged spinor matter in background magnetic field}},\ }\href
  {https://doi.org/10.1103/PhysRevD.91.085012} {\bibfield  {journal} {\bibinfo
  {journal} {Phys. Rev. D}\ }\textbf {\bibinfo {volume} {91}},\ \bibinfo
  {pages} {085012} (\bibinfo {year} {2015}{\natexlab{a}})},\ \Eprint
  {https://arxiv.org/abs/1411.2460} {arXiv:1411.2460 [hep-th]} \BibitemShut
  {NoStop}%
\bibitem [{\citenamefont {Sitenko}(2015{\natexlab{b}})}]{Sitenko:2015eza}%
  \BibitemOpen
  \bibfield  {author} {\bibinfo {author} {\bibfnamefont {Y.~A.}\ \bibnamefont
  {Sitenko}},\ }\bibfield  {title} {\bibinfo {title} {{Influence of quantized
  massive matter fields on the Casimir effect}},\ }\href
  {https://doi.org/10.1142/S0217732315500996} {\bibfield  {journal} {\bibinfo
  {journal} {Mod. Phys. Lett. A}\ }\textbf {\bibinfo {volume} {30}},\ \bibinfo
  {pages} {1550099} (\bibinfo {year} {2015}{\natexlab{b}})},\ \Eprint
  {https://arxiv.org/abs/1506.05034} {arXiv:1506.05034 [hep-th]} \BibitemShut
  {NoStop}%
\bibitem [{\citenamefont {Sitenko}\ and\ \citenamefont
  {Yushchenko}(2015)}]{Sitenko:2015wzd}%
  \BibitemOpen
  \bibfield  {author} {\bibinfo {author} {\bibfnamefont {Y.~A.}\ \bibnamefont
  {Sitenko}}\ and\ \bibinfo {author} {\bibfnamefont {S.~A.}\ \bibnamefont
  {Yushchenko}},\ }\bibfield  {title} {\bibinfo {title} {{Pressure from the
  vacuum of confined spinor matter}},\ }\href
  {https://doi.org/10.1142/S0217751X15501845} {\bibfield  {journal} {\bibinfo
  {journal} {Int. J. Mod. Phys. A}\ }\textbf {\bibinfo {volume} {30}},\
  \bibinfo {pages} {1550184} (\bibinfo {year} {2015})},\ \Eprint
  {https://arxiv.org/abs/1512.01397} {arXiv:1512.01397 [hep-th]} \BibitemShut
  {NoStop}%
\bibitem [{\citenamefont {Nakayama}\ and\ \citenamefont
  {Suzuki}(2023{\natexlab{a}})}]{Nakayama:2022fvh}%
  \BibitemOpen
  \bibfield  {author} {\bibinfo {author} {\bibfnamefont {K.}~\bibnamefont
  {Nakayama}}\ and\ \bibinfo {author} {\bibfnamefont {K.}~\bibnamefont
  {Suzuki}},\ }\bibfield  {title} {\bibinfo {title} {{Dirac/Weyl-node-induced
  oscillating Casimir effect}},\ }\href
  {https://doi.org/10.1016/j.physletb.2023.138017} {\bibfield  {journal}
  {\bibinfo  {journal} {Phys. Lett. B}\ }\textbf {\bibinfo {volume} {843}},\
  \bibinfo {pages} {138017} (\bibinfo {year} {2023}{\natexlab{a}})},\ \Eprint
  {https://arxiv.org/abs/2207.14078} {arXiv:2207.14078 [cond-mat.mes-hall]}
  \BibitemShut {NoStop}%
\bibitem [{\citenamefont {Rohim}\ \emph {et~al.}(2024)\citenamefont {Rohim},
  \citenamefont {Romadani},\ and\ \citenamefont {Salim~Adam}}]{Rohim:2023tmy}%
  \BibitemOpen
  \bibfield  {author} {\bibinfo {author} {\bibfnamefont {A.}~\bibnamefont
  {Rohim}}, \bibinfo {author} {\bibfnamefont {A.}~\bibnamefont {Romadani}},\
  and\ \bibinfo {author} {\bibfnamefont {A.}~\bibnamefont {Salim~Adam}},\
  }\bibfield  {title} {\bibinfo {title} {{Casimir effect of Lorentz-violating
  charged Dirac field in background magnetic field}},\ }\href
  {https://doi.org/10.1093/ptep/ptae016} {\bibfield  {journal} {\bibinfo
  {journal} {Prog. Theor. Exp. Phys.}\ }\textbf {\bibinfo {volume} {2024}},\
  \bibinfo {pages} {033B01} (\bibinfo {year} {2024})},\ \Eprint
  {https://arxiv.org/abs/2307.04448} {arXiv:2307.04448 [hep-th]} \BibitemShut
  {NoStop}%
\bibitem [{\citenamefont {Erdas}(2023)}]{Erdas:2023wzy}%
  \BibitemOpen
  \bibfield  {author} {\bibinfo {author} {\bibfnamefont {A.}~\bibnamefont
  {Erdas}},\ }\bibfield  {title} {\bibinfo {title} {{Magnetic corrections to
  the fermionic Casimir effect in Horava-Lifshitz theories}},\ }\href
  {https://doi.org/10.1142/S0217751X23501178} {\bibfield  {journal} {\bibinfo
  {journal} {Int. J. Mod. Phys. A}\ }\textbf {\bibinfo {volume} {38}},\
  \bibinfo {pages} {2350117} (\bibinfo {year} {2023})},\ \Eprint
  {https://arxiv.org/abs/2307.06228} {arXiv:2307.06228 [hep-th]} \BibitemShut
  {NoStop}%
\bibitem [{\citenamefont {Flachi}\ \emph {et~al.}(2025)\citenamefont {Flachi},
  \citenamefont {Nitta}, \citenamefont {Takada},\ and\ \citenamefont
  {Yoshii}}]{Flachi:2024ztd}%
  \BibitemOpen
  \bibfield  {author} {\bibinfo {author} {\bibfnamefont {A.}~\bibnamefont
  {Flachi}}, \bibinfo {author} {\bibfnamefont {M.}~\bibnamefont {Nitta}},
  \bibinfo {author} {\bibfnamefont {S.}~\bibnamefont {Takada}},\ and\ \bibinfo
  {author} {\bibfnamefont {R.}~\bibnamefont {Yoshii}},\ }\bibfield  {title}
  {\bibinfo {title} {{Fermion Casimir effect and magnetic Larkin-Ovchinnikov
  phases}},\ }\href {https://doi.org/10.1103/PhysRevD.111.016003} {\bibfield
  {journal} {\bibinfo  {journal} {Phys. Rev. D}\ }\textbf {\bibinfo {volume}
  {111}},\ \bibinfo {pages} {016003} (\bibinfo {year} {2025})},\ \Eprint
  {https://arxiv.org/abs/2410.18771} {arXiv:2410.18771 [hep-th]} \BibitemShut
  {NoStop}%
\bibitem [{\citenamefont {Cougo-Pinto}\ \emph
  {et~al.}(1999{\natexlab{b}})\citenamefont {Cougo-Pinto}, \citenamefont
  {Farina}, \citenamefont {Negrao},\ and\ \citenamefont
  {Tort}}]{Cougo-Pinto:1998jun}%
  \BibitemOpen
  \bibfield  {author} {\bibinfo {author} {\bibfnamefont {M.~V.}\ \bibnamefont
  {Cougo-Pinto}}, \bibinfo {author} {\bibfnamefont {C.}~\bibnamefont {Farina}},
  \bibinfo {author} {\bibfnamefont {M.~R.}\ \bibnamefont {Negrao}},\ and\
  \bibinfo {author} {\bibfnamefont {A.~C.}\ \bibnamefont {Tort}},\ }\bibfield
  {title} {\bibinfo {title} {{Bosonic Casimir effect in an external magnetic
  field}},\ }\href {https://doi.org/10.1088/0305-4470/32/24/310} {\bibfield
  {journal} {\bibinfo  {journal} {J. Phys. A}\ }\textbf {\bibinfo {volume}
  {32}},\ \bibinfo {pages} {4457} (\bibinfo {year} {1999}{\natexlab{b}})},\
  \Eprint {https://arxiv.org/abs/hep-th/9809214} {arXiv:hep-th/9809214}
  \BibitemShut {NoStop}%
\bibitem [{\citenamefont {Cougo-Pinto}\ \emph {et~al.}(2000)\citenamefont
  {Cougo-Pinto}, \citenamefont {Farina}, \citenamefont {Negrao},\ and\
  \citenamefont {Tort}}]{Cougo-Pinto:1998mdw}%
  \BibitemOpen
  \bibfield  {author} {\bibinfo {author} {\bibfnamefont {M.~V.}\ \bibnamefont
  {Cougo-Pinto}}, \bibinfo {author} {\bibfnamefont {C.}~\bibnamefont {Farina}},
  \bibinfo {author} {\bibfnamefont {M.~R.}\ \bibnamefont {Negrao}},\ and\
  \bibinfo {author} {\bibfnamefont {A.~C.}\ \bibnamefont {Tort}},\ }\bibfield
  {title} {\bibinfo {title} {{Magnetic permeability of constrained scalar QED
  vacuum}},\ }\href {https://doi.org/10.1016/S0370-2693(00)00547-5} {\bibfield
  {journal} {\bibinfo  {journal} {Phys. Lett. B}\ }\textbf {\bibinfo {volume}
  {483}},\ \bibinfo {pages} {144} (\bibinfo {year} {2000})},\ \Eprint
  {https://arxiv.org/abs/hep-th/9809216} {arXiv:hep-th/9809216} \BibitemShut
  {NoStop}%
\bibitem [{\citenamefont {Cougo-Pinto}\ \emph
  {et~al.}(1998{\natexlab{a}})\citenamefont {Cougo-Pinto}, \citenamefont
  {Farina}, \citenamefont {Negrao},\ and\ \citenamefont
  {Tort}}]{Cougo-Pinto:1998fpo}%
  \BibitemOpen
  \bibfield  {author} {\bibinfo {author} {\bibfnamefont {M.~V.}\ \bibnamefont
  {Cougo-Pinto}}, \bibinfo {author} {\bibfnamefont {C.}~\bibnamefont {Farina}},
  \bibinfo {author} {\bibfnamefont {M.~R.}\ \bibnamefont {Negrao}},\ and\
  \bibinfo {author} {\bibfnamefont {A.}~\bibnamefont {Tort}},\ }\bibfield
  {title} {\bibinfo {title} {{Casimir effect at finite temperature of charged
  scalar field in an external magnetic field}},\ }\href@noop {} {\  (\bibinfo
  {year} {1998}{\natexlab{a}})},\ \Eprint
  {https://arxiv.org/abs/hep-th/9810033} {arXiv:hep-th/9810033} \BibitemShut
  {NoStop}%
\bibitem [{\citenamefont {Cougo-Pinto}\ \emph
  {et~al.}(1998{\natexlab{b}})\citenamefont {Cougo-Pinto}, \citenamefont
  {Farina},\ and\ \citenamefont {Negrao}}]{Cougo-Pinto:1998zge}%
  \BibitemOpen
  \bibfield  {author} {\bibinfo {author} {\bibfnamefont {M.~V.}\ \bibnamefont
  {Cougo-Pinto}}, \bibinfo {author} {\bibfnamefont {C.}~\bibnamefont
  {Farina}},\ and\ \bibinfo {author} {\bibfnamefont {M.~R.}\ \bibnamefont
  {Negrao}},\ }\bibfield  {title} {\bibinfo {title} {{Magnetic properties of
  confined bosonic vacuum at finite temperature}},\ }\href@noop {} {\
  (\bibinfo {year} {1998}{\natexlab{b}})},\ \Eprint
  {https://arxiv.org/abs/hep-th/9811095} {arXiv:hep-th/9811095} \BibitemShut
  {NoStop}%
\bibitem [{\citenamefont {Erdas}\ and\ \citenamefont
  {Seltzer}(2013)}]{Erdas:2013jga}%
  \BibitemOpen
  \bibfield  {author} {\bibinfo {author} {\bibfnamefont {A.}~\bibnamefont
  {Erdas}}\ and\ \bibinfo {author} {\bibfnamefont {K.~P.}\ \bibnamefont
  {Seltzer}},\ }\bibfield  {title} {\bibinfo {title} {{Finite temperature
  Casimir effect for charged massless scalars in a magnetic field}},\ }\href
  {https://doi.org/10.1103/PhysRevD.88.105007} {\bibfield  {journal} {\bibinfo
  {journal} {Phys. Rev. D}\ }\textbf {\bibinfo {volume} {88}},\ \bibinfo
  {pages} {105007} (\bibinfo {year} {2013})},\ \Eprint
  {https://arxiv.org/abs/1304.6417} {arXiv:1304.6417 [hep-th]} \BibitemShut
  {NoStop}%
\bibitem [{\citenamefont {Erdas}\ and\ \citenamefont
  {Seltzer}(2014)}]{Erdas:2013dha}%
  \BibitemOpen
  \bibfield  {author} {\bibinfo {author} {\bibfnamefont {A.}~\bibnamefont
  {Erdas}}\ and\ \bibinfo {author} {\bibfnamefont {K.~P.}\ \bibnamefont
  {Seltzer}},\ }\bibfield  {title} {\bibinfo {title} {{Finite temperature
  Casimir effect for massive scalars in a magnetic field}},\ }\href
  {https://doi.org/10.1142/S0217751X14500912} {\bibfield  {journal} {\bibinfo
  {journal} {Int. J. Mod. Phys. A}\ }\textbf {\bibinfo {volume} {29}},\
  \bibinfo {pages} {1450091} (\bibinfo {year} {2014})},\ \Eprint
  {https://arxiv.org/abs/1312.1432} {arXiv:1312.1432 [hep-th]} \BibitemShut
  {NoStop}%
\bibitem [{\citenamefont {Sitenko}\ and\ \citenamefont
  {Yushchenko}(2014)}]{Sitenko:2014wwa}%
  \BibitemOpen
  \bibfield  {author} {\bibinfo {author} {\bibfnamefont {Y.~A.}\ \bibnamefont
  {Sitenko}}\ and\ \bibinfo {author} {\bibfnamefont {S.~A.}\ \bibnamefont
  {Yushchenko}},\ }\bibfield  {title} {\bibinfo {title} {{The Casimir effect
  with quantized charged scalar matter in background magnetic field}},\ }\href
  {https://doi.org/10.1142/S0217751X14500523} {\bibfield  {journal} {\bibinfo
  {journal} {Int. J. Mod. Phys. A}\ }\textbf {\bibinfo {volume} {29}},\
  \bibinfo {pages} {1450052} (\bibinfo {year} {2014})},\ \Eprint
  {https://arxiv.org/abs/1401.6950} {arXiv:1401.6950 [hep-th]} \BibitemShut
  {NoStop}%
\bibitem [{\citenamefont {Erdas}(2016)}]{Erdas:2015yac}%
  \BibitemOpen
  \bibfield  {author} {\bibinfo {author} {\bibfnamefont {A.}~\bibnamefont
  {Erdas}},\ }\bibfield  {title} {\bibinfo {title} {{Magnetic field corrections
  to the repulsive Casimir effect at finite temperature}},\ }\href
  {https://doi.org/10.1142/S0217751X16500184} {\bibfield  {journal} {\bibinfo
  {journal} {Int. J. Mod. Phys. A}\ }\textbf {\bibinfo {volume} {31}},\
  \bibinfo {pages} {07} (\bibinfo {year} {2016})},\ \Eprint
  {https://arxiv.org/abs/1511.05940} {arXiv:1511.05940 [hep-th]} \BibitemShut
  {NoStop}%
\bibitem [{\citenamefont {Erdas}(2020)}]{Erdas:2020ilo}%
  \BibitemOpen
  \bibfield  {author} {\bibinfo {author} {\bibfnamefont {A.}~\bibnamefont
  {Erdas}},\ }\bibfield  {title} {\bibinfo {title} {{Casimir effect of a
  Lorentz-violating scalar in magnetic field}},\ }\href
  {https://doi.org/10.1142/S0217751X20502097} {\bibfield  {journal} {\bibinfo
  {journal} {Int. J. Mod. Phys. A}\ }\textbf {\bibinfo {volume} {35}},\
  \bibinfo {pages} {2050209} (\bibinfo {year} {2020})},\ \Eprint
  {https://arxiv.org/abs/2005.07830} {arXiv:2005.07830 [hep-th]} \BibitemShut
  {NoStop}%
\bibitem [{\citenamefont {Haridev}\ and\ \citenamefont
  {Samantray}(2022)}]{Haridev:2021jwi}%
  \BibitemOpen
  \bibfield  {author} {\bibinfo {author} {\bibfnamefont {S.~R.}\ \bibnamefont
  {Haridev}}\ and\ \bibinfo {author} {\bibfnamefont {P.}~\bibnamefont
  {Samantray}},\ }\bibfield  {title} {\bibinfo {title} {{Revisiting vacuum
  energy in compact spacetimes}},\ }\href
  {https://doi.org/10.1016/j.physletb.2022.137489} {\bibfield  {journal}
  {\bibinfo  {journal} {Phys. Lett. B}\ }\textbf {\bibinfo {volume} {835}},\
  \bibinfo {pages} {137489} (\bibinfo {year} {2022})},\ \Eprint
  {https://arxiv.org/abs/2106.12171} {arXiv:2106.12171 [hep-th]} \BibitemShut
  {NoStop}%
\bibitem [{\citenamefont {Erdas}(2021)}]{Erdas:2021xvv}%
  \BibitemOpen
  \bibfield  {author} {\bibinfo {author} {\bibfnamefont {A.}~\bibnamefont
  {Erdas}},\ }\bibfield  {title} {\bibinfo {title} {{Thermal effects on the
  Casimir energy of a Lorentz-violating scalar in magnetic field}},\ }\href
  {https://doi.org/10.1142/S0217751X21501554} {\bibfield  {journal} {\bibinfo
  {journal} {Int. J. Mod. Phys. A}\ }\textbf {\bibinfo {volume} {36}},\
  \bibinfo {pages} {2150155} (\bibinfo {year} {2021})},\ \Eprint
  {https://arxiv.org/abs/2103.12823} {arXiv:2103.12823 [hep-th]} \BibitemShut
  {NoStop}%
\bibitem [{\citenamefont {Erdas}(2024)}]{Erdas:2024gkq}%
  \BibitemOpen
  \bibfield  {author} {\bibinfo {author} {\bibfnamefont {A.}~\bibnamefont
  {Erdas}},\ }\bibfield  {title} {\bibinfo {title} {{Casimir effect of a doubly
  Lorentz-violating scalar in magnetic field}},\ }\href@noop {} {\  (\bibinfo
  {year} {2024})},\ \Eprint {https://arxiv.org/abs/2408.13188}
  {arXiv:2408.13188 [hep-th]} \BibitemShut {NoStop}%
\bibitem [{\citenamefont {Fujii}\ \emph
  {et~al.}(2024{\natexlab{a}})\citenamefont {Fujii}, \citenamefont {Nakayama},\
  and\ \citenamefont {Suzuki}}]{Fujii:2024fzy}%
  \BibitemOpen
  \bibfield  {author} {\bibinfo {author} {\bibfnamefont {D.}~\bibnamefont
  {Fujii}}, \bibinfo {author} {\bibfnamefont {K.}~\bibnamefont {Nakayama}},\
  and\ \bibinfo {author} {\bibfnamefont {K.}~\bibnamefont {Suzuki}},\
  }\bibfield  {title} {\bibinfo {title} {{Dual chiral density wave induced
  oscillating Casimir effect}},\ }\href
  {https://doi.org/10.1103/PhysRevD.110.014039} {\bibfield  {journal} {\bibinfo
   {journal} {Phys. Rev. D}\ }\textbf {\bibinfo {volume} {110}},\ \bibinfo
  {pages} {014039} (\bibinfo {year} {2024}{\natexlab{a}})},\ \Eprint
  {https://arxiv.org/abs/2402.17638} {arXiv:2402.17638 [hep-ph]} \BibitemShut
  {NoStop}%
\bibitem [{\citenamefont {Dautry}\ and\ \citenamefont
  {Nyman}(1979)}]{Dautry:1979bk}%
  \BibitemOpen
  \bibfield  {author} {\bibinfo {author} {\bibfnamefont {F.}~\bibnamefont
  {Dautry}}\ and\ \bibinfo {author} {\bibfnamefont {E.~M.}\ \bibnamefont
  {Nyman}},\ }\bibfield  {title} {\bibinfo {title} {{Pion condensation and the
  $\sigma$-model in liquid neutron matter}},\ }\href
  {https://doi.org/10.1016/0375-9474(79)90518-9} {\bibfield  {journal}
  {\bibinfo  {journal} {Nucl. Phys. A}\ }\textbf {\bibinfo {volume} {319}},\
  \bibinfo {pages} {323} (\bibinfo {year} {1979})}\BibitemShut {NoStop}%
\bibitem [{\citenamefont {Tatsumi}\ and\ \citenamefont
  {Nakano}()}]{Tatsumi:2004dx}%
  \BibitemOpen
  \bibfield  {author} {\bibinfo {author} {\bibfnamefont {T.}~\bibnamefont
  {Tatsumi}}\ and\ \bibinfo {author} {\bibfnamefont {E.}~\bibnamefont
  {Nakano}},\ }\bibfield  {title} {\bibinfo {title} {{Dual chiral density wave
  in quark matter}},\ }\href@noop {} {\ }\Eprint
  {https://arxiv.org/abs/hep-ph/0408294} {arXiv:hep-ph/0408294} \BibitemShut
  {NoStop}%
\bibitem [{\citenamefont {Nakano}\ and\ \citenamefont
  {Tatsumi}(2005)}]{Nakano:2004cd}%
  \BibitemOpen
  \bibfield  {author} {\bibinfo {author} {\bibfnamefont {E.}~\bibnamefont
  {Nakano}}\ and\ \bibinfo {author} {\bibfnamefont {T.}~\bibnamefont
  {Tatsumi}},\ }\bibfield  {title} {\bibinfo {title} {{Chiral symmetry and
  density waves in quark matter}},\ }\href
  {https://doi.org/10.1103/PhysRevD.71.114006} {\bibfield  {journal} {\bibinfo
  {journal} {Phys. Rev. D}\ }\textbf {\bibinfo {volume} {71}},\ \bibinfo
  {pages} {114006} (\bibinfo {year} {2005})},\ \Eprint
  {https://arxiv.org/abs/hep-ph/0411350} {arXiv:hep-ph/0411350} \BibitemShut
  {NoStop}%
\bibitem [{\citenamefont {Frolov}\ \emph {et~al.}(2010)\citenamefont {Frolov},
  \citenamefont {Zhukovsky},\ and\ \citenamefont {Klimenko}}]{Frolov:2010wn}%
  \BibitemOpen
  \bibfield  {author} {\bibinfo {author} {\bibfnamefont {I.~E.}\ \bibnamefont
  {Frolov}}, \bibinfo {author} {\bibfnamefont {V.~C.}\ \bibnamefont
  {Zhukovsky}},\ and\ \bibinfo {author} {\bibfnamefont {K.~G.}\ \bibnamefont
  {Klimenko}},\ }\bibfield  {title} {\bibinfo {title} {{Chiral density waves in
  quark matter within the Nambu--Jona-Lasinio model in an external magnetic
  field}},\ }\href {https://doi.org/10.1103/PhysRevD.82.076002} {\bibfield
  {journal} {\bibinfo  {journal} {Phys. Rev. D}\ }\textbf {\bibinfo {volume}
  {82}},\ \bibinfo {pages} {076002} (\bibinfo {year} {2010})},\ \Eprint
  {https://arxiv.org/abs/1007.2984} {arXiv:1007.2984 [hep-ph]} \BibitemShut
  {NoStop}%
\bibitem [{\citenamefont {Tatsumi}\ \emph {et~al.}(2015)\citenamefont
  {Tatsumi}, \citenamefont {Nishiyama},\ and\ \citenamefont
  {Karasawa}}]{Tatsumi:2014wka}%
  \BibitemOpen
  \bibfield  {author} {\bibinfo {author} {\bibfnamefont {T.}~\bibnamefont
  {Tatsumi}}, \bibinfo {author} {\bibfnamefont {K.}~\bibnamefont {Nishiyama}},\
  and\ \bibinfo {author} {\bibfnamefont {S.}~\bibnamefont {Karasawa}},\
  }\bibfield  {title} {\bibinfo {title} {{Novel Lifshitz point for chiral
  transition in the magnetic field}},\ }\href
  {https://doi.org/10.1016/j.physletb.2015.02.033} {\bibfield  {journal}
  {\bibinfo  {journal} {Phys. Lett. B}\ }\textbf {\bibinfo {volume} {743}},\
  \bibinfo {pages} {66} (\bibinfo {year} {2015})},\ \Eprint
  {https://arxiv.org/abs/1405.2155} {arXiv:1405.2155 [hep-ph]} \BibitemShut
  {NoStop}%
\bibitem [{\citenamefont {Nishiyama}\ \emph {et~al.}(2015)\citenamefont
  {Nishiyama}, \citenamefont {Karasawa},\ and\ \citenamefont
  {Tatsumi}}]{Nishiyama:2015fba}%
  \BibitemOpen
  \bibfield  {author} {\bibinfo {author} {\bibfnamefont {K.}~\bibnamefont
  {Nishiyama}}, \bibinfo {author} {\bibfnamefont {S.}~\bibnamefont
  {Karasawa}},\ and\ \bibinfo {author} {\bibfnamefont {T.}~\bibnamefont
  {Tatsumi}},\ }\bibfield  {title} {\bibinfo {title} {{Hybrid chiral condensate
  in the external magnetic field}},\ }\href
  {https://doi.org/10.1103/PhysRevD.92.036008} {\bibfield  {journal} {\bibinfo
  {journal} {Phys. Rev. D}\ }\textbf {\bibinfo {volume} {92}},\ \bibinfo
  {pages} {036008} (\bibinfo {year} {2015})},\ \Eprint
  {https://arxiv.org/abs/1505.01928} {arXiv:1505.01928 [nucl-th]} \BibitemShut
  {NoStop}%
\bibitem [{\citenamefont {Carignano}\ \emph {et~al.}(2015)\citenamefont
  {Carignano}, \citenamefont {Ferrer}, \citenamefont {de~la Incera},\ and\
  \citenamefont {Paulucci}}]{Carignano:2015kda}%
  \BibitemOpen
  \bibfield  {author} {\bibinfo {author} {\bibfnamefont {S.}~\bibnamefont
  {Carignano}}, \bibinfo {author} {\bibfnamefont {E.~J.}\ \bibnamefont
  {Ferrer}}, \bibinfo {author} {\bibfnamefont {V.}~\bibnamefont {de~la
  Incera}},\ and\ \bibinfo {author} {\bibfnamefont {L.}~\bibnamefont
  {Paulucci}},\ }\bibfield  {title} {\bibinfo {title} {{Crystalline chiral
  condensates as a component of compact stars}},\ }\href
  {https://doi.org/10.1103/PhysRevD.92.105018} {\bibfield  {journal} {\bibinfo
  {journal} {Phys. Rev. D}\ }\textbf {\bibinfo {volume} {92}},\ \bibinfo
  {pages} {105018} (\bibinfo {year} {2015})},\ \Eprint
  {https://arxiv.org/abs/1505.05094} {arXiv:1505.05094 [nucl-th]} \BibitemShut
  {NoStop}%
\bibitem [{\citenamefont {Ferrer}\ and\ \citenamefont {de~la
  Incera}(2018)}]{Ferrer:2015iop}%
  \BibitemOpen
  \bibfield  {author} {\bibinfo {author} {\bibfnamefont {E.~J.}\ \bibnamefont
  {Ferrer}}\ and\ \bibinfo {author} {\bibfnamefont {V.}~\bibnamefont {de~la
  Incera}},\ }\bibfield  {title} {\bibinfo {title} {{Novel topological effects
  in dense QCD in a magnetic field}},\ }\href
  {https://doi.org/10.1016/j.nuclphysb.2018.04.009} {\bibfield  {journal}
  {\bibinfo  {journal} {Nucl. Phys. B}\ }\textbf {\bibinfo {volume} {931}},\
  \bibinfo {pages} {192} (\bibinfo {year} {2018})},\ \Eprint
  {https://arxiv.org/abs/1512.03972} {arXiv:1512.03972 [nucl-th]} \BibitemShut
  {NoStop}%
\bibitem [{\citenamefont {Ferrer}\ and\ \citenamefont {de~la
  Incera}(2017)}]{Ferrer:2016toh}%
  \BibitemOpen
  \bibfield  {author} {\bibinfo {author} {\bibfnamefont {E.~J.}\ \bibnamefont
  {Ferrer}}\ and\ \bibinfo {author} {\bibfnamefont {V.}~\bibnamefont {de~la
  Incera}},\ }\bibfield  {title} {\bibinfo {title} {{Dissipationless Hall
  Current in Dense Quark Matter in a Magnetic Field}},\ }\href
  {https://doi.org/10.1016/j.physletb.2017.02.066} {\bibfield  {journal}
  {\bibinfo  {journal} {Phys. Lett. B}\ }\textbf {\bibinfo {volume} {769}},\
  \bibinfo {pages} {208} (\bibinfo {year} {2017})},\ \Eprint
  {https://arxiv.org/abs/1611.00660} {arXiv:1611.00660 [nucl-th]} \BibitemShut
  {NoStop}%
\bibitem [{\citenamefont {Abuki}(2018)}]{Abuki:2018iqp}%
  \BibitemOpen
  \bibfield  {author} {\bibinfo {author} {\bibfnamefont {H.}~\bibnamefont
  {Abuki}},\ }\bibfield  {title} {\bibinfo {title} {{Chiral crystallization in
  an external magnetic background: Chiral spiral versus real kink crystal}},\
  }\href {https://doi.org/10.1103/PhysRevD.98.054006} {\bibfield  {journal}
  {\bibinfo  {journal} {Phys. Rev. D}\ }\textbf {\bibinfo {volume} {98}},\
  \bibinfo {pages} {054006} (\bibinfo {year} {2018})},\ \Eprint
  {https://arxiv.org/abs/1808.05767} {arXiv:1808.05767 [hep-ph]} \BibitemShut
  {NoStop}%
\bibitem [{\citenamefont {Ferrer}\ and\ \citenamefont {de~la
  Incera}(2020)}]{Ferrer:2019zfp}%
  \BibitemOpen
  \bibfield  {author} {\bibinfo {author} {\bibfnamefont {E.~J.}\ \bibnamefont
  {Ferrer}}\ and\ \bibinfo {author} {\bibfnamefont {V.}~\bibnamefont {de~la
  Incera}},\ }\bibfield  {title} {\bibinfo {title} {{Absence of Landau-Peierls
  Instability in the Magnetic Dual Chiral Density Wave Phase of Dense QCD}},\
  }\href {https://doi.org/10.1103/PhysRevD.102.014010} {\bibfield  {journal}
  {\bibinfo  {journal} {Phys. Rev. D}\ }\textbf {\bibinfo {volume} {102}},\
  \bibinfo {pages} {014010} (\bibinfo {year} {2020})},\ \Eprint
  {https://arxiv.org/abs/1902.06810} {arXiv:1902.06810 [nucl-th]} \BibitemShut
  {NoStop}%
\bibitem [{\citenamefont {Feng}\ \emph {et~al.}(2020)\citenamefont {Feng},
  \citenamefont {Ferrer},\ and\ \citenamefont {Portillo}}]{Feng:2020vxn}%
  \BibitemOpen
  \bibfield  {author} {\bibinfo {author} {\bibfnamefont {B.}~\bibnamefont
  {Feng}}, \bibinfo {author} {\bibfnamefont {E.~J.}\ \bibnamefont {Ferrer}},\
  and\ \bibinfo {author} {\bibfnamefont {I.}~\bibnamefont {Portillo}},\
  }\bibfield  {title} {\bibinfo {title} {{Lack of Debye and Meissner screening
  in strongly magnetized quark matter at intermediate densities}},\ }\href
  {https://doi.org/10.1103/PhysRevD.101.056012} {\bibfield  {journal} {\bibinfo
   {journal} {Phys. Rev. D}\ }\textbf {\bibinfo {volume} {101}},\ \bibinfo
  {pages} {056012} (\bibinfo {year} {2020})},\ \Eprint
  {https://arxiv.org/abs/2001.02617} {arXiv:2001.02617 [hep-ph]} \BibitemShut
  {NoStop}%
\bibitem [{\citenamefont {Ferrer}\ and\ \citenamefont {de~la
  Incera}(2023)}]{Ferrer:2020ulz}%
  \BibitemOpen
  \bibfield  {author} {\bibinfo {author} {\bibfnamefont {E.~J.}\ \bibnamefont
  {Ferrer}}\ and\ \bibinfo {author} {\bibfnamefont {V.}~\bibnamefont {de~la
  Incera}},\ }\bibfield  {title} {\bibinfo {title} {{Axion-polaritons in the
  magnetic dual chiral density wave phase of dense QCD}},\ }\href
  {https://doi.org/10.1016/j.nuclphysb.2023.116307} {\bibfield  {journal}
  {\bibinfo  {journal} {Nucl. Phys. B}\ }\textbf {\bibinfo {volume} {994}},\
  \bibinfo {pages} {116307} (\bibinfo {year} {2023})},\ \Eprint
  {https://arxiv.org/abs/2010.02314} {arXiv:2010.02314 [hep-ph]} \BibitemShut
  {NoStop}%
\bibitem [{\citenamefont {Ferrer}\ \emph {et~al.}(2021)\citenamefont {Ferrer},
  \citenamefont {de~la Incera},\ and\ \citenamefont {Sanson}}]{Ferrer:2021vuy}%
  \BibitemOpen
  \bibfield  {author} {\bibinfo {author} {\bibfnamefont {E.~J.}\ \bibnamefont
  {Ferrer}}, \bibinfo {author} {\bibfnamefont {V.}~\bibnamefont {de~la
  Incera}},\ and\ \bibinfo {author} {\bibfnamefont {P.}~\bibnamefont
  {Sanson}},\ }\bibfield  {title} {\bibinfo {title} {{Quark matter contribution
  to the heat capacity of magnetized neutron stars}},\ }\href
  {https://doi.org/10.1103/PhysRevD.103.123013} {\bibfield  {journal} {\bibinfo
   {journal} {Phys. Rev. D}\ }\textbf {\bibinfo {volume} {103}},\ \bibinfo
  {pages} {123013} (\bibinfo {year} {2021})},\ \Eprint
  {https://arxiv.org/abs/2101.04032} {arXiv:2101.04032 [nucl-th]} \BibitemShut
  {NoStop}%
\bibitem [{\citenamefont {Anzuini}\ and\ \citenamefont
  {Melatos}(2021)}]{Anzuini:2021gwi}%
  \BibitemOpen
  \bibfield  {author} {\bibinfo {author} {\bibfnamefont {F.}~\bibnamefont
  {Anzuini}}\ and\ \bibinfo {author} {\bibfnamefont {A.}~\bibnamefont
  {Melatos}},\ }\bibfield  {title} {\bibinfo {title} {{Gradient expansion
  technique for inhomogeneous, magnetized quark matter}},\ }\href
  {https://doi.org/10.1140/epja/s10050-021-00505-9} {\bibfield  {journal}
  {\bibinfo  {journal} {Eur. Phys. J. A}\ }\textbf {\bibinfo {volume} {57}},\
  \bibinfo {pages} {220} (\bibinfo {year} {2021})},\ \Eprint
  {https://arxiv.org/abs/2106.04744} {arXiv:2106.04744 [hep-ph]} \BibitemShut
  {NoStop}%
\bibitem [{\citenamefont {Gyory}\ and\ \citenamefont {de~la
  Incera}(2022)}]{Gyory:2022hnv}%
  \BibitemOpen
  \bibfield  {author} {\bibinfo {author} {\bibfnamefont {W.}~\bibnamefont
  {Gyory}}\ and\ \bibinfo {author} {\bibfnamefont {V.}~\bibnamefont {de~la
  Incera}},\ }\bibfield  {title} {\bibinfo {title} {{Phase transitions and
  resilience of the magnetic dual chiral density wave phase at finite
  temperature and density}},\ }\href
  {https://doi.org/10.1103/PhysRevD.106.016011} {\bibfield  {journal} {\bibinfo
   {journal} {Phys. Rev. D}\ }\textbf {\bibinfo {volume} {106}},\ \bibinfo
  {pages} {016011} (\bibinfo {year} {2022})},\ \Eprint
  {https://arxiv.org/abs/2203.14209} {arXiv:2203.14209 [nucl-th]} \BibitemShut
  {NoStop}%
\bibitem [{\citenamefont {Ferrer}\ and\ \citenamefont
  {Hackebill}(2023)}]{Ferrer:2022afu}%
  \BibitemOpen
  \bibfield  {author} {\bibinfo {author} {\bibfnamefont {E.~J.}\ \bibnamefont
  {Ferrer}}\ and\ \bibinfo {author} {\bibfnamefont {A.}~\bibnamefont
  {Hackebill}},\ }\bibfield  {title} {\bibinfo {title} {{Speed of sound for
  hadronic and quark phases in a magnetic field}},\ }\href
  {https://doi.org/10.1016/j.nuclphysa.2023.122608} {\bibfield  {journal}
  {\bibinfo  {journal} {Nucl. Phys. A}\ }\textbf {\bibinfo {volume} {1031}},\
  \bibinfo {pages} {122608} (\bibinfo {year} {2023})},\ \Eprint
  {https://arxiv.org/abs/2203.16576} {arXiv:2203.16576 [hep-ph]} \BibitemShut
  {NoStop}%
\bibitem [{\citenamefont {Ghalati}\ and\ \citenamefont
  {Sadooghi}(2023)}]{Ghalati:2023npr}%
  \BibitemOpen
  \bibfield  {author} {\bibinfo {author} {\bibfnamefont {H.~M.}\ \bibnamefont
  {Ghalati}}\ and\ \bibinfo {author} {\bibfnamefont {N.}~\bibnamefont
  {Sadooghi}},\ }\bibfield  {title} {\bibinfo {title} {{Magnetic dual chiral
  density wave phase in rotating cold quark matter}},\ }\href
  {https://doi.org/10.1103/PhysRevD.108.054032} {\bibfield  {journal} {\bibinfo
   {journal} {Phys. Rev. D}\ }\textbf {\bibinfo {volume} {108}},\ \bibinfo
  {pages} {054032} (\bibinfo {year} {2023})},\ \Eprint
  {https://arxiv.org/abs/2306.04472} {arXiv:2306.04472 [nucl-th]} \BibitemShut
  {NoStop}%
\bibitem [{\citenamefont {Ferrer}\ \emph {et~al.}(2024)\citenamefont {Ferrer},
  \citenamefont {Gyory},\ and\ \citenamefont {de~la Incera}}]{Ferrer:2023xvl}%
  \BibitemOpen
  \bibfield  {author} {\bibinfo {author} {\bibfnamefont {E.~J.}\ \bibnamefont
  {Ferrer}}, \bibinfo {author} {\bibfnamefont {W.}~\bibnamefont {Gyory}},\ and\
  \bibinfo {author} {\bibfnamefont {V.}~\bibnamefont {de~la Incera}},\
  }\bibfield  {title} {\bibinfo {title} {{Thermal phonon fluctuations and
  stability of the magnetic dual chiral density wave phase in dense QCD}},\
  }\href {https://doi.org/10.1103/PhysRevD.109.036023} {\bibfield  {journal}
  {\bibinfo  {journal} {Phys. Rev. D}\ }\textbf {\bibinfo {volume} {109}},\
  \bibinfo {pages} {036023} (\bibinfo {year} {2024})},\ \Eprint
  {https://arxiv.org/abs/2307.05621} {arXiv:2307.05621 [nucl-th]} \BibitemShut
  {NoStop}%
\bibitem [{\citenamefont {Ferrer}\ and\ \citenamefont {de~la
  Incera}(2024)}]{Ferrer:2024xwu}%
  \BibitemOpen
  \bibfield  {author} {\bibinfo {author} {\bibfnamefont {E.~J.}\ \bibnamefont
  {Ferrer}}\ and\ \bibinfo {author} {\bibfnamefont {V.}~\bibnamefont {de~la
  Incera}},\ }\bibfield  {title} {\bibinfo {title} {{Axion-polaritons in quark
  stars: a possible solution to the missing pulsar problem}},\ }\href
  {https://doi.org/10.1140/epjc/s10052-024-12486-2} {\bibfield  {journal}
  {\bibinfo  {journal} {Eur. Phys. J. C}\ }\textbf {\bibinfo {volume} {84}},\
  \bibinfo {pages} {133} (\bibinfo {year} {2024})}\BibitemShut {NoStop}%
\bibitem [{\citenamefont {Ferrer}\ and\ \citenamefont {de~la
  Incera}(2021)}]{Ferrer:2021mpq}%
  \BibitemOpen
  \bibfield  {author} {\bibinfo {author} {\bibfnamefont {E.~J.}\ \bibnamefont
  {Ferrer}}\ and\ \bibinfo {author} {\bibfnamefont {V.}~\bibnamefont {de~la
  Incera}},\ }\bibfield  {title} {\bibinfo {title} {{Magnetic Dual Chiral
  Density Wave: A Candidate Quark Matter Phase for the Interior of Neutron
  Stars}},\ }\href {https://doi.org/10.3390/universe7120458} {\bibfield
  {journal} {\bibinfo  {journal} {Universe}\ }\textbf {\bibinfo {volume} {7}},\
  \bibinfo {pages} {458} (\bibinfo {year} {2021})},\ \Eprint
  {https://arxiv.org/abs/2201.04032} {arXiv:2201.04032 [hep-ph]} \BibitemShut
  {NoStop}%
\bibitem [{\citenamefont {Nambu}\ and\ \citenamefont
  {Jona-Lasinio}(1961{\natexlab{a}})}]{Nambu:1961tp}%
  \BibitemOpen
  \bibfield  {author} {\bibinfo {author} {\bibfnamefont {Y.}~\bibnamefont
  {Nambu}}\ and\ \bibinfo {author} {\bibfnamefont {G.}~\bibnamefont
  {Jona-Lasinio}},\ }\bibfield  {title} {\bibinfo {title} {{Dynamical Model of
  Elementary Particles Based on an Analogy with Superconductivity. I}},\ }\href
  {https://doi.org/10.1103/PhysRev.122.345} {\bibfield  {journal} {\bibinfo
  {journal} {Phys. Rev.}\ }\textbf {\bibinfo {volume} {122}},\ \bibinfo {pages}
  {345} (\bibinfo {year} {1961}{\natexlab{a}})}\BibitemShut {NoStop}%
\bibitem [{\citenamefont {Nambu}\ and\ \citenamefont
  {Jona-Lasinio}(1961{\natexlab{b}})}]{Nambu:1961fr}%
  \BibitemOpen
  \bibfield  {author} {\bibinfo {author} {\bibfnamefont {Y.}~\bibnamefont
  {Nambu}}\ and\ \bibinfo {author} {\bibfnamefont {G.}~\bibnamefont
  {Jona-Lasinio}},\ }\bibfield  {title} {\bibinfo {title} {{Dynamical Model of
  Elementary Particles Based on an Analogy with Superconductivity. II}},\
  }\href {https://doi.org/10.1103/PhysRev.124.246} {\bibfield  {journal}
  {\bibinfo  {journal} {Phys. Rev.}\ }\textbf {\bibinfo {volume} {124}},\
  \bibinfo {pages} {246} (\bibinfo {year} {1961}{\natexlab{b}})}\BibitemShut
  {NoStop}%
\bibitem [{\citenamefont {Vogl}\ and\ \citenamefont
  {Weise}(1991)}]{Vogl:1991qt}%
  \BibitemOpen
  \bibfield  {author} {\bibinfo {author} {\bibfnamefont {U.}~\bibnamefont
  {Vogl}}\ and\ \bibinfo {author} {\bibfnamefont {W.}~\bibnamefont {Weise}},\
  }\bibfield  {title} {\bibinfo {title} {{The Nambu and Jona-Lasinio model: Its
  implications for hadrons and nuclei}},\ }\href
  {https://doi.org/10.1016/0146-6410(91)90005-9} {\bibfield  {journal}
  {\bibinfo  {journal} {Prog. Part. Nucl. Phys.}\ }\textbf {\bibinfo {volume}
  {27}},\ \bibinfo {pages} {195} (\bibinfo {year} {1991})}\BibitemShut
  {NoStop}%
\bibitem [{\citenamefont {Klevansky}(1992)}]{Klevansky:1992qe}%
  \BibitemOpen
  \bibfield  {author} {\bibinfo {author} {\bibfnamefont {S.~P.}\ \bibnamefont
  {Klevansky}},\ }\bibfield  {title} {\bibinfo {title} {{The
  Nambu--Jona-Lasinio model of quantum chromodynamics}},\ }\href
  {https://doi.org/10.1103/RevModPhys.64.649} {\bibfield  {journal} {\bibinfo
  {journal} {Rev. Mod. Phys.}\ }\textbf {\bibinfo {volume} {64}},\ \bibinfo
  {pages} {649} (\bibinfo {year} {1992})}\BibitemShut {NoStop}%
\bibitem [{\citenamefont {Hatsuda}\ and\ \citenamefont
  {Kunihiro}(1994)}]{Hatsuda:1994pi}%
  \BibitemOpen
  \bibfield  {author} {\bibinfo {author} {\bibfnamefont {T.}~\bibnamefont
  {Hatsuda}}\ and\ \bibinfo {author} {\bibfnamefont {T.}~\bibnamefont
  {Kunihiro}},\ }\bibfield  {title} {\bibinfo {title} {{QCD phenomenology based
  on a chiral effective Lagrangian}},\ }\href
  {https://doi.org/10.1016/0370-1573(94)90022-1} {\bibfield  {journal}
  {\bibinfo  {journal} {Phys. Rep.}\ }\textbf {\bibinfo {volume} {247}},\
  \bibinfo {pages} {221} (\bibinfo {year} {1994})},\ \Eprint
  {https://arxiv.org/abs/hep-ph/9401310} {arXiv:hep-ph/9401310} \BibitemShut
  {NoStop}%
\bibitem [{\citenamefont {Buballa}(2005)}]{Buballa:2003qv}%
  \BibitemOpen
  \bibfield  {author} {\bibinfo {author} {\bibfnamefont {M.}~\bibnamefont
  {Buballa}},\ }\bibfield  {title} {\bibinfo {title} {{NJL-model analysis of
  dense quark matter}},\ }\href {https://doi.org/10.1016/j.physrep.2004.11.004}
  {\bibfield  {journal} {\bibinfo  {journal} {Phys. Rep.}\ }\textbf {\bibinfo
  {volume} {407}},\ \bibinfo {pages} {205} (\bibinfo {year} {2005})},\ \Eprint
  {https://arxiv.org/abs/hep-ph/0402234} {arXiv:hep-ph/0402234} \BibitemShut
  {NoStop}%
\bibitem [{\citenamefont {Ferrer}\ \emph {et~al.}(1999)\citenamefont {Ferrer},
  \citenamefont {Gusynin},\ and\ \citenamefont {de~la Incera}}]{Ferrer:1999gs}%
  \BibitemOpen
  \bibfield  {author} {\bibinfo {author} {\bibfnamefont {E.~J.}\ \bibnamefont
  {Ferrer}}, \bibinfo {author} {\bibfnamefont {V.~P.}\ \bibnamefont
  {Gusynin}},\ and\ \bibinfo {author} {\bibfnamefont {V.}~\bibnamefont {de~la
  Incera}},\ }\bibfield  {title} {\bibinfo {title} {{Boundary effects in the
  magnetic catalysis of chiral symmetry breaking}},\ }\href
  {https://doi.org/10.1016/S0370-2693(99)00470-0} {\bibfield  {journal}
  {\bibinfo  {journal} {Phys. Lett. B}\ }\textbf {\bibinfo {volume} {455}},\
  \bibinfo {pages} {217} (\bibinfo {year} {1999})},\ \Eprint
  {https://arxiv.org/abs/hep-ph/9901446} {arXiv:hep-ph/9901446} \BibitemShut
  {NoStop}%
\bibitem [{\citenamefont {Abreu}\ \emph {et~al.}(2019)\citenamefont {Abreu},
  \citenamefont {Corr\^ea}, \citenamefont {Linhares},\ and\ \citenamefont
  {Malbouisson}}]{Abreu:2019czp}%
  \BibitemOpen
  \bibfield  {author} {\bibinfo {author} {\bibfnamefont {L.~M.}\ \bibnamefont
  {Abreu}}, \bibinfo {author} {\bibfnamefont {E.~B.~S.}\ \bibnamefont
  {Corr\^ea}}, \bibinfo {author} {\bibfnamefont {C.~A.}\ \bibnamefont
  {Linhares}},\ and\ \bibinfo {author} {\bibfnamefont {A.~P.~C.}\ \bibnamefont
  {Malbouisson}},\ }\bibfield  {title} {\bibinfo {title} {{Finite-volume and
  magnetic effects on the phase structure of the three-flavor
  Nambu\textendash{}Jona-Lasinio model}},\ }\href
  {https://doi.org/10.1103/PhysRevD.99.076001} {\bibfield  {journal} {\bibinfo
  {journal} {Phys. Rev. D}\ }\textbf {\bibinfo {volume} {99}},\ \bibinfo
  {pages} {076001} (\bibinfo {year} {2019})},\ \Eprint
  {https://arxiv.org/abs/1903.09249} {arXiv:1903.09249 [hep-ph]} \BibitemShut
  {NoStop}%
\bibitem [{\citenamefont {Abreu}\ \emph {et~al.}(2021)\citenamefont {Abreu},
  \citenamefont {Nery},\ and\ \citenamefont {Corr\^ea}}]{Abreu:2020uxc}%
  \BibitemOpen
  \bibfield  {author} {\bibinfo {author} {\bibfnamefont {L.~M.}\ \bibnamefont
  {Abreu}}, \bibinfo {author} {\bibfnamefont {E.~S.}\ \bibnamefont {Nery}},\
  and\ \bibinfo {author} {\bibfnamefont {E.~B.~S.}\ \bibnamefont {Corr\^ea}},\
  }\bibfield  {title} {\bibinfo {title} {{Boundary effects on constituent quark
  masses and on chiral susceptibility in a four-fermion interaction model}},\
  }\href {https://doi.org/10.1016/j.physa.2021.125885} {\bibfield  {journal}
  {\bibinfo  {journal} {Physica A Stat. Mech. Appl.}\ }\textbf {\bibinfo
  {volume} {572}},\ \bibinfo {pages} {125885} (\bibinfo {year} {2021})},\
  \Eprint {https://arxiv.org/abs/2004.11237} {arXiv:2004.11237 [hep-ph]}
  \BibitemShut {NoStop}%
\bibitem [{\citenamefont {Abreu}\ \emph {et~al.}(2022)\citenamefont {Abreu},
  \citenamefont {Corr\^ea},\ and\ \citenamefont {Nery}}]{Abreu:2021btt}%
  \BibitemOpen
  \bibfield  {author} {\bibinfo {author} {\bibfnamefont {L.~M.}\ \bibnamefont
  {Abreu}}, \bibinfo {author} {\bibfnamefont {E.~B.~S.}\ \bibnamefont
  {Corr\^ea}},\ and\ \bibinfo {author} {\bibfnamefont {E.~S.}\ \bibnamefont
  {Nery}},\ }\bibfield  {title} {\bibinfo {title} {{Properties of neutral
  mesons in a hot and magnetized quark matter: Size-dependent effects}},\
  }\href {https://doi.org/10.1103/PhysRevD.105.056010} {\bibfield  {journal}
  {\bibinfo  {journal} {Phys. Rev. D}\ }\textbf {\bibinfo {volume} {105}},\
  \bibinfo {pages} {056010} (\bibinfo {year} {2022})},\ \Eprint
  {https://arxiv.org/abs/2109.09593} {arXiv:2109.09593 [hep-ph]} \BibitemShut
  {NoStop}%
\bibitem [{\citenamefont {Abreu}\ \emph {et~al.}(2023)\citenamefont {Abreu},
  \citenamefont {Corr\^ea},\ and\ \citenamefont {Nery}}]{Abreu:2022cgm}%
  \BibitemOpen
  \bibfield  {author} {\bibinfo {author} {\bibfnamefont {L.~M.}\ \bibnamefont
  {Abreu}}, \bibinfo {author} {\bibfnamefont {E.~B.~S.}\ \bibnamefont
  {Corr\^ea}},\ and\ \bibinfo {author} {\bibfnamefont {E.~S.}\ \bibnamefont
  {Nery}},\ }\bibfield  {title} {\bibinfo {title} {{Inverse magnetic catalysis
  and size-dependent effects on the chiral symmetry restoration}},\ }\href
  {https://doi.org/10.1140/epja/s10050-023-01078-5} {\bibfield  {journal}
  {\bibinfo  {journal} {Eur. Phys. J. A}\ }\textbf {\bibinfo {volume} {59}},\
  \bibinfo {pages} {157} (\bibinfo {year} {2023})},\ \Eprint
  {https://arxiv.org/abs/2211.11083} {arXiv:2211.11083 [hep-ph]} \BibitemShut
  {NoStop}%
\bibitem [{\citenamefont {Corr\^ea}(2023)}]{Correa:2023ebh}%
  \BibitemOpen
  \bibfield  {author} {\bibinfo {author} {\bibfnamefont {E.~B.~S.}\
  \bibnamefont {Corr\^ea}},\ }\bibfield  {title} {\bibinfo {title} {{Phase
  transition in a four-fermion interaction model under boundary conditions and
  electromagnetic effects}},\ }\href
  {https://doi.org/10.1103/PhysRevD.108.076002} {\bibfield  {journal} {\bibinfo
   {journal} {Phys. Rev. D}\ }\textbf {\bibinfo {volume} {108}},\ \bibinfo
  {pages} {076002} (\bibinfo {year} {2023})}\BibitemShut {NoStop}%
\bibitem [{\citenamefont {Ebert}\ and\ \citenamefont
  {Klimenko}(2010)}]{Ebert:2010eq}%
  \BibitemOpen
  \bibfield  {author} {\bibinfo {author} {\bibfnamefont {D.}~\bibnamefont
  {Ebert}}\ and\ \bibinfo {author} {\bibfnamefont {K.~G.}\ \bibnamefont
  {Klimenko}},\ }\bibfield  {title} {\bibinfo {title} {{Cooper pairing and
  finite-size effects in a Nambu--Jona-Lasinio-type four-fermion model}},\
  }\href {https://doi.org/10.1103/PhysRevD.82.025018} {\bibfield  {journal}
  {\bibinfo  {journal} {Phys. Rev. D}\ }\textbf {\bibinfo {volume} {82}},\
  \bibinfo {pages} {025018} (\bibinfo {year} {2010})},\ \Eprint
  {https://arxiv.org/abs/1005.0699} {arXiv:1005.0699 [hep-ph]} \BibitemShut
  {NoStop}%
\bibitem [{\citenamefont {Lifshitz}(1956)}]{Lifshitz:1956zz}%
  \BibitemOpen
  \bibfield  {author} {\bibinfo {author} {\bibfnamefont {E.~M.}\ \bibnamefont
  {Lifshitz}},\ }\bibfield  {title} {\bibinfo {title} {{The theory of molecular
  attractive forces between solids}},\ }\href@noop {} {\bibfield  {journal}
  {\bibinfo  {journal} {Sov. Phys. JETP}\ }\textbf {\bibinfo {volume} {2}},\
  \bibinfo {pages} {73} (\bibinfo {year} {1956})}\BibitemShut {NoStop}%
\bibitem [{\citenamefont {Fujii}\ \emph
  {et~al.}(2024{\natexlab{b}})\citenamefont {Fujii}, \citenamefont {Nakayama},\
  and\ \citenamefont {Suzuki}}]{Fujii:2024ixq}%
  \BibitemOpen
  \bibfield  {author} {\bibinfo {author} {\bibfnamefont {D.}~\bibnamefont
  {Fujii}}, \bibinfo {author} {\bibfnamefont {K.}~\bibnamefont {Nakayama}},\
  and\ \bibinfo {author} {\bibfnamefont {K.}~\bibnamefont {Suzuki}},\
  }\bibfield  {title} {\bibinfo {title} {{Lifshitz formulas for finite-density
  Casimir effect}},\ }\href@noop {} {\  (\bibinfo {year}
  {2024}{\natexlab{b}})},\ \Eprint {https://arxiv.org/abs/2408.08384}
  {arXiv:2408.08384 [quant-ph]} \BibitemShut {NoStop}%
\bibitem [{\citenamefont {Actor}\ \emph {et~al.}(2000)\citenamefont {Actor},
  \citenamefont {Bender},\ and\ \citenamefont {Reingruber}}]{Actor:1999nb}%
  \BibitemOpen
  \bibfield  {author} {\bibinfo {author} {\bibfnamefont {A.}~\bibnamefont
  {Actor}}, \bibinfo {author} {\bibfnamefont {I.}~\bibnamefont {Bender}},\ and\
  \bibinfo {author} {\bibfnamefont {J.}~\bibnamefont {Reingruber}},\ }\bibfield
   {title} {\bibinfo {title} {{Casimir effect on a finite lattice}},\ }\href
  {https://doi.org/10.1002/(SICI)1521-3978(200004)48:4<303::AID-PROP303>3.0.CO;2-J}
  {\bibfield  {journal} {\bibinfo  {journal} {Fortschr. Phys.}\ }\textbf
  {\bibinfo {volume} {48}},\ \bibinfo {pages} {303} (\bibinfo {year} {2000})},\
  \Eprint {https://arxiv.org/abs/quant-ph/9908058} {arXiv:quant-ph/9908058
  [quant-ph]} \BibitemShut {NoStop}%
%%CITATION = QUANT-PH/9908058;%%
\bibitem [{\citenamefont {Pawellek}()}]{Pawellek:2013sda}%
  \BibitemOpen
  \bibfield  {author} {\bibinfo {author} {\bibfnamefont {M.}~\bibnamefont
  {Pawellek}},\ }\bibfield  {title} {\bibinfo {title} {{Finite-sites
  corrections to the Casimir energy on a periodic lattice}},\ }\href@noop {} {\
  }\Eprint {https://arxiv.org/abs/1303.4708} {arXiv:1303.4708 [hep-th]}
  \BibitemShut {NoStop}%
%%CITATION = ARXIV:1303.4708;%%
\bibitem [{\citenamefont {Ishikawa}\ \emph {et~al.}(2020)\citenamefont
  {Ishikawa}, \citenamefont {Nakayama},\ and\ \citenamefont
  {Suzuki}}]{Ishikawa:2020ezm}%
  \BibitemOpen
  \bibfield  {author} {\bibinfo {author} {\bibfnamefont {T.}~\bibnamefont
  {Ishikawa}}, \bibinfo {author} {\bibfnamefont {K.}~\bibnamefont {Nakayama}},\
  and\ \bibinfo {author} {\bibfnamefont {K.}~\bibnamefont {Suzuki}},\
  }\bibfield  {title} {\bibinfo {title} {{Casimir effect for lattice
  fermions}},\ }\href {https://doi.org/10.1016/j.physletb.2020.135713}
  {\bibfield  {journal} {\bibinfo  {journal} {Phys. Lett. B}\ }\textbf
  {\bibinfo {volume} {809}},\ \bibinfo {pages} {135713} (\bibinfo {year}
  {2020})},\ \Eprint {https://arxiv.org/abs/2005.10758} {arXiv:2005.10758
  [hep-lat]} \BibitemShut {NoStop}%
\bibitem [{\citenamefont {Ishikawa}\ \emph {et~al.}(2021)\citenamefont
  {Ishikawa}, \citenamefont {Nakayama},\ and\ \citenamefont
  {Suzuki}}]{Ishikawa:2020icy}%
  \BibitemOpen
  \bibfield  {author} {\bibinfo {author} {\bibfnamefont {T.}~\bibnamefont
  {Ishikawa}}, \bibinfo {author} {\bibfnamefont {K.}~\bibnamefont {Nakayama}},\
  and\ \bibinfo {author} {\bibfnamefont {K.}~\bibnamefont {Suzuki}},\
  }\bibfield  {title} {\bibinfo {title} {{Lattice-fermionic Casimir effect and
  topological insulators}},\ }\href
  {https://doi.org/10.1103/PhysRevResearch.3.023201} {\bibfield  {journal}
  {\bibinfo  {journal} {Phys. Rev. Res.}\ }\textbf {\bibinfo {volume} {3}},\
  \bibinfo {pages} {023201} (\bibinfo {year} {2021})},\ \Eprint
  {https://arxiv.org/abs/2012.11398} {arXiv:2012.11398 [hep-lat]} \BibitemShut
  {NoStop}%
\bibitem [{\citenamefont {Nakayama}\ and\ \citenamefont
  {Suzuki}(2023{\natexlab{b}})}]{Nakayama:2022ild}%
  \BibitemOpen
  \bibfield  {author} {\bibinfo {author} {\bibfnamefont {K.}~\bibnamefont
  {Nakayama}}\ and\ \bibinfo {author} {\bibfnamefont {K.}~\bibnamefont
  {Suzuki}},\ }\bibfield  {title} {\bibinfo {title} {{Remnants of the
  nonrelativistic Casimir effect on the lattice}},\ }\href
  {https://doi.org/10.1103/PhysRevResearch.5.L022054} {\bibfield  {journal}
  {\bibinfo  {journal} {Phys. Rev. Res.}\ }\textbf {\bibinfo {volume} {5}},\
  \bibinfo {pages} {L022054} (\bibinfo {year} {2023}{\natexlab{b}})},\ \Eprint
  {https://arxiv.org/abs/2204.12032} {arXiv:2204.12032 [quant-ph]} \BibitemShut
  {NoStop}%
\bibitem [{\citenamefont {Nakata}\ and\ \citenamefont
  {Suzuki}(2023)}]{Nakata:2022pen}%
  \BibitemOpen
  \bibfield  {author} {\bibinfo {author} {\bibfnamefont {K.}~\bibnamefont
  {Nakata}}\ and\ \bibinfo {author} {\bibfnamefont {K.}~\bibnamefont
  {Suzuki}},\ }\bibfield  {title} {\bibinfo {title} {{Magnonic Casimir Effect
  in Ferrimagnets}},\ }\href {http://doi.org/10.1103/PhysRevLett.130.096702}
  {\bibfield  {journal} {\bibinfo  {journal} {Phys. Rev. Lett.}\ }\textbf
  {\bibinfo {volume} {130}},\ \bibinfo {pages} {096702} (\bibinfo {year}
  {2023})},\ \Eprint {https://arxiv.org/abs/2205.13802} {arXiv:2205.13802
  [quant-ph]} \BibitemShut {NoStop}%
\bibitem [{\citenamefont {Mandlecha}\ and\ \citenamefont
  {Gavai}(2022)}]{Mandlecha:2022cll}%
  \BibitemOpen
  \bibfield  {author} {\bibinfo {author} {\bibfnamefont {Y.~V.}\ \bibnamefont
  {Mandlecha}}\ and\ \bibinfo {author} {\bibfnamefont {R.~V.}\ \bibnamefont
  {Gavai}},\ }\bibfield  {title} {\bibinfo {title} {{Lattice fermionic Casimir
  effect in a slab bag and universality}},\ }\href
  {https://doi.org/10.1016/j.physletb.2022.137558} {\bibfield  {journal}
  {\bibinfo  {journal} {Phys. Lett. B}\ }\textbf {\bibinfo {volume} {835}},\
  \bibinfo {pages} {137558} (\bibinfo {year} {2022})},\ \Eprint
  {https://arxiv.org/abs/2207.00889} {arXiv:2207.00889 [hep-lat]} \BibitemShut
  {NoStop}%
\bibitem [{\citenamefont {Swingle}\ and\ \citenamefont
  {Van~Raamsdonk}(2023)}]{Swingle:2022vie}%
  \BibitemOpen
  \bibfield  {author} {\bibinfo {author} {\bibfnamefont {B.}~\bibnamefont
  {Swingle}}\ and\ \bibinfo {author} {\bibfnamefont {M.}~\bibnamefont
  {Van~Raamsdonk}},\ }\bibfield  {title} {\bibinfo {title} {{Enhanced negative
  energy with a massless Dirac field}},\ }\href
  {https://doi.org/10.1007/JHEP08(2023)183} {\bibfield  {journal} {\bibinfo
  {journal} {JHEP}\ }\textbf {\bibinfo {volume} {08}},\ \bibinfo {pages}
  {183}},\ \Eprint {https://arxiv.org/abs/2212.02609} {arXiv:2212.02609
  [hep-th]} \BibitemShut {NoStop}%
\bibitem [{\citenamefont {Nakata}\ and\ \citenamefont
  {Suzuki}(2024)}]{Nakata:2023keh}%
  \BibitemOpen
  \bibfield  {author} {\bibinfo {author} {\bibfnamefont {K.}~\bibnamefont
  {Nakata}}\ and\ \bibinfo {author} {\bibfnamefont {K.}~\bibnamefont
  {Suzuki}},\ }\bibfield  {title} {\bibinfo {title} {{Non-Hermitian Casimir
  effect of magnons}},\ }\href {https://doi.org/10.1038/s44306-024-00017-4}
  {\bibfield  {journal} {\bibinfo  {journal} {npj Spintronics}\ }\textbf
  {\bibinfo {volume} {2}},\ \bibinfo {pages} {11} (\bibinfo {year} {2024})},\
  \Eprint {https://arxiv.org/abs/2305.09231} {arXiv:2305.09231 [quant-ph]}
  \BibitemShut {NoStop}%
\bibitem [{\citenamefont {Flores}\ \emph {et~al.}()\citenamefont {Flores},
  \citenamefont {Ireland}, \citenamefont {Jamhour}, \citenamefont {Lasasso},
  \citenamefont {Kurth},\ and\ \citenamefont {Leinbach}}]{Flores:2023whr}%
  \BibitemOpen
  \bibfield  {author} {\bibinfo {author} {\bibfnamefont {E.}~\bibnamefont
  {Flores}}, \bibinfo {author} {\bibfnamefont {C.}~\bibnamefont {Ireland}},
  \bibinfo {author} {\bibfnamefont {N.}~\bibnamefont {Jamhour}}, \bibinfo
  {author} {\bibfnamefont {V.}~\bibnamefont {Lasasso}}, \bibinfo {author}
  {\bibfnamefont {N.}~\bibnamefont {Kurth}},\ and\ \bibinfo {author}
  {\bibfnamefont {M.}~\bibnamefont {Leinbach}},\ }\bibfield  {title} {\bibinfo
  {title} {{Casimir force in discrete scalar fields I: 1D and 2D cases}},\
  }\href@noop {} {\ }\Eprint {https://arxiv.org/abs/2309.00624}
  {arXiv:2309.00624 [quant-ph]} \BibitemShut {NoStop}%
\bibitem [{\citenamefont {Nakayama}\ and\ \citenamefont
  {Suzuki}(2024)}]{Nakayama:2023zvm}%
  \BibitemOpen
  \bibfield  {author} {\bibinfo {author} {\bibfnamefont {K.}~\bibnamefont
  {Nakayama}}\ and\ \bibinfo {author} {\bibfnamefont {K.}~\bibnamefont
  {Suzuki}},\ }\bibfield  {title} {\bibinfo {title} {{Casimir effect in axion
  electrodynamics with lattice regularizations}},\ }\href
  {https://doi.org/10.1103/PhysRevD.109.065002} {\bibfield  {journal} {\bibinfo
   {journal} {Phys. Rev. D}\ }\textbf {\bibinfo {volume} {109}},\ \bibinfo
  {pages} {065002} (\bibinfo {year} {2024})},\ \Eprint
  {https://arxiv.org/abs/2310.18092} {arXiv:2310.18092 [hep-th]} \BibitemShut
  {NoStop}%
\bibitem [{\citenamefont {Beenakker}(2024)}]{Beenakker:2024yhq}%
  \BibitemOpen
  \bibfield  {author} {\bibinfo {author} {\bibfnamefont {C.~W.~J.}\
  \bibnamefont {Beenakker}},\ }\bibfield  {title} {\bibinfo {title}
  {{Topologically protected Casimir effect for lattice fermions}},\ }\href
  {https://doi.org/10.1103/PhysRevResearch.6.023058} {\bibfield  {journal}
  {\bibinfo  {journal} {Phys. Rev. Res.}\ }\textbf {\bibinfo {volume} {6}},\
  \bibinfo {pages} {023058} (\bibinfo {year} {2024})},\ \Eprint
  {https://arxiv.org/abs/2402.02477} {arXiv:2402.02477 [quant-ph]} \BibitemShut
  {NoStop}%
\bibitem [{\citenamefont {Fukushima}\ \emph {et~al.}(2019)\citenamefont
  {Fukushima}, \citenamefont {Imaki},\ and\ \citenamefont
  {Qiu}}]{Fukushima:2019sjn}%
  \BibitemOpen
  \bibfield  {author} {\bibinfo {author} {\bibfnamefont {K.}~\bibnamefont
  {Fukushima}}, \bibinfo {author} {\bibfnamefont {S.}~\bibnamefont {Imaki}},\
  and\ \bibinfo {author} {\bibfnamefont {Z.}~\bibnamefont {Qiu}},\ }\bibfield
  {title} {\bibinfo {title} {{Anomalous Casimir effect in axion
  electrodynamics}},\ }\href {https://doi.org/10.1103/PhysRevD.100.045013}
  {\bibfield  {journal} {\bibinfo  {journal} {Phys. Rev. D}\ }\textbf {\bibinfo
  {volume} {100}},\ \bibinfo {pages} {045013} (\bibinfo {year} {2019})},\
  \Eprint {https://arxiv.org/abs/1906.08975} {arXiv:1906.08975 [hep-th]}
  \BibitemShut {NoStop}%
\bibitem [{\citenamefont {Brevik}(2021)}]{Brevik:2021ivj}%
  \BibitemOpen
  \bibfield  {author} {\bibinfo {author} {\bibfnamefont {I.}~\bibnamefont
  {Brevik}},\ }\bibfield  {title} {\bibinfo {title} {{Axion Electrodynamics and
  the Axionic Casimir Effect}},\ }\href
  {https://doi.org/10.3390/universe7050133} {\bibfield  {journal} {\bibinfo
  {journal} {Universe}\ }\textbf {\bibinfo {volume} {7}},\ \bibinfo {pages}
  {133} (\bibinfo {year} {2021})},\ \Eprint {https://arxiv.org/abs/2202.11152}
  {arXiv:2202.11152 [hep-ph]} \BibitemShut {NoStop}%
\bibitem [{\citenamefont {Canfora}\ \emph {et~al.}(2022)\citenamefont
  {Canfora}, \citenamefont {Dudal}, \citenamefont {Oosthuyse}, \citenamefont
  {Pais},\ and\ \citenamefont {Rosa}}]{Canfora:2022xcx}%
  \BibitemOpen
  \bibfield  {author} {\bibinfo {author} {\bibfnamefont {F.}~\bibnamefont
  {Canfora}}, \bibinfo {author} {\bibfnamefont {D.}~\bibnamefont {Dudal}},
  \bibinfo {author} {\bibfnamefont {T.}~\bibnamefont {Oosthuyse}}, \bibinfo
  {author} {\bibfnamefont {P.}~\bibnamefont {Pais}},\ and\ \bibinfo {author}
  {\bibfnamefont {L.}~\bibnamefont {Rosa}},\ }\bibfield  {title} {\bibinfo
  {title} {{The Casimir effect in chiral media using path integral
  techniques}},\ }\href {https://doi.org/10.1007/JHEP09(2022)095} {\bibfield
  {journal} {\bibinfo  {journal} {JHEP}\ }\textbf {\bibinfo {volume} {09}},\
  \bibinfo {pages} {095}},\ \Eprint {https://arxiv.org/abs/2207.09175}
  {arXiv:2207.09175 [hep-th]} \BibitemShut {NoStop}%
\bibitem [{\citenamefont {Oosthuyse}\ and\ \citenamefont
  {Dudal}(2023)}]{Oosthuyse:2023mbs}%
  \BibitemOpen
  \bibfield  {author} {\bibinfo {author} {\bibfnamefont {T.}~\bibnamefont
  {Oosthuyse}}\ and\ \bibinfo {author} {\bibfnamefont {D.}~\bibnamefont
  {Dudal}},\ }\bibfield  {title} {\bibinfo {title} {{Interplay between chiral
  media and perfect electromagnetic conductor plates: Repulsive vs. attractive
  Casimir force transitions}},\ }\href
  {https://doi.org/10.21468/SciPostPhys.15.5.213} {\bibfield  {journal}
  {\bibinfo  {journal} {SciPost Phys.}\ }\textbf {\bibinfo {volume} {15}},\
  \bibinfo {pages} {213} (\bibinfo {year} {2023})},\ \Eprint
  {https://arxiv.org/abs/2301.12870} {arXiv:2301.12870 [hep-th]} \BibitemShut
  {NoStop}%
\bibitem [{\citenamefont {Favitta}\ \emph {et~al.}(2023)\citenamefont
  {Favitta}, \citenamefont {Brevik},\ and\ \citenamefont
  {Chaichian}}]{Favitta:2023hlx}%
  \BibitemOpen
  \bibfield  {author} {\bibinfo {author} {\bibfnamefont {A.~M.}\ \bibnamefont
  {Favitta}}, \bibinfo {author} {\bibfnamefont {I.~H.}\ \bibnamefont
  {Brevik}},\ and\ \bibinfo {author} {\bibfnamefont {M.~M.}\ \bibnamefont
  {Chaichian}},\ }\bibfield  {title} {\bibinfo {title} {{Axion electrodynamics:
  Green\textquoteright{}s functions, zero-point energy and optical activity}},\
  }\href {https://doi.org/10.1016/j.aop.2023.169396} {\bibfield  {journal}
  {\bibinfo  {journal} {Ann. Phys. (Amsterdam)}\ }\textbf {\bibinfo {volume}
  {455}},\ \bibinfo {pages} {169396} (\bibinfo {year} {2023})},\ \Eprint
  {https://arxiv.org/abs/2302.13129} {arXiv:2302.13129 [hep-th]} \BibitemShut
  {NoStop}%
\bibitem [{\citenamefont {Ema}\ \emph {et~al.}(2023)\citenamefont {Ema},
  \citenamefont {Hazumi}, \citenamefont {Iizuka}, \citenamefont {Mukaida},\
  and\ \citenamefont {Nakayama}}]{Ema:2023kvw}%
  \BibitemOpen
  \bibfield  {author} {\bibinfo {author} {\bibfnamefont {Y.}~\bibnamefont
  {Ema}}, \bibinfo {author} {\bibfnamefont {M.}~\bibnamefont {Hazumi}},
  \bibinfo {author} {\bibfnamefont {H.}~\bibnamefont {Iizuka}}, \bibinfo
  {author} {\bibfnamefont {K.}~\bibnamefont {Mukaida}},\ and\ \bibinfo {author}
  {\bibfnamefont {K.}~\bibnamefont {Nakayama}},\ }\bibfield  {title} {\bibinfo
  {title} {{Zero Casimir force in axion electrodynamics and the search for a
  new force}},\ }\href {https://doi.org/10.1103/PhysRevD.108.016009} {\bibfield
   {journal} {\bibinfo  {journal} {Phys. Rev. D}\ }\textbf {\bibinfo {volume}
  {108}},\ \bibinfo {pages} {016009} (\bibinfo {year} {2023})},\ \Eprint
  {https://arxiv.org/abs/2302.14676} {arXiv:2302.14676 [hep-ph]} \BibitemShut
  {NoStop}%
\bibitem [{\citenamefont {Chernodub}\ \emph {et~al.}(2018)\citenamefont
  {Chernodub}, \citenamefont {Goy}, \citenamefont {Molochkov},\ and\
  \citenamefont {Nguyen}}]{Chernodub:2018pmt}%
  \BibitemOpen
  \bibfield  {author} {\bibinfo {author} {\bibfnamefont {M.~N.}\ \bibnamefont
  {Chernodub}}, \bibinfo {author} {\bibfnamefont {V.~A.}\ \bibnamefont {Goy}},
  \bibinfo {author} {\bibfnamefont {A.~V.}\ \bibnamefont {Molochkov}},\ and\
  \bibinfo {author} {\bibfnamefont {H.~H.}\ \bibnamefont {Nguyen}},\ }\bibfield
   {title} {\bibinfo {title} {{Casimir Effect in Yang-Mills Theory in
  $D=2+1$}},\ }\href {https://doi.org/10.1103/PhysRevLett.121.191601}
  {\bibfield  {journal} {\bibinfo  {journal} {Phys. Rev. Lett.}\ }\textbf
  {\bibinfo {volume} {121}},\ \bibinfo {pages} {191601} (\bibinfo {year}
  {2018})},\ \Eprint {https://arxiv.org/abs/1805.11887} {arXiv:1805.11887
  [hep-lat]} \BibitemShut {NoStop}%
%%CITATION = ARXIV:1805.11887;%%
\bibitem [{\citenamefont {Chernodub}\ \emph {et~al.}(2019)\citenamefont
  {Chernodub}, \citenamefont {Goy},\ and\ \citenamefont
  {Molochkov}}]{Chernodub:2018aix}%
  \BibitemOpen
  \bibfield  {author} {\bibinfo {author} {\bibfnamefont {M.~N.}\ \bibnamefont
  {Chernodub}}, \bibinfo {author} {\bibfnamefont {V.~A.}\ \bibnamefont {Goy}},\
  and\ \bibinfo {author} {\bibfnamefont {A.~V.}\ \bibnamefont {Molochkov}},\
  }\bibfield  {title} {\bibinfo {title} {{Phase structure of lattice Yang-Mills
  theory on ${\mathbb T}^2 \times {\mathbb R}^2$}},\ }\href
  {https://doi.org/10.1103/PhysRevD.99.074021} {\bibfield  {journal} {\bibinfo
  {journal} {Phys. Rev. D}\ }\textbf {\bibinfo {volume} {99}},\ \bibinfo
  {pages} {074021} (\bibinfo {year} {2019})},\ \Eprint
  {https://arxiv.org/abs/1811.01550} {arXiv:1811.01550 [hep-lat]} \BibitemShut
  {NoStop}%
%%CITATION = ARXIV:1811.01550;%%
\bibitem [{\citenamefont {Kitazawa}\ \emph {et~al.}(2019)\citenamefont
  {Kitazawa}, \citenamefont {Mogliacci}, \citenamefont {Kolb\'e},\ and\
  \citenamefont {Horowitz}}]{Kitazawa:2019otp}%
  \BibitemOpen
  \bibfield  {author} {\bibinfo {author} {\bibfnamefont {M.}~\bibnamefont
  {Kitazawa}}, \bibinfo {author} {\bibfnamefont {S.}~\bibnamefont {Mogliacci}},
  \bibinfo {author} {\bibfnamefont {I.}~\bibnamefont {Kolb\'e}},\ and\ \bibinfo
  {author} {\bibfnamefont {W.~A.}\ \bibnamefont {Horowitz}},\ }\bibfield
  {title} {\bibinfo {title} {{Anisotropic pressure induced by finite-size
  effects in SU(3) Yang-Mills theory}},\ }\href
  {https://doi.org/10.1103/PhysRevD.99.094507} {\bibfield  {journal} {\bibinfo
  {journal} {Phys. Rev. D}\ }\textbf {\bibinfo {volume} {99}},\ \bibinfo
  {pages} {094507} (\bibinfo {year} {2019})},\ \Eprint
  {https://arxiv.org/abs/1904.00241} {arXiv:1904.00241 [hep-lat]} \BibitemShut
  {NoStop}%
%%CITATION = ARXIV:1904.00241;%%
\bibitem [{\citenamefont {Chernodub}\ \emph {et~al.}(2023)\citenamefont
  {Chernodub}, \citenamefont {Goy}, \citenamefont {Molochkov},\ and\
  \citenamefont {Tanashkin}}]{Chernodub:2023dok}%
  \BibitemOpen
  \bibfield  {author} {\bibinfo {author} {\bibfnamefont {M.~N.}\ \bibnamefont
  {Chernodub}}, \bibinfo {author} {\bibfnamefont {V.~A.}\ \bibnamefont {Goy}},
  \bibinfo {author} {\bibfnamefont {A.~V.}\ \bibnamefont {Molochkov}},\ and\
  \bibinfo {author} {\bibfnamefont {A.~S.}\ \bibnamefont {Tanashkin}},\
  }\bibfield  {title} {\bibinfo {title} {{Boundary states and non-Abelian
  Casimir effect in lattice Yang-Mills theory}},\ }\href
  {https://doi.org/10.1103/PhysRevD.108.014515} {\bibfield  {journal} {\bibinfo
   {journal} {Phys. Rev. D}\ }\textbf {\bibinfo {volume} {108}},\ \bibinfo
  {pages} {014515} (\bibinfo {year} {2023})},\ \Eprint
  {https://arxiv.org/abs/2302.00376} {arXiv:2302.00376 [hep-lat]} \BibitemShut
  {NoStop}%
\bibitem [{\citenamefont {Cao}\ and\ \citenamefont
  {Huang}(2016)}]{Cao:2016fby}%
  \BibitemOpen
  \bibfield  {author} {\bibinfo {author} {\bibfnamefont {G.}~\bibnamefont
  {Cao}}\ and\ \bibinfo {author} {\bibfnamefont {A.}~\bibnamefont {Huang}},\
  }\bibfield  {title} {\bibinfo {title} {{Solitonic modulation and Lifshitz
  point in an external magnetic field within Nambu\textendash{}Jona-Lasinio
  model}},\ }\href {https://doi.org/10.1103/PhysRevD.93.076007} {\bibfield
  {journal} {\bibinfo  {journal} {Phys. Rev. D}\ }\textbf {\bibinfo {volume}
  {93}},\ \bibinfo {pages} {076007} (\bibinfo {year} {2016})},\ \Eprint
  {https://arxiv.org/abs/1601.03493} {arXiv:1601.03493 [nucl-th]} \BibitemShut
  {NoStop}%
\bibitem [{\citenamefont {Son}\ and\ \citenamefont
  {Stephanov}(2008)}]{Son:2007ny}%
  \BibitemOpen
  \bibfield  {author} {\bibinfo {author} {\bibfnamefont {D.~T.}\ \bibnamefont
  {Son}}\ and\ \bibinfo {author} {\bibfnamefont {M.~A.}\ \bibnamefont
  {Stephanov}},\ }\bibfield  {title} {\bibinfo {title} {{Axial anomaly and
  magnetism of nuclear and quark matter}},\ }\href
  {https://doi.org/10.1103/PhysRevD.77.014021} {\bibfield  {journal} {\bibinfo
  {journal} {Phys. Rev. D}\ }\textbf {\bibinfo {volume} {77}},\ \bibinfo
  {pages} {014021} (\bibinfo {year} {2008})},\ \Eprint
  {https://arxiv.org/abs/0710.1084} {arXiv:0710.1084 [hep-ph]} \BibitemShut
  {NoStop}%
\bibitem [{\citenamefont {Brauner}\ and\ \citenamefont
  {Yamamoto}(2017)}]{Brauner:2016pko}%
  \BibitemOpen
  \bibfield  {author} {\bibinfo {author} {\bibfnamefont {T.}~\bibnamefont
  {Brauner}}\ and\ \bibinfo {author} {\bibfnamefont {N.}~\bibnamefont
  {Yamamoto}},\ }\bibfield  {title} {\bibinfo {title} {{Chiral soliton lattice
  and charged pion condensation in strong magnetic fields}},\ }\href
  {https://doi.org/10.1007/JHEP04(2017)132} {\bibfield  {journal} {\bibinfo
  {journal} {JHEP}\ }\textbf {\bibinfo {volume} {04}},\ \bibinfo {pages}
  {132}},\ \Eprint {https://arxiv.org/abs/1609.05213} {arXiv:1609.05213
  [hep-ph]} \BibitemShut {NoStop}%
\bibitem [{\citenamefont {Armitage}\ \emph {et~al.}(2018)\citenamefont
  {Armitage}, \citenamefont {Mele},\ and\ \citenamefont
  {Vishwanath}}]{Armitage:2017cjs}%
  \BibitemOpen
  \bibfield  {author} {\bibinfo {author} {\bibfnamefont {N.~P.}\ \bibnamefont
  {Armitage}}, \bibinfo {author} {\bibfnamefont {E.~J.}\ \bibnamefont {Mele}},\
  and\ \bibinfo {author} {\bibfnamefont {A.}~\bibnamefont {Vishwanath}},\
  }\bibfield  {title} {\bibinfo {title} {{Weyl and Dirac Semimetals in three
  dimensional solids}},\ }\href {http://doi.org/10.1103/RevModPhys.90.015001}
  {\bibfield  {journal} {\bibinfo  {journal} {Rev. Mod. Phys.}\ }\textbf
  {\bibinfo {volume} {90}},\ \bibinfo {pages} {015001} (\bibinfo {year}
  {2018})},\ \Eprint {https://arxiv.org/abs/1705.01111} {arXiv:1705.01111
  [cond-mat.str-el]} \BibitemShut {NoStop}%
\bibitem [{\citenamefont {Lv}\ \emph {et~al.}(2021)\citenamefont {Lv},
  \citenamefont {Qian},\ and\ \citenamefont {Ding}}]{Lv:2021oam}%
  \BibitemOpen
  \bibfield  {author} {\bibinfo {author} {\bibfnamefont {B.~Q.}\ \bibnamefont
  {Lv}}, \bibinfo {author} {\bibfnamefont {T.}~\bibnamefont {Qian}},\ and\
  \bibinfo {author} {\bibfnamefont {H.}~\bibnamefont {Ding}},\ }\bibfield
  {title} {\bibinfo {title} {{Experimental perspective on three-dimensional
  topological semimetals}},\ }\href
  {https://doi.org/10.1103/RevModPhys.93.025002} {\bibfield  {journal}
  {\bibinfo  {journal} {Rev. Mod. Phys.}\ }\textbf {\bibinfo {volume} {93}},\
  \bibinfo {pages} {025002} (\bibinfo {year} {2021})}\BibitemShut {NoStop}%
\end{thebibliography}%

\end{document}